\documentclass[fleqn,10pt]{wlscirep}
\usepackage[utf8]{inputenc}
\usepackage[T1]{fontenc}
\usepackage{diagbox}
\usepackage{makecell}
\usepackage{lscape}
\title{On the current failure---but bright future---of topology-driven biological network alignment}

\author[1]{Siyue Wang}
\author[1]{Xiaoyin Chen}
\author[1]{Brent J. Frederisy}
\author[1]{Benedict A. Mbakogu}
\author[1]{Amy D. Kanne}
\author[1]{Pasha Khosravi}
\author[1,*]{Wayne B. Hayes}
\affil[1]{Department of Computer Science, University of California, Irvine CA 92697-3435, USA}
\affil[*]{corresponding: whayes@uci.edu}

\begin{document}

\begin{abstract}
Since the function of a protein is defined by its interaction partners, and since we expect similar interaction patterns across species, the alignment of protein-protein interaction (PPI) networks between species, based on network topology alone, should uncover functionally related proteins across species.
Surprisingly, despite the publication of more than {\em fifty} algorithms aimed at performing PPI network alignment, few have demonstrated a statistically significant link between network topology and functional similarity, and {\em none} have demonstrated that orthologs can be recovered using network topology alone.
We find that the major contributing factors are to this surprising failure are: (i) edge densities in most currently available experimental PPI networks are {\em demonstrably} too low to expect topological network alignment to succeed; (ii) in the few cases where the edge densities are high enough, some measures of topological similarity easily uncover functionally similar proteins while others do not; and (iii) most network alignment algorithms to date perform poorly at optimizing even their own topological objective functions, hampering their ability to use topology effectively.  We demonstrate that SANA---the {\it Simulated Annealing Network Aligner}---significantly outperforms existing aligners at optimizing their own objective functions, even achieving near-optimal solutions when optimal solution is known. We offer the first demonstration of global network alignments based on topology alone that align functionally similar proteins with $p$-values in some cases below $10^{-300}$. We predict that topological network alignment has a bright future as edge densities increase towards the value where good alignments become possible. We demonstrate that when enough common topology is present at high enough edge densities---for example in the recent, partly synthetic networks of the {\it Integrated Interaction Database}---topological network alignment {\em easily} recovers most orthologs, paving the way towards high-throughput functional prediction based on topology-driven network alignment.

\noindent{\\ \bf keywords}: network alignment, graph theory, information theory, orthologs, protein-protein interaction, stochastic sampling, optimization, Gene Ontology, GO term prediction
\end{abstract}
\maketitle
\section{Introduction}\label{sec:Intro}

\noindent ``No protein is an island unto itself.''
    
    -- {\it Physical Biology of the Cell}, 2nd Ed. (2013), Phillips {\it et al.} \cite[Figure 4.20]{phillips2013physical}.

\vspace{2mm}
\noindent ``A complex network of protein interactions underlies cell function.''
    
    -- {\it Molecular Biology of the Cell}, 6th Ed., (2015), Alberts {\it et al.} \cite[p. 166]{alberts2015molecular}.


\subsection{Motivation}\label{sec:motivation}

The statements above---taken from reputable textbooks less than a decade old---arise from one of the canonical assumptions of modern biology: that a protein's function is intimately tied to its set of interaction partners, as well as the broader context of the network in which they are all embedded. Thus, to understand cellular function requires understanding its network of protein-protein interactions (PPIs).
Luckily, just as with individual proteins, the PPI network of one species is not an island unto itself: proteins come from genes, and there is significant genetic similarity across species. For example, the genomes of humans and chimps are about 96--97\% identical \cite{sequencing2005initial}; about $80\%$ of mouse genes have a direct 1-to-1 ortholog in human \cite{pennacchio2003insights}; and life-critical genes can be near-identical even across species as far apart as yeast and human \cite{pearson2013introduction}. Given the significant genetic sequence similarity between species, and since proteins arise from genes, we expect that proteins across different species that arise from orthologous genes will perform similar functions---and since function is {\em defined} by one's interaction partners, orthologous proteins across closely-related species---or even across distantly related species that perform critical ancestral functions---should often share interaction partners. In the parlance of network analysis, {\em we expect that PPI networks across species will exhibit a significant amount of topological network similarity}.

Most computational prediction of protein function today is sequence-based. Sequence data are ubiquitous and cheap to acquire, and there is a robust and sophisticated ecosystem of algorithms for their comparison and alignment. Some methods attempt to predict function directly from sequence (eg., \cite{kulmanov2020deepgoplus}), while others use longer chains of reasoning including sequence, structure, binding domains, interfaces, and compatible interaction partners (eg., \cite{zhang2017cofactor}). Existing functional information anywhere along this path may implicate the original protein with a similar function via the ``guilt-by-association'' principle. However, this road from sequence analysis to function prediction is long, complex, and error-prone: there are examples of proteins with no sequence similarity having near-identical function \cite{furuse1998claudin,schlicker2006new}; examples of proteins with identical sequence having multiple, completely different functions\cite{kabsch1984use,morrone2011denatured}; and even examples of structural but not functional similarity \cite{madsen1999psoriasis}.

In contrast, PPI interactions can be {\em directly} measured via yeast-two-hybrid and other methods \cite{VidalPPI01}, allowing direct detection of interactions and allowing a much shorter path to predictions based on guilt-by-association. Finally, since functional similarity can exist even in the absence of sequence similarity, network-based methods may be able to predict function in cases where sequence cannot. Furthermore, functions are naturally encoded using a network schema---witness how the word ``pathway'' is commonly used to describe how a set of entities coordinate their interactions to perform some function. 

Given the growing amount of network data, and the expected similarity of PPI networks across species, it is surprising that, despite the existence of over {\em fifty} published papers attempting to solve the PPI network alignment problem (see reviews such as \cite{Milenkovic:2013:GNA:2506583.2508968,clark2014comparison,MamanoHayesSANA}), none of them have been able to use topology {\em alone} to recover cross-species links between proteins with known similar function such as orthologs, and only a few have been able to show a statistically significant link between Gene Ontology (GO)-based function and topology-based network alignment---unless sequence information is also used\cite{GRAAL,MIGRAAL,faisal2014global,DavisPrzulj2015TopoFunction,gaudelet2018higher,malod2019functional}. As a result, almost all network alignment algorithms today use sequence similarity between protein pairs to guide network alignments. Even more worrisome, when trying to balance the objectives of aligning proteins using network topology vs. aligning proteins that have similar sequences, a negative correlation has been widely observed: topology-weighted network alignments are far less able to align functionally similar proteins than sequence-similarity-weighted network alignments.
The connection between topological network similarity and functional similarity has been so tenuous for so long that many authors now refer explicitly to a ``sequence-topology trade-off'', effectively abandoning the promise of independently using the expected common network topology across species to uncover functional similarity \cite{Ulign,gligorijevic2015fuse,NATALIE,NATALIE2,kalecky2018primalign,alberich2019alignet}.

While it is clear that the best functional predictions will ultimately come from integrating all relevant sources of data---sequence, structure, and network connections---each source of data being integrated should first pass a stringent “litmus test” demonstrating its ability, alone or in conjunction with other data, to improve the result. The widespread belief in the ``sequence-topology trade-off'' makes it clear that network topology has not yet passed this litmus test. Our goal in this paper is to explain this surprising failure.

\subsection{The Sequence-Topology ``Trade-off''}\label{sec:seq-topo-trade-off}

To date, the only success in identifying biologically relevant common subnetworks of interactions across species using topology {\em alone} have used {\it graphlets} and their derivatives \cite{DavisPrzulj2015TopoFunction,gaudelet2018higher,malod2019functional}, which have also seen modest success in detecting functional similarity in network alignments\cite{GRAAL,H-GRAAL,MIGRAAL,faisal2014global}. To our knowledge, in order to gain biological relevance, {\em every} other network alignment algorithm applied to PPI networks has had to guide the alignment with objective functions that include protein-pair sequence similarities, in order to encourage sequence---and thus functionally---similar protein pairs to align to each other.
This fact has led to a widespread belief in a so-called ``sequence-topology trade-off''\cite{Ulign,alberich2019alignet}, so that virtually every alignment algorithm either imposes sequence-based restrictions\cite{NATALIE,NATALIE2,kalecky2018primalign,GEDEVO,CytoGEDEVO,zhu2017gmalign,elmsallati2018index,xie2016adaptive,PROPER,yasar2018iterative,tame}, or has a literal trade-off in an objective function of the form $\alpha T(a) + (1-\alpha) S(a)$, where $a$ is an alignment, $T(a)$ measures the topological similarity exposed by the alignment according to some topological measure, $S(a)$ is an objective based on sequence similarity, and $\alpha$ is a balancing parameter that makes the trade-off explicit\cite{GHOST,alkan2015sipan,HubAlign,Ulign,MIGRAAL,LGRAAL,Isorank,mir2017index,optnetalign,PISwap,modulealign,NETAL,WAVE,MAGNA,MAGNA++,GREAT,gong2015global,SPINAL}. Virtually all methods that use the explicit trade-off have found that functional similarity is positively correlated with the weight given to sequence, and negatively correlated with the weight given to topology \cite{GHOST,clark2014comparison,HubAlign,PROPER,Ulign}.

\subsection{Possible reasons for the failure}\label{sec:whyFailure}

The failure of purely topology-driven network alignments to uncover functional similarity may be due to several factors. In order of decreasing pessimism, possible reasons include:
\begin{enumerate}
    \item[R0:] Perhaps there is little or no correlation between common network topology and common protein function/orthology across species. In addition to being severely at odds with most of modern biology, in the Supplementary we explicitly demonstrate the astronomically low probability of this case (cf. Table \ref{tab:interologs}).
    
    \item[R1:] Perhaps there is a strong topology-function relationship in the {\em true} PPI networks, but our current data are too noisy or incomplete \cite{ideker2012differential}, or network coverage too uneven \cite{rolland2014proteome} for {\em any} algorithm to detect the underlying common topology. We address this hypothesis by leveraging Information Theory (cf. \S\ref{sec:infoTheory}).

    \item[R2:] Perhaps current data {\em does} exhibit, to some extent, the expected underlying topology-function relationships, but currently available measures of topological network similarity fail to capture this relationship. For example, some objective functions may {\it saturate} before others (cf. \S\ref{sec:saturation}), meaning they are inherently incapable of detecting relevant topological features even if those features exist. We address this hypothesis by measuring the ability of various objective functions to recover known functional relationships (cf. \S\ref{sec:whichMeasures}).

    \item[R3:] Perhaps current topological measures {\em are} capable of exposing the expected functional similarities, but existing algorithms do not adequately optimize their topological objective functions due to the difficulty in optimally solving the NP-complete problems that arise in topological network alignment. We address this in \SS \ref{sec:SANA-is-optimal}, \ref{sec:WeBeat}.
\end{enumerate}
Hypothesis R0 may be dismissed virtually out-of-hand: it is assumed by modern medicine\cite{thomas2012use}, contradicts our understanding of molecular evolution\cite{lockhart1994recovering}, is at odds with recent explicit demonstrations of functional identity between orthologs of distantly related species\cite{kachroo2015systematic}, and is inconsistent with the astronomically significant $p$-values of interologs (common interactions between species---cf. Supplementary Table \ref{tab:interologs}).

\subsection{Contribution}\label{sec:contribution}
In this paper, we attempt to disentangle the latter three hypotheses, but in the order R1, R3, and finally R2. In particular, we walk the reader through the following observations:

{\bf Regarding R1}: In \S \ref{sec:infoTheory} we propose a novel hypothesis based on Information Theory that places a lower bound on the amount of information required to uniquely specify one alignment (for example, a ``correct'' one) out of all possible alignments (cf. Equation \ref{eq:searchSpace}). We then suggest ways to estimate the formal information {\em content} of PPI input networks---essentially, their edge density---to test if there is enough information to satisfy the requirement. Though quantifying network information content is extremely difficult and requires significant theoretical development, we show that a preliminary closed-form estimate agrees with existing literature as well as our own computational experiments to within a few percent in edge density. We note that no pair of existing BioGRID networks have edge densities that satisfy the hypothesis, suggesting that {\em no} algorithm based on topology alone is capable of producing high-quality global network alignments between existing networks. Finally, we show that the recent, partially synthetic IID networks \cite{kotlyar2018iid} have edge densities well above the hypothesized bound, and that excellent alignments can easily be found using several existing measures of topological similarity such as EC \cite{GRAAL} or $S^3$ from MAGNA\cite{MAGNA}.

{\bf Regarding R3}: We demonstrate that nine of the most recent, competitive global network aligners all produce sub-optimal alignments according to their own objective functions, by showing that SANA---the {\it Simulated Annealing Network Aligner}---provides far more optimal values of these objectives than the original aligners do themselves (cf. Figure \ref{fig:avgsif}). We further show that, given enough simulated annealing time (an hour is enough for existing PPI networks), SANA is able to produce alignments with near-optimal values of the objective function in all cases where the optimal value is known. 

{\bf Regarding R2}: Since SANA is a random search algorithm, each alignment is different even though all alignments achieve almost the same (and presumably near-optimal) scores of the desired objective function. This means that SANA can be used to {\em randomly sample} alignments from the frontier consisting of near-optimal alignments according to any given objective function. This allows us to compare the alignments produced by different topological objectives, confident that the alignments being compared are among the best possible for each chosen topological objective.
We compare these objectives for their ability to (a) align proteins that share a statistically significant number of GO terms (\S\ref{sec:GOrec}), and (b) recover known orthologs between species in current PPI data (\S\ref{sec:orthologs}). Finally, we argue that no topological measure is {\em always} best, by demonstrating that different measures of topological similarity are applicable in different circumstances (cf. \S \ref{sec:EC-vs-S3}).


\section{Preliminaries}

\subsection{Pairwise Global Network Alignment (PGNA)}\label{sec:PGNA}

\begin{figure}[tb]
  \centering
  \includegraphics[width=0.5\linewidth]{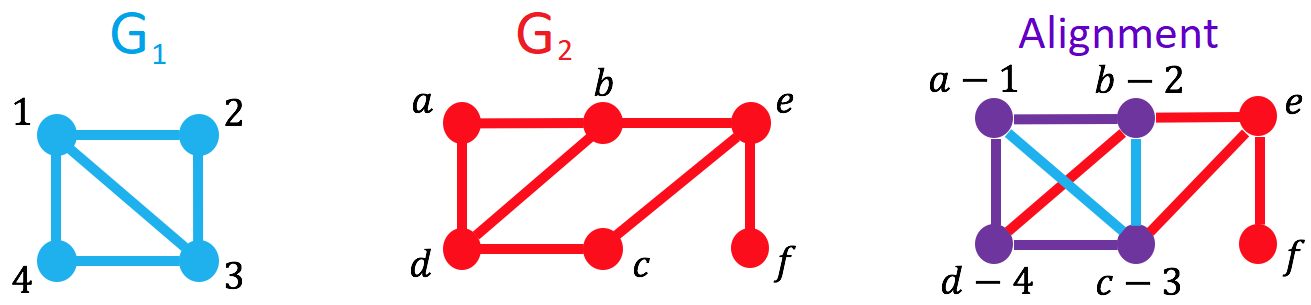}
  \caption{\small A schematic depiction of a 1-to-1 Pairwise Global Network Alignment (PGNA). The input graphs are $G_1$ (blue, with fewer nodes), and $G_2$ (red, with more nodes). The alignment can be depicted itself as a network: aligned nodes and edges are purple (depicting a mix of red and blue), while unaligned nodes and edges retain their original color. Note there are no blue nodes in the alignment because, being a 1-to-1 mapping, every blue node must be aligned to a red one, producing purple. Two commonly used topological measures are EC and $S^3$, which in our schematic can be computed as
  $EC=|\mbox{purple edges in the alignment}|/|\mbox{all edges of }G_1|,$ 
  while $S^3=|\mbox{purple edges in the alignment}|/|\mbox{edges of all colors between purple nodes}|.$ Thus, the depicted alignment has EC$=\frac{3}{5}$ and $S^3=\frac{3}{6}.$
  }
  \label{fig:NetAlign}
\end{figure}

For the purposes of this paper, we define an alignment as a 1-to-1 mapping between the nodes of exactly two networks. More formally, let $G_1=(V_1,E_1)$ and $G_2=(V_2,E_2)$ be two graphs (networks) with node sets $V_i$ and edge sets $E_i,i=1,2$, and assume without loss of generality that $|V_1|\leq|V_2|$. A {\it Pairwise Global Network Alignment} (PGNA) is a 1-to-1 mapping $a$ from $V_1$ to $V_2$. (We show later how to extend it to a many-to-many alignment.) Given a particular alignment $a$, assume we have some {\em objective function} $f(a)$ that measures the ``quality'' of $a$ in some fashion. Our goal is to find an alignment $a^*$ that maximizes the objective:
\begin{equation}
    f(a^*)\ge f(a)\;\forall a. \label{eq:optimalA}
\end{equation}
Figure \ref{fig:NetAlign} depicts a schematic diagram of a small PGNA, and schematically depicts how the commonly used topological measures EC and $S^3$ are computed. In Figure \ref{fig:NetAlign}, we use colors to depict the original networks: blue for $G_1$, and red for $G_2$. Though there are many possible alignments between them, Figure \ref{fig:NetAlign} depicts the visually obvious one where the square 1-2-3-4 of $G_1$ has simply been moved to the right until it is ``on top of'' the a-b-c-d square in $G_2$. The resulting alignment is depicted at the far right, where aligned nodes and edges are depicted in purple (a mix of blue and red), and unaligned nodes and edges retain their original color. Note that this is not the {\em best} alignment: rotating $G_1$ 90 degrees in either direction causes its ``diagonal'' edge to align with the corresponding diagonal edge in $G_2$; this causes the EC to increase from 3/5 to 4/5, and the $S^3$ score to increase from 3/6 to 4/5---the denominator of $S^3$ decreases by one since we gained one purple edge but we {\em lost} two: one blue, one red.

\subsection{Measuring topological similarity}\label{sec:measures}
\begin{table*}[hbt]
\begin{center}
\begin{tabular}{|| p{3.0cm} p{3.5cm} p{9cm} ||} 
 \hline
 Objective Function & Used By & Brief Description  \\ [0.5ex] 
 \hline\hline
 EC\cite{GRAAL} & GRAAL, MAGNA++, OptNetAlign, SANA & The percent of edges in the smaller network that align to edges in the bigger one.\\ 
 \hline
 Symmetric Substructure Score ($S^3$)\cite{MAGNA} & MAGNA(++), OptNetAlign, SANA & Similar to EC, but symmetric and measured only on the subgraph induced by the node alignment. \\
 \hline
 GDV Signature Sim.\cite{milenkovic2008uncovering} & GRAAL, MAGNA++, SANA & Node-based similarity measure comparing local graphlet counts around a node. \\ 
 \hline
 L-GRAAL GDV Sim.\cite{LGRAAL} & L-GRAAL, SANA & L-GRAAL's modified GDV similarity, normalized differently and using only graphlets with 4 or fewer nodes.\\
 \hline
 Spectral Signature Sim.\cite{GHOST} & GHOST, SANA & Based on the distribution of eigenvectors of normalized Laplacian matrices of surrounding subgraphs of each node.\\
 \hline
 Edge Graphlet Vector Sim.\cite{GREAT} & GREAT, SANA & Similar to Graphlet similarity, but scores pairs of edges instead of pairs of nodes. \\
 \hline
 Importance\cite{HubAlign} & HubAlign, ModuleAlign, SANA & Ranks nodes recursively based on their degree and the Importance of adjacent nodes; inspired by Google's PageRank\cite{page1999pagerank}.\\
 \hline
 Greedy EC + Degree Diff. & PROPER, INDEX, SANA & Greedily built ``local EC,'' ties broken by degree difference.\\
 \hline
 GDV-Weighted EC (WEC)\cite{WAVE} & WAVE, SANA & Overlapping edges receive a score based on the average of the graphlet degree difference scores of their endpoints.\\
 \hline
\end{tabular}
\end{center}
\caption{Objective Functions and Alignment Algorithms Compared.}
\label{tab:objectives}
\end{table*}

There are many ways to measure topological similarity. In the context of network alignment, similarity is defined as a function of the alignment: just as with sequence similarity, we can only measure topological ``similarity’’ once an alignment is specified, and the goal is to find an alignment which maximizes the topological similarity.
Table \ref{tab:objectives} lists (to our knowledge) all measures of topological similarity introduced in the past 15 years in the context of pairwise PPI network alignment, along with the algorithms that introduced and/or used them; these are the measures we will be studying below for their ability to recover functionally similar regions and/or orthologs between species.

\subsection{SANA: The Simulated Annealing Network Aligner}\label{sec:SANA}
An alignment algorithm consists of two orthogonal components: first, a {\it objective function} $M$ that measures the topological similarity $M(A)$ observed in a given alignment $A:V_1\rightarrow V_2$; and second, a {\it search algorithm} designed to search the space of all possible alignments looking for ones that score well according to $M$.

Many existing publications convolve the search algorithm and objective function, making it difficult to discern which is more responsible for the quality (good or bad) of a given alignment. Our algorithm, SANA\cite{MamanoHayesSANA}, {\it Simulated Annealing Network Aligner}, clearly separates the two and uses simulated annealing (SA\cite{kirkpatrick1983optimization}) as the search algorithm. (SANA is available on GitHub at \url{https://github.com/waynebhayes/SANA}.) SA has a rich history of successful application to NP-complete problems across a wide array of application domains\cite{mitra1985convergence,romeo1986efficient,autosa,szu1987FSA,meise1998convergence,Strens2003,suman2006survey,dowsland2012simulated,AguiareOliveiraJunior2012,zhan2016list}. One important aspect of SA is the choice of {\it temperature schedule}; SANA automatically determines effective temperature limits using an algorithm detailed elsewhere\cite{SANAtemperature}. The only significant unknown is the amount of CPU time one should spend traversing the temperature range; for now we determine this empirically, by performing longer-and-longer runs, each from scratch, until the final score stops increasing. For the networks in this paper, we have empirically determined that 1 hour is sufficient on the machines used in this study (a cluster of 100 identical machines, each with 96GB of RAM and a 3.33GHz, 24-core Intel X5680 CPU).

\section{Testing Hypotheses R1 and R3}\label{sec:testingR1R3}

Recall  briefly our hypotheses for the current failure of network alignment: {\bf R1} is that networks contain too little topological information for good alignments to be found; {\bf R2} is that current topological measures of similarity are incapable of finding {\em biologically} relevant alignments because they are somehow measuring the ``wrong’’ type of topology; and {\bf R3} hypothesizes that current alignment algorithms, even if optimizing biologically relevant measures of topology, are producing alignments that are far from their chosen measure’s optimal value, artificially degrading the objective function's ability to find relevant biology.

\subsection{Addressing R1 using Information theory and edge density}\label{sec:infoTheory}

Let $\mathcal{A}$ be the set of all possible alignments---i.e., $\mathcal{A}$ defines the {\it search space}. If $G_1$ and $G_2$ have $n_1\le n_2$ nodes, respectively, the total number of possible 1-to-1 alignments is
\begin{equation}
    |\mathcal{A}|={n_2 \choose n_1}n_1! = \frac{n_2!}{(n_2-n_1)!}. \label{eq:searchSpace}
\end{equation}
If we give each element of $\mathcal{A}$ a unique integer identifier between 0 and $|\mathcal{A}|-1$, then elementary Information Theory (see Methods--Information theory, \S\ref{sec:InfoTheoryIntro}) tells us that the minimum amount of information required to uniquely specify one alignment, measured in bits\cite{shannon1948mathematical}, is
\begin{equation}
    \mbox{Netbits($n_1,n_2$)}
    = \log_2(|\mathcal{A}|) 
    = \log_2\left(\frac{n_2!}{(n_2-n_1)!}\right)
    = \log_2\left(\prod_{k=n_2-n_1+1}^{n_2} k \right)
    = \sum_{k=n_2-n_1+1}^{n_2} \log_2 k.
    \label{eq:Netbits}    
\end{equation}
For typical PPI networks, the log of the values of $k$ in the sum are easily computed, and so the exact value of the information requirement is easily computed (see Methods--Equation \ref{eq:searchSpaceBits}).

For particular values of $n_1$ and $n_2$, we refer to Equation \ref{eq:Netbits} as Netbits($n_1,n_2$); it specifies the absolute minimum amount of input information required to uniquely identify an alignment of $n_1$ nodes to $n_2$ nodes---for example, one that we may deem as the ``correct'' alignment. In a topological alignment of two networks, the only input information we have is the two networks, which can be viewed as two lists of edges. The way we {\em leverage} that information is encoded in (i) the topological objective function that measures the similarity exposed between the two networks by any given alignment, and (ii) the algorithm used to build the alignment, guided at least in part by the topological objective function. We will discuss the algorithms and objectives below, but for now we are tasked with determining how much raw information exists in two networks---and whether it is enough information to satisfy Equation \ref{eq:Netbits}.

While specifying our information {\em requirement} in Equation \ref{eq:Netbits} is fairly straightforward, quantifying the information {\em content} in our pair of networks---formally called a ``message'' in the parlance of Information Theory---is quite nontrivial; entire modern research areas are built around quantifying information, including the areas of data compression\cite{johnson2003introduction}, cryptography \cite{adamek2011foundations}, and virtually all forms of digital communication \cite{sakrison1968communication}. For now, we conjecture that the amount of information encoded in an {\em arbitrary} network, without any restrictions on its structure, is approximately equal to its number of edges. This hypothesis arises from the fact that the adjacency matrix for a network consists of bits, with each edge corresponding to a single 1-bit in the matrix. Recent work has shown that networks roughly the size of our PPI networks can be encoded with as few as 1.8 bits per edge, on average, with the authors \cite{dhulipala2016compressing} also claiming that there is ``much room for improvement''. According to Shannon Information Theory\cite{shannon1948mathematical}, no object---including a graph---can be encoded using fewer bits than its information content, so the {\em actual} storage requirement for a particular scheme to encode $G$ (in this case, about 1.8 bits per edge) is an {\em upper} bound on $G$'s information content. The above results suggest that we are justified in claiming that each edge in a network provides approximately one bit of information that can be leveraged towards narrowing the search space of a network alignment. This in turn suggests that, when aligning two networks, they will require $\approx$Netbits($n_1,n_2$) edges if we are to isolate a specific alignment $a^*$---and if there is some idea of a ``correct'' alignment, then this likely represents the approximate number of {\em common} edges between the node pairs deemed as the correct alignment.

\begin{figure}[tb]
    \centering\small
    \includegraphics[scale=0.99]{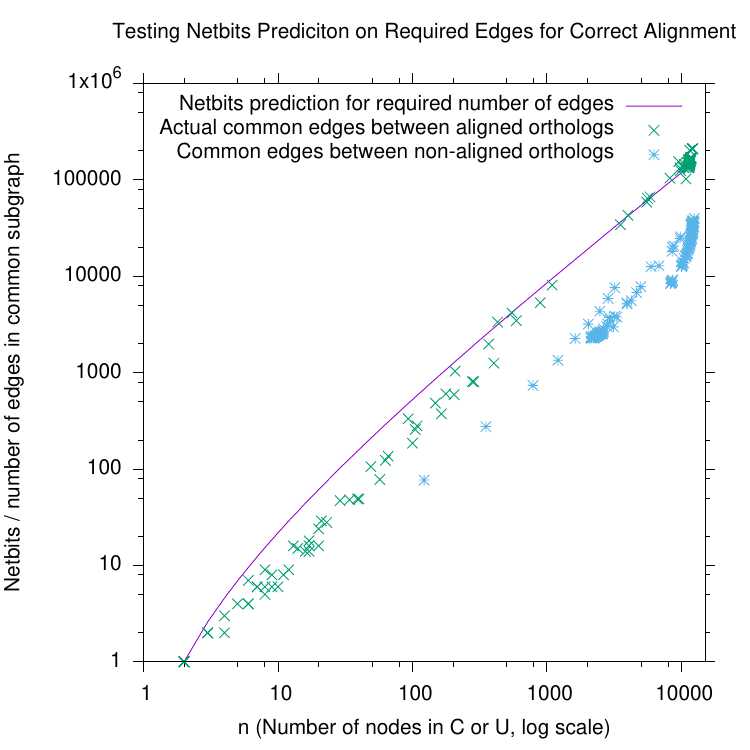}
    \includegraphics[scale=0.99]{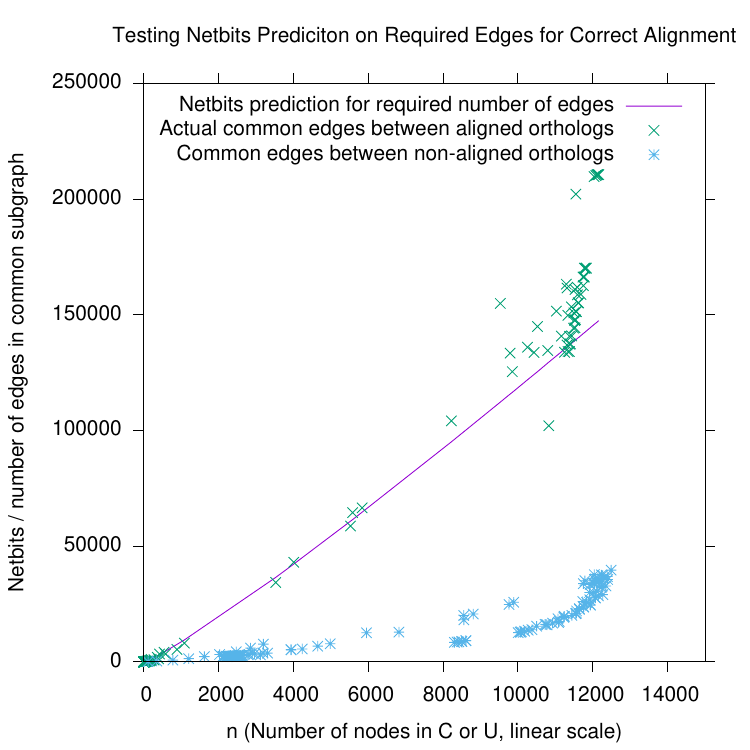}
    \caption{\textbf{Testing Information Theory ``Netbits'' Prediction by aligning rat-mouse orthologs from the {\it Integrated Interaction Database} (IID\cite{kotlyar2018iid})}
    We take the PPI networks of rat ({\it R. norvegicus}) and mouse ({\it M. musculus}) from IID, induced on the 1-to-1 orthologs between them. Then, for various values of $F\in[0,1]$, we independently choose a fraction $F$ of edges from each network and align them with SANA for 1 hour optimizing $S^3$. Then, from the {\it Common Connected Subgraph}, we extract the set of 1-to-1 orthologs, $C$, that were correctly aligned, and $U$, the set that were incorrectly aligned.
    We then plot Netbits($|C|$) (purple curve), the actual number of common edges induced on $C$ (green x's), and the actual number of common edges induced on $U$ (blue x's), as a function of either $|C|$ or $|U|$, as appropriate.
    (Note the left and right Figures plot the same data: log scale on the left, linear on the right.)
    We see that the Netbits curve closely follows the number of common edges between correctly aligned orthologs, whereas incorrectly aligned orthologs have far fewer common edges than the Netbits value.
    }
    \label{fig:IIDrat-mouse}
\end{figure}

While the previous paragraph is far from rigorous and will require significant development before details can be made more precise, the premise can easily be tested empirically. Namely, to empirically test Equation \ref{eq:Netbits}, we need realistic-looking PPI networks with (i) enough similarity to expect good alignments to exist, and (ii) enough edges to satisfy the Netbits threshold. To that end, we leverage IID---the {\it Integrated Interaction Database}\cite{kotlyar2018iid}. Each non-human mammalian PPI network in IID has been augmented by transferring high-confidence experimental interactions from the human PPI network to the corresponding interologs connecting 1-to-1 orthologs in the non-human mammal. As a result, all the mammalian PPI networks in IID are highly similar and have more than enough edges to satisfy the Netbits criterion (Equation \ref{eq:Netbits}).
In particular, the IID networks of rat and mouse contain 13,510 1-to-1 orthologs between them with degree at least 1 in both networks. Taking the subgraphs induced on these 13,510 nodes gives two highly similar PPI networks, which we call $R$ and $M$, each with 13,510 nodes, with the 1-to-1 orthologs providing a ``correct'' mapping between them;
$R$ has 233,289 edges, while $M$ has 237,380---both well above the Netbits value of 165,898 for $n_1=n_2=13,510$. Starting with $R$ and $M$, we choose a fraction $F\in[0,1]$ and---independently for each network---randomly remove edges until only a fraction $F$ of the original edges remain. This gives networks $R_F$ and $M_F$, which we align with SANA for 1 hour, optimizing $S^3$. (We choose $S^3$, the {\it Symmetric Substructure Score}, since the input IID networks are highly symmetric, by design.) Then, looking at the {\it Common Connected Subgraph} from the alignment, we extract the set of aligned node pairs that are orthologs---ie., the set of aligned node pairs that were {\em correctly} aligned. This set is called $C$, while the complement set in the alignment is $U$, the set of aligned node pairs that are not orthologs.
Figure \ref{fig:IIDrat-mouse} plots the number of edges induced on $C$ or $U$ as a function of the cardinality of the respective set.
The purple curve is Netbits($n$)---where $n$ is $|C|$ or $|U|$ as appropriate---ie., the predicted number of bits required to correctly align $n$ nodes. The green x's plot the actual number of common edges induced on $C$---that is, the number of common edges between correctly aligned orthologs (i.e., correctly recovered {\it interologs} according to IID). The blue x's depict the number of edges induced on $U$---that is, between {\em incorrectly} aligned orthologs in the same alignments. (Note that for each alignment, we have the identity $|U|+|C|=13,510$, so that each alignment has one green x and one blue x, but they are at different horizontal locations.) As we can see in the left Figure (using a log scale), when the number of correctly aligned orthologs is a small fraction of the total (ie., green x's with $n\lesssim 1000$), we can get correct alignments with slightly fewer edges than the Netbits prediction; but when a large fraction of the orthologs are correctly aligned (best observed in the right Figure that uses a linear scale), alignments with fewer common edges than predicted by Netbits are rare---ie., most alignments with a large fraction of correctly aligned ortholog pairs have a number of common edges at or above the Netbits threshold. In contrast, the total number of edges induced between incorrectly aligned orthologs is {\em always} well below the Netbits threshold. 

Incidentally, the green x's below the purple curve in Figure \ref{fig:IIDrat-mouse} are not terribly surprising: while the general idea is to suggest that each edge provides ``approximately'' 1 bit of information, the general structural properties of the graph can impart significant additional information.  For example, consider a tree consisting of a root node and three linear ``arms'' extending outwards, each with a different, arbitrary length. Such a three-armed tree can have an arbitrary number of nodes, each of which is the sole occupant of its automorphism orbit---meaning that when aligning the tree to itself, there is only one alignment that achieves an $S^3$ score of 1, and that alignment correctly aligns every node to itself. Thus, the $S^3$ measure in this case is able to isolate the single, unique {\em correct} alignment in an arbitrarily large search space with $n$ nodes but only $n-1$ edges---{\em far} fewer edges than Netbits($n,n$) for large $n$. Thus, Equation (\ref{eq:Netbits}) is clearly only an approximation on the number of required edges for a correct alignment, though Figure \ref{fig:IIDrat-mouse} suggests it tends to become statistically more likely to be a good approximation to a lower bound as $n$ increases.

To conclude, we have offered a theoretical argument, supported by quantitative empirical evidence in good agreement with the theory (cf. Figure \ref{fig:IIDrat-mouse}), for the observed widespread failure of topology-only global network alignments: it is due, at least in part, to insufficient edge densities in existing experimental PPI networks. In the absence of artificially inflated edge densities in databases such as IID, these low edge densities make it impossible for {\em any} algorithm based on topology alone to reliably produce global network alignments with robust biological value.


\subsection{Addressing R3: Inadequate optimization of chosen topological objective functions}

Our ultimate goal is to produce the best possible topological alignments according to the topological measures of Table \ref{tab:objectives}, in order to measure how well each of them is able to uncover functional similarity. To ensure a fair comparison between the objective functions of Table \ref{tab:objectives}, we need to show that SANA provides a level playing field.
To do this, we demonstrate in \S\ref{sec:SANA-is-optimal} that SANA provides {\it near-optimal} solutions---that is, alignments with near-optimal values of the objective function---in all cases the optimal value of the objective is known. One could not ask for a more level playing field than allowing each objective function to reach its (near-)optimal value.
Second, to demonstrate that no other published algorithms are capable of providing this level playing field, in \S\ref{sec:WeBeat} we demonstrate that SANA's near-optimal alignments dramatically out-score each of the algorithms of Table \ref{tab:objectives} at simultaneously optimizing EC, $S^3$, and their own topological objectives.

\subsubsection{SANA achieves near-optimal solutions when the optimal solution is known}\label{sec:SANA-is-optimal}
\begin{table*}[hbt]
\centering
\begin{tabular}{|l|r|l|r|r|r|r|r|r|}
\hline
$G_1$ & nodes & $G_2$ & nodes & $T_{H-GRAAL}$ & $T_{SANA}$ & score$_{H-GRAAL}$ & score$_{SANA}$ & SANA/Optimal\\
\hline
syeast0&1004 & syeast05&1004 & 2.87  & 20 & 0.946163 & 0.946139 &  0.99997\\
syeast0&1004 & yeast&2390 & 53.43 & 60 & 0.86285  & 0.86276 & 0.9999\\
yeast&2390 & human&9194 & 5567.85 & 120 & 0.883 & 0.882 & 0.998\\
RNorvegicus&1657 & HSapiens&13276 & 566.03 & 120 &0.918 & 0.917 &  0.998\\
CElegans&3134 & AThaliana&5897 & 5516.58 & 120 & 0.959 & 0.956 & 0.996 \\
\hline
\end{tabular}
\caption{\small\textbf{Achieving provably near-optimal solutions:} We compare SANA's ability to optimize the local (node-pairwise) orbit degree vector similarity \cite{Milenkovic2008} to that of the Hungarian algorithm as implemented by H-GRAAL\cite{H-GRAAL}, which provides provably optimal solutions when the objective consists only of node-pair similarities. $T_X$ is the run-time in minutes for $X=H-GRAAL,SANA$; similarly for $score_X$. Time does not include the pre-computation of the orbit degree similarity matrix, which is identical for both programs. The far right column is SANA's score as a fraction of the optimal score produced by H-GRAAL \cite{H-GRAAL}. Networks are from the H-GRAAL suite, which include ``synthetic yeast' (syeast) from Krogan \cite{Krogan2006}, yeast2 from Collins \cite{Collins2007yeast2}, human1 from Radijovac \cite{Radivojac2008} and BioGRID networks from 2013.
(We do not attempt more recent BioGRID networks since H-GRAAL would require months, rather than days, on modern BioGRID networks.)
}
\label{tab:H-GRAAL}
\end{table*}

In general, given an appropriate temperature schedule and enough CPU time, simulated annealing converges to an optimal solution with probability one \cite{ingber1989VFSA}.
To provide evidence that SANA is performing at this level, we offer two lines of evidence. First, Table \ref{tab:H-GRAAL} demonstrates that SANA produces alignments that score extremely close to optimal ones computed by the Hungarian algorithm\cite{H-GRAAL}, while taking only a fraction of the time on large networks. For edge-based NP-complete cost functions such as EC and $S^3$, Figures \ref{fig:SC-HS-self} and \ref{fig:sat_graph} (Supplementary) show that SANA easily aligns any existing PPI network with a subgraph of itself---effectively providing quick solutions to the subgraph isomorphism problem---a well-known NP-complete problem\cite{GareyJohnson79}.




\subsubsection{SANA out-scores other aligners even at optimizing their own objectives}\label{sec:WeBeat}

\begin{figure}[tb]
    \centering
    \includegraphics[scale=0.6]{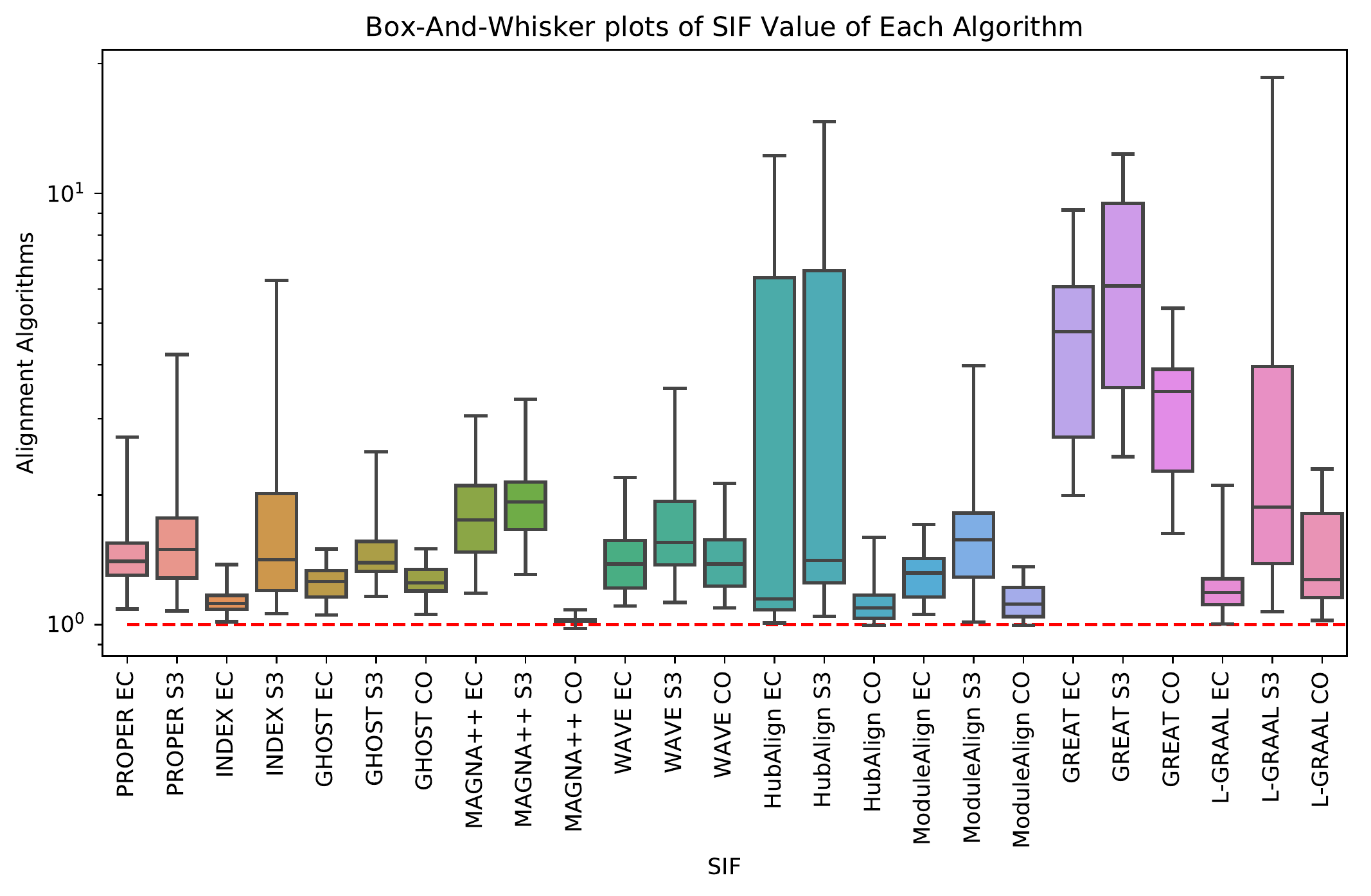}
    \caption{\small{\bf SANA universally outperforms other aligners at optimizing topological objectives}: For each competing search algorithm, we plot the {\it SANA Improvement Factor}, or {\it SIF}, which is the ratio of SANA's score to that of the other aligner, for the measures EC, $S^3$, and the competitor's ``custom'' objective (``CO'') if one exists. Note that the vertical axis is logarithmic. 
    The dotted red line represents a SIF ratio of 1---ie., that SANA and the other algorithm achieved the same score.
    The height of each box marks the lower and upper quartile of SIF across the 28 pairs of species (in some cases fewer---see text), while the line inside the box marks the median SIF. The whiskers mark the minimum and maximum SIF values. Note that even the bottom whiskers rarely dip below 1, meaning even SANA's {\em worst} performance is usually better than that of the competing aligners. 
    }
    \label{fig:avgsif}
\end{figure}



We compare SANA's ability to simultaneously optimize EC, $S^3$, and the objectives of Table \ref{tab:objectives} to that of the aligners listed in the same table. (See Methods \S\ref{sec:WeBeatMethods} for details.) 
Since SANA almost universally outperforms all other search algorithms in all measures, we introduce the metric \textit{SANA Improvement Factor}, or \textit{SIF}, which is the ratio of SANA's value of any objective $f$ vs. the value of $f$ achieved by the other alignment algorithm, when both are applied to the same network pair:
$$ SIF(f,G_1,G_2)=\dfrac{\mbox{SANA's score on objective function $f$ aligning $G_1,G_2$}}{\mbox{Other Search Algorithm's score on $f$ aligning $G_1,G_2$}}.$$
Figure \ref{fig:avgsif} depicts SIF scores against the aligners of Table \ref{tab:objectives} across all 28 pairs of BioGRID networks (cf. \S \ref{sec:WeBeatMethods}). As we can see, SANA almost universally outperforms other aligners at simultaneously optimizing EC, $S^3$, and the competing aligner's own custom objective.
In particular, out of a total of 188 tests, SANA was only marginally beaten in the following handful of cases:
HubAlign beat SANA by 0.3\% in Importance in one out of 28 BioGRID pairs (RN-CE)---but on the same pair, SANA outperformed HubAlign's EC and $S^3$ scores by 0.9\% and 158\% respectively; for its own objective function, ModuleAlign outperformed SANA in 2 out of 28 network pairs: by 0.3\% for RN-HS and 0.2\% for CE-DM---while on the same pairs SANA outperformed ModuleAlign's scores by 23\% in EC and 160\% in $S^3$ for RN-HS, and by 12\% in EC and 50\% in $S^3$ for CE-DM; MAGNA++ outperformed SANA by 2\% in one out of 28 pairs (RN-MM), while on the same pairs SANA outperformed MAGNA++'s EC and $S^3$ scores by 103\% and 105\% respectively.

\subsubsection{Summary: SANA provides a near-optimal level-playing field for objective function comparison}
Together, Table \ref{tab:H-GRAAL} and Figures \ref{fig:avgsif}, \ref{fig:SC-HS-self} and \ref{fig:sat_graph} give us confidence that SANA provides a level playing field for comparing objective functions because it produces alignments with near-optimal scores of the topological objective. More explicitly,
\begin{itemize}
    \item Table \ref{tab:H-GRAAL} demonstrates that for node-based objective functions for which the Hungarian Algorithm can provide exactly optimal solutions, SANA comes within a fraction of a percent of the optimal value in a reasonable amount of CPU time.
    
    \item Supplementary Figures \ref{fig:SC-HS-self} and \ref{fig:sat_graph} demonstrate that for edge-based objective functions which are NP-complete, SANA correctly aligns each of the depicted PPI networks to itself in a few minutes.
    
    \item Figure \ref{fig:avgsif} tells us that SANA almost universally outperforms all other aligners at optimizing their own objective functions. In the handful of cases that the other aligner performs better, it does so by only a minuscule amount in {\em one} of the three objectives being optimized, while SANA soundly beats it in the other two.
\end{itemize}


\section{Addressing R2: Measuring functional relevance of topological objective functions}\label{sec:whichMeasures}

\subsection{Functional relevance}\label{sec:functionalRelevance}

Given a network alignment $a$, let $f:a\rightarrow \mathcal{R}$ be some measure of topological similarity exposed by $a$, and let $g:a\rightarrow \mathcal{R}$ be some measure of functional similarity across nodes aligned by $a$. 
We say that an alignment $a$ is ``functionally relevant'' if it scores higher in $g$ than some user-defined threshold.

Let $f^*$ be the highest possible topological score across all alignments and let $A^*$ be the set of alignments that achieve that score. Similarly, let $g^*$ be the highest possible functional score, and let $B^*$ be the set of alignments that achieve it.
Since the goal of topological network alignment is to uncover functionally similar nodes across networks, we want $f$ to have the following properties: i) there should be a high correlation between the values of $f(a)$ and $g(a)$, so that increasing $f$ also tends to increase $g$ (a high Spearman correlation would be best, though a high Pearson correlation will suffice); (ii) there should be substantial overlap between the sets $A^*$ and $B^*$---or, at the very least, between alignments that score close to $f^*$ and those that score close to $g^*$; and (iii) the fraction of alignments in $A^*$ that are also in $B^*$---formally, the ratio $|A^* \cap B^*|/|A^*|$, or its ``near-optimal'' equivalent---should not be vanishingly small. If these three properties are satisfied, then $f$ is functionally relevant measure of topological similarity, because optimizing $f$ will move us towards optimizing $g$, and (near-)optimal alignments under $f$ will have a good chance of exposing high functional similarity among aligned nodes.

Conversely, if any one of the above properties fails, then $f$ is not a functionally relevant measure of topological similarity. Respectively: (i) if $f$ does not correlate with $g$, then increasing $f$ will not increase $g$---the latter being necessary to expose alignments with high functional similarity; (ii) even when $f$ and $g$ are {\em correlated}, if topologically optimal alignments are not close to functionally optimal ones, then there is no benefit to aggressively pushing $f$ towards its optimal value; (iii) if $A^*$ contains functionally (near-)optimal alignments but they are greatly outnumbered by alignments with low functional score, then a randomly chosen alignment from $A^*$ has only a small chance of having high functional similarity. In any of these cases, $f$ fails the test of being a ``functionally relevant'' topological objective function.

\subsection{Objective Function Saturation}\label{sec:saturation}
It is the failure of point (iii) above that we refer to as ``objective function saturation'': assume that (1) the value of $g^*$ is known, (2) alignments that score $g^*$ also have optimal topological score $f^*$, but (3) there exist many alignments with optimal topological score $f^*$ that nonetheless score significantly below $g^*$ in functional similarity. Then, we say that $f$ {\it saturates}: like a sponge that is far too small to soak up a spill, it reaches its full capacity long before the job is done.

Saturation can be caused by at least two fundamental issues: (a) not enough topological input information, resulting in the inability of {\em any} topology-based method to isolate ``good'' individual alignments; and (b) the topological objective function having insufficient discriminative power even when enough information is available. The first case is addressed by our appeal to Information Theory; we address the second in our detailed comparison of objectives in the following sections. (Supplementary Section \ref{sec:saturationSupp} provides detailed schematic examples of saturation.)

\subsection{Evaluating the functional relevance of topological measures}

We now compare the measures of topological similarity proposed in Table \ref{tab:objectives}---taken from a wide range of publications---for their ability to produce alignments that highlight functionally similar proteins across species. To that end, ideally we would like to test each of them for the three properties in \S\ref{sec:functionalRelevance}.

As has been noted by others\cite{GRAAL,H-GRAAL} and discussed in several places above, there may exist a large (but difficult to estimate) number of alignments with optimal or near-optimal topological scores. We have provided evidence (cf. Figures \ref{fig:IIDrat-mouse}, \ref{fig:sat_graph}; Table \ref{tab:H-GRAAL}; sections \ref{sec:SANA-is-optimal}, \ref{sec:saturation}) that SANA tends to provide alignments with near-optimal topological scores when given enough CPU time (an hour seems enough on existing PPI networks).
Since simulated annealing is a random search algorithm, SANA may produce a different alignment each time it is run. Since it produces near-{\em optimal} alignments, it can thus be viewed as providing a {\bf\em random sample of alignments with near-optimal scores} according to the objective function used. (We cannot claim the sample is unbiased since we currently know nothing about the distribution. That is, even if we achieve an alignment in $A^*$ we cannot currently claim we are sampling $A^*$ in an unbiased manner.) Thus, SANA effectively allows us to simultaneously test all three properties of functional relevance, viz.: (i) since SANA searches the alignment space randomly, if the value of $f$ and $g$ are uncorrelated, then increasing $f$ will not tend to increase $g$; (ii) if alignments that are near-optimal in $g$ do not have scores that are near-optimal in $f$, then SANA's ability to push the value of $f$ to near-optimality will have little functional benefit; (iii) even if near-optimal scores in $g$ have near-optimal scores in $f$ but $f$ saturates, then a randomly chosen near-optimal alignment according to $f$ is unlikely to uncover one of the relatively few alignments that are well-scoring in $g$.

\begin{table*}[hbt]
\centering\small
\caption{Major networks from BioGRID (v3.4.164, August/September 2018), sorted by mean degree.}
\label{tab:BioGRID}
\begin{tabular}{|rrrrrllll|}
\hline
nodes & edges & Eq(\ref{eq:Netbits}) & ratio & degree & density & name & Abbr. & species \\
\hline
5984    &104962 &66455  &1.58    &35.08    &0.00586     &baker's yeast &SC &Saccharomyces cerevisiae\\
17200   &282181 &217200 &1.30    &32.81    &0.00191     &human &HS &Homo sapiens\\
8728    &46364  &101678 &0.46    &10.62    &0.00122     &fruit fly &DM &Drosophila melanogaster\\
9364    &34725  &110037 &0.32    & 7.42    &0.00079     &cress &AT &Arabidopsis thaliana\\
2811    &8931   &28156  &0.32    & 6.35    &0.00226     &fission yeast &SP &Schizosaccharomyces pombe\\
6777    &18108  &76477  &0.24    & 5.34    &0.00079     &mouse &MM &Mus musculus\\
3194    &5572   &32581  &0.17    & 3.49    &0.00109     &worm &CE &Caenorhabditis elegans\\
2391    &3554   &23392  &0.15    & 2.97    &0.00124     &rat &RN &Rattus norvegicus\\
\hline
\end{tabular}
\end{table*}

In the following sections, we evaluate a large set of alignments, across
objective functions from Table \ref{tab:objectives} and BioGRID pairs from Table \ref{tab:BioGRID}. Each of SANA's output alignments consists of pairs of proteins providing a 1-to-1 mapping from proteins in $G_1$ to those in $G_2$.
For each pair of BioGRID networks from Table \ref{tab:BioGRID}, and for each objective in Table \ref{tab:objectives}, we ran SANA 100 times for 1 hour each. Across all cases, the objective function values agreed to less than 1\%, 3\%, and 7\% in 51\%, 90\%, and 100\% of cases, respectively, suggesting that 50\% of SANA's alignments achieved a topological score within 1\% of optimal, a further 40\% were within 3\% of optimal, and the remaining 10\% were within 7\% of optimal.

\subsection{Recovery of common Gene Ontology terms}\label{sec:GOrec}

\begin{figure}[tb]
    \centering
    \includegraphics[width=0.49 \textwidth]{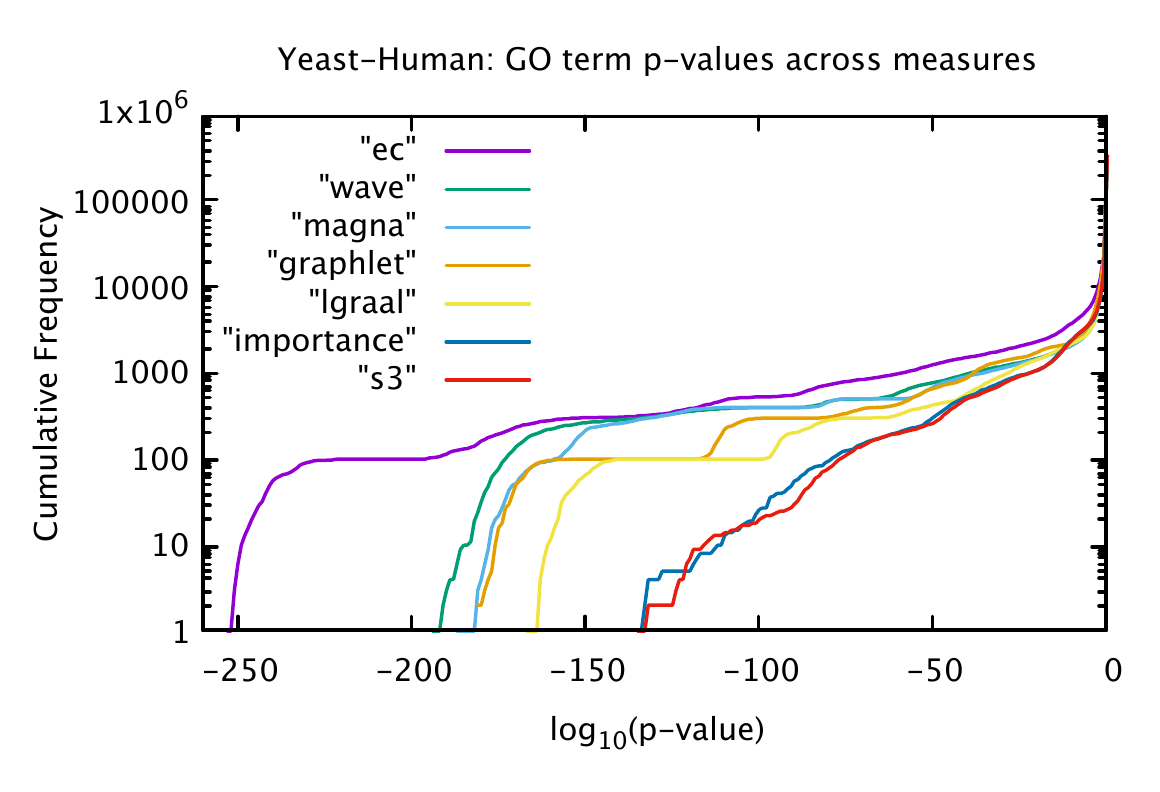}
    \includegraphics[width=0.49 \textwidth]{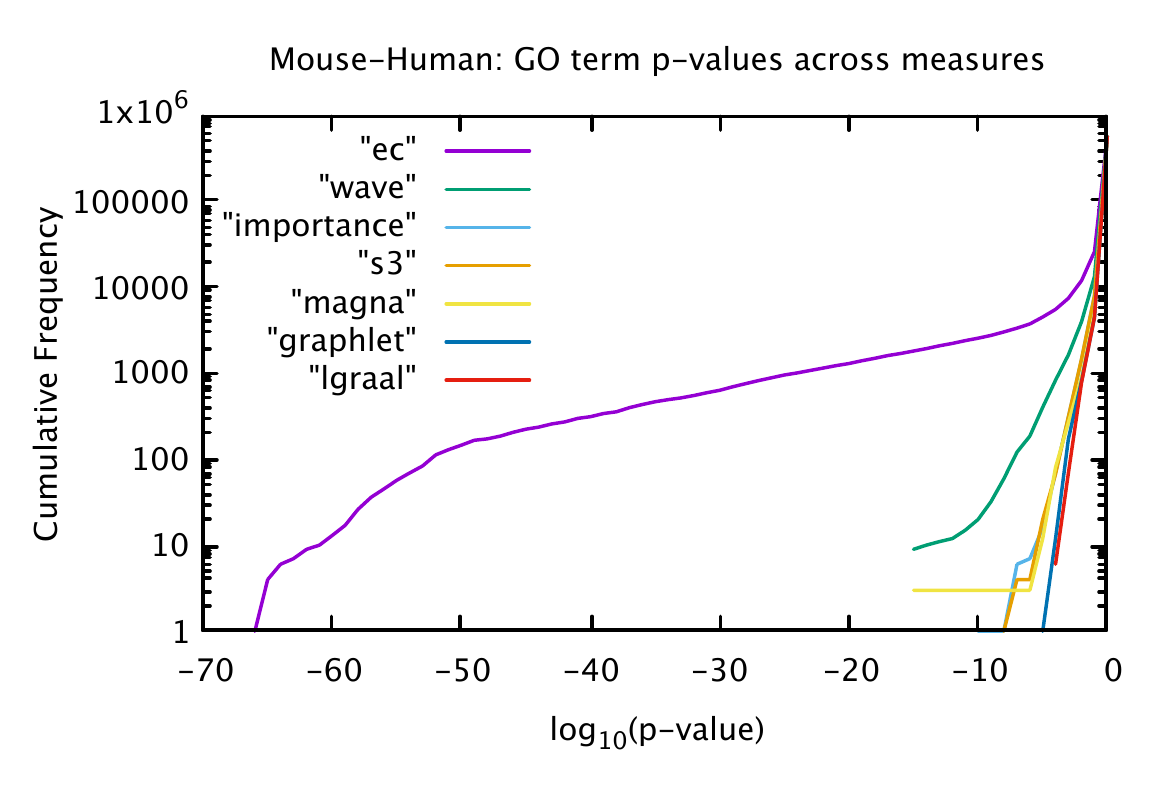}
    \caption{Cumulative distribution of $p$-values of individual GO terms across 100 alignments of yeast-human (left) and mouse-human (right), for alignments optimizing each objective function that was computable for these two network pairs.
    For each GO term $g$, we use the tail of the hypergeometric distribution to compute the $p$-value that the observed number of aligned protein pairs would share $g$ in a random alignment (see Methods, \S \ref{sec:GOp-value}).
    Note that all scales are logarithmic.
    The legend of each figure lists the objectives best-to-worst for that pair.
    We observe that there are vast differences in the functional similarity uncovered by optimizing the different objective functions.
    In particular, EC is by far the best measure for uncovering functional similarity between aligned proteins in BioGRID 3.4.156. Its yeast-human alignments contain over 3800 GO terms with $p$-values below $10^{-10}$, over 500 with $p$-values below $10^{-100}$, and over 100 below $10^{-200}$.
    The second most effective objective is WAVE's WEC (a graphlet-weighted version of EC), with over 2,100 GO terms with $p$-values below $10^{-10}$, almost 400 with $p$-values below $10^{-100}$, but none below $10^{-200}$. 
    The mouse-human alignments have the same top two measures in the same order (EC and WAVE), though in this case EC hugely dominates all measures, with WAVE as a distant second.
    The other five measures (Importance, $S^3$, MAGNA, graphlet, and L-GRAAL) show no solid ordering between them here, nor in other pairs of networks (cf. Supplementary Information).
    }
    \label{fig:GOhist}
\end{figure}

The Gene Ontology \cite{GO} describes our understanding of genes and proteins. It consists of thousands of descriptive annotations called {\it GO terms} which are arranged into a hierarchical description of molecular functions (MF), biological processes (BP), and cellular components (CC).
Genes and gene products, such as proteins, are ``annotated'' with various GO terms as we learn what those genes/proteins do.
GO terms high in the hierarchy can be vague and can annotate thousands of proteins, while more specific GO terms lower in the hierarchy tend to annotate fewer proteins. Well-understood genes and proteins can have many GO annotations, while lesser understood proteins may have few or no GO annotations, or be annotated with only high-level, vague GO terms.

To evaluate the functional relevance of each topological similarity measure $f$ from Table \ref{tab:objectives}, we performed 100 independent runs of SANA optimizing $f$ for 1 hour, across all 28 network pairs from Table \ref{tab:BioGRID}. Given an alignment, for each GO term $g$ we counted the number, $k_g$, of aligned protein pairs that share $g$, and then evaluated the statistical significance of $k_g$ by computing the probability that a random alignment would have $k_g$ or more protein pairs sharing $g$. This provides a $p$-value for each single GO term, in each single alignment (cf \S\ref{sec:GOp-value}). Since we performed 100 alignments for each of the 28 network pairs for each of topological objectives, we have 100 samples of each GO term's statistical significance in each pair of networks, for each objective. Figure \ref{fig:GOhist} presents the cumulative distributions of the $p$-values of 7 measures, for the network pairs yeast-human (left) and mouse-human (right) of BioGRID 3.4.164. (We choose yeast-human because these are by far the most complete PPI networks; and mouse-human because mouse is the most complete mammalian PPI network after human.) We observe that the topological measures vary quite substantially in their ability to align a statistically significant number of protein pairs that share GO terms: EC seems by far the best, with WAVE's ``WEC'' (a graphlet-based weighted EC) also performing significantly better than other measures.


\begin{figure}[tb]
    \centering
    \includegraphics[width=0.49 \textwidth]{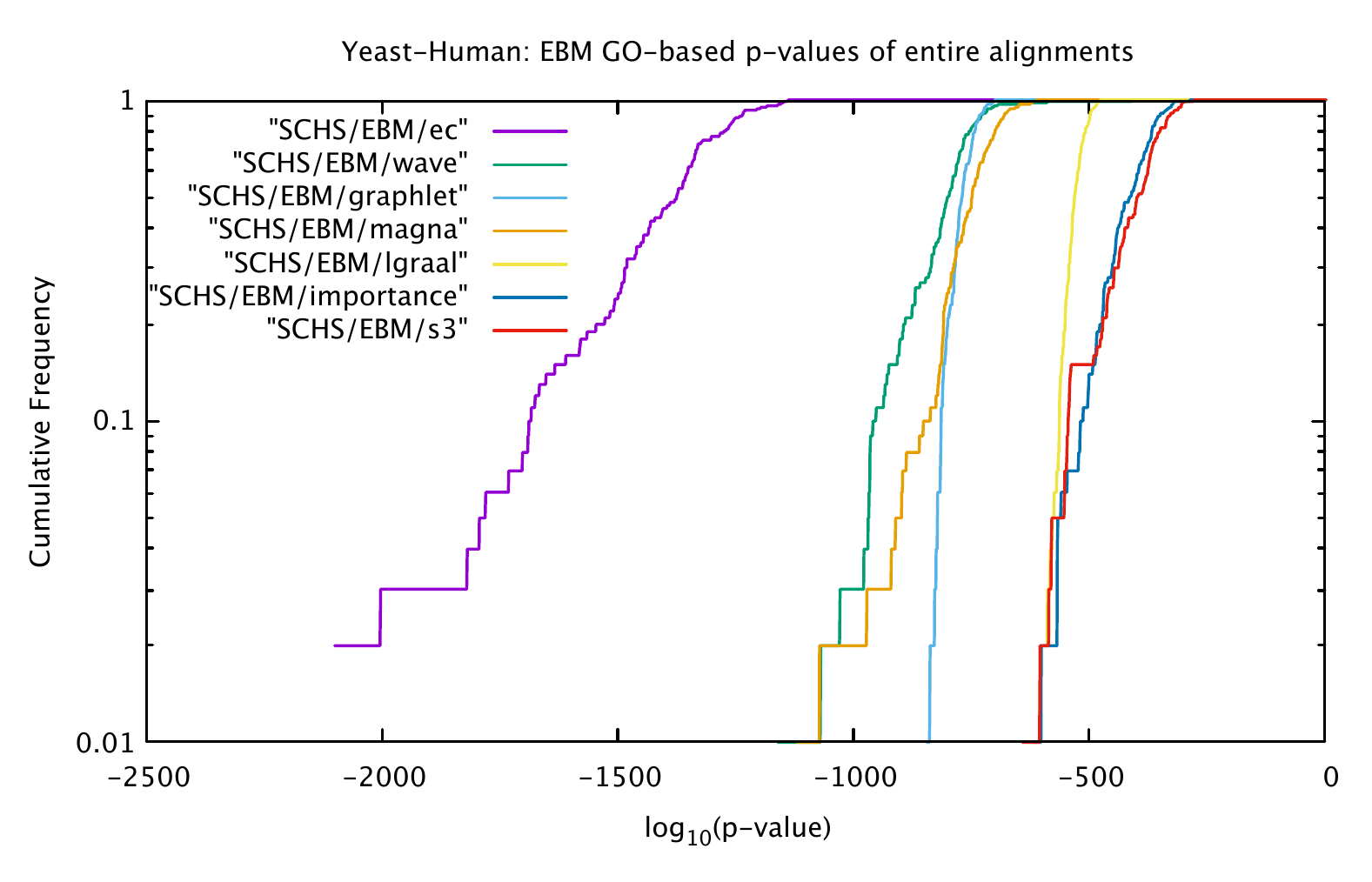}
    \includegraphics[width=0.49 \textwidth]{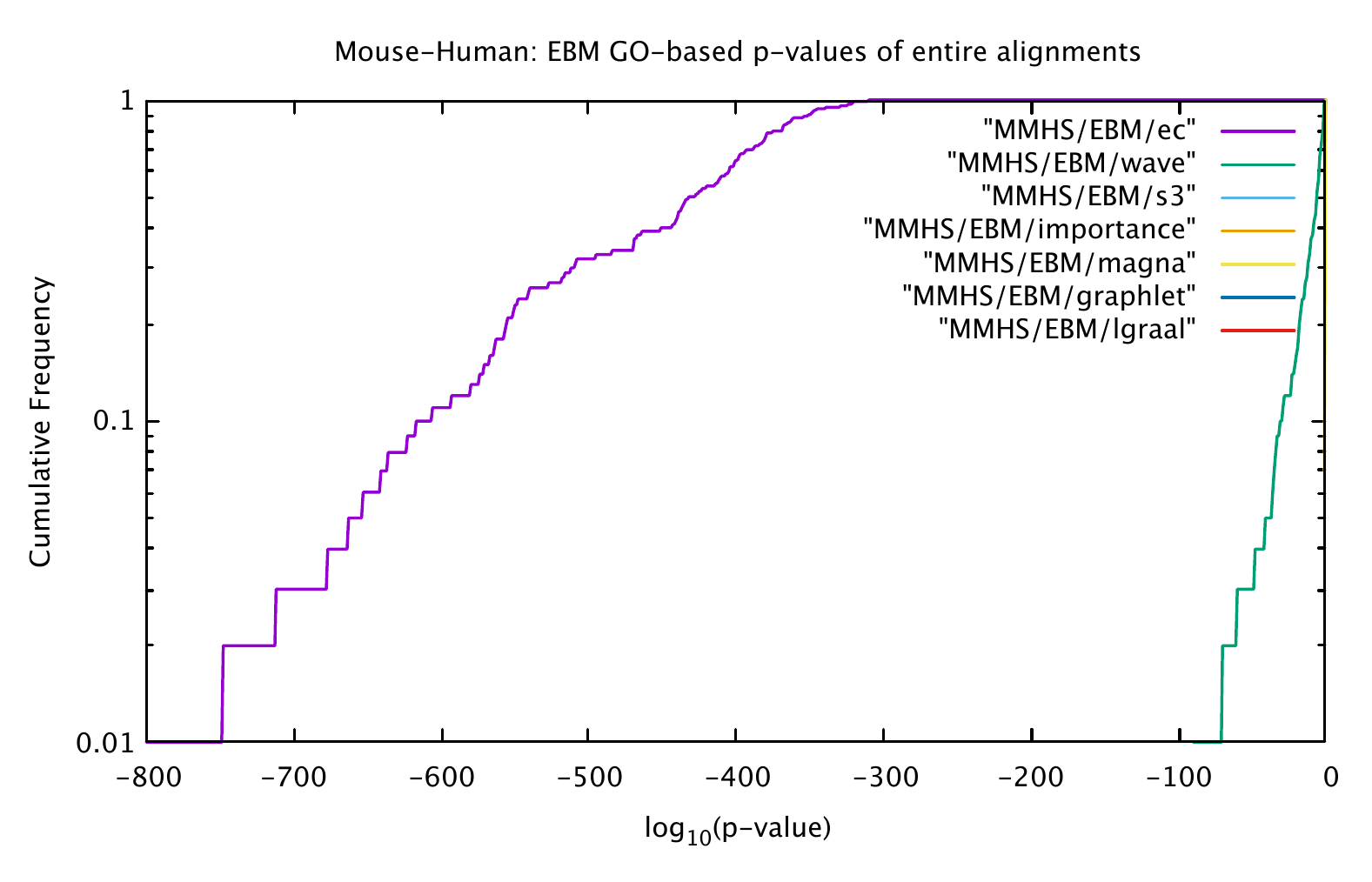}
    \caption{Similar to Figure \ref{fig:GOhist}, except all the $p$-values for individual GO terms have been combined using the {\em Empirical Brown's Method} \cite{poole2016combining}, producing a single, holistic $p$-value for each pairwise alignment. There are 100 alignments for each network pair and for each objective function. We see again that EC is the dominant measure, producing individual alignments with GO-term-based $p$-values universally below $10^{-1000}$ for yeast-human and  $10^{-300}$ for mouse-human. In the case of yeast-human, all other measures have $p$-values that, though quite significant, are hundreds of orders of magnitude less so than EC. In the case of mouse-human, all objectives except EC and WAVE have so many GO terms with a p-value of 1 (ie., totally insignificant due to zero proteins sharing that GO term despite there being a non-zero number that {\em could} share it) that their whole-alignment p-values are so close to 1 that they are invisible against the border on the right side of the plot. (More network pairs listed in the Supplementary.)
    }
    \label{fig:EBMhist}
\end{figure}

While Figure \ref{fig:GOhist} displays the distribution of $p$-values of individual GO terms, computing the $p$-value of an entire alignment based on GO terms is nontrivial since most GO terms in the hierarchy are not independent of each other. While in principle it should be possible analytically to derive the inter-relationships between GO terms, we have instead employed the {\it Empirical Brown's Method}\cite{poole2016combining} (see Methods, \S\ref{sec:GOp-value}) to compute $p$-values for each alignment, accounting for the non-independence across GO terms. Figure \ref{fig:EBMhist} is similar to Figure \ref{fig:GOhist}, but for entire alignments rather than individual GO terms. As we can see, the EC measure again dominates, and the strongest $p$-values are for the most dense networks, with the best EC-driven alignments between yeast and human having $p$-values of $10^{-2000}$, while even the {\em worst} yeast-human alignments by this measure have $p$-values of about $10^{-1200}$. As can be seen in the Supplementary Info, these two network pairs are not atypical: fully {\em half} of all network pairs have strong $p$-values when alignment is driven by EC.

\subsection{Topology-based recovery of thousands of orthologs between major BioGRID species}\label{sec:orthologs}

Aligning a large set of proteins annotated with a particular GO term to another large set annotated by the same GO term provides plenty of statistical evidence that topology is able to align functionally similar {\em regions} to each other. However, while aligning a whole region of the PPI network in one species to a similar region in the other species is encouraging, it provides little {\em predictive} power about individual proteins. In addition, correctly aligning {\em one specific pair} of proteins is far more stringent than aligning one set to another set but allowing arbitrary permutations among members of each set. Thus, in this section we highlight the ability of network topology to recover the ``correct'' alignment between specific pairs of orthologous proteins between species.

Consider Table \ref{tab:orthologs}: collectively, across all 28 pairs of species pairs and all objectives, we recovered almost 3,000 known orthologous protein pairs; as seen in the bottom row of Table \ref{tab:orthologs}, this recovery rate has a cumulative $p$-value of about $10^{-144}$.
(The objectives for GHOST and GREAT were too expensive to compute for some of the largest network pairs---cf. the blank entries in Table \ref{tab:orthologs}).
Observing the different columns of Table \ref{tab:orthologs}, we see that different topological objectives result in different ortholog recovery rates; the columns are sorted by the last row (total recovered orthologs across all species). Furthermore, comparing the ``uniq'' column with ``sum'' demonstrates that there is not much overlap between the orthologs recovered by different objectives, so that each objective appears to have value in that they recover near-orthogonal sets of orthologs. As discussed in section \ref{sec:EC-vs-S3}, this suggests that no single topological measure is ``best'' at recovering orthologs. Interestingly, this conclusion is quite different from the GO-term based $p$-values discussed above (cf. \S\ref{sec:GOp-value}), where EC was dominant. One clear difference is that GO-term based evaluations are more global, while ortholog recovery is more local. Clearly more research is needed before we have a full understanding of these results.

\begin{table}[htb]
\small    \centering
\begin{tabular}{|l|r||r|r|r|r|r|r|r|r|r||r|r|r|r|}
\hline
\backslashbox[15mm]{nets}{obj.}       
&\thead{orth.\\pairs} &EC     &$S^3$  &Imp.   &wave   &grt    &ghst   &mag    &gr     &lg     &sum    &uniq   &rate   &$p\;\;\;$ \\
\hline
SC-HS   &1245   &{\bf 72}       &52     &64     &18     &-      &-      &19     &8      &5      &238    &172    &13.8\% &$10^{-44}$\\
MM-HS   &6286   &{\bf 101}      &37     &34     &77     &-      &-      &26     &5      &7      &287    &282    &4.5\% &$10^{-17}$\\
SP-SC   &1242   &{\bf 71}       &38     &36     &65     &-      &-      &33     &1      &3      &247    &227    &18.3\% &$10^{-16}$\\
SP-HS   &945    &{\bf 34}       &5      &2      &7      &-      &-      &5      &0      &0      &53     &53     &5.6\% &$10^{-15}$\\
SC-DM   &955    &{\bf 43}       &36     &38     &23     &-      &-      &7      &1      &1      &149    &126    &13.2\% &$10^{-12}$\\
DM-HS   &3511   &{\bf 52}       &28     &31     &46     &-      &-      &14     &1      &1      &173    &161    &4.6\% &$10^{-08}$\\
RN-HS   &2223   &{\bf 38}       &15     &14     &5      &-      &-      &6      &1      &1      &80     &79     &3.6\% &$10^{-07}$\\
RN-MM   &2267   &{\bf 68}       &43     &39     &28     &39     &37     &29     &13     &5      &301    &276    &12.2\% &$10^{-06}$\\
CE-HS   &1137   &{\bf 24}       &7      &5      &8      &-      &-      &1      &0      &0      &45     &45     &4.0\% &$10^{-06}$\\
SP-DM   &806    &{\bf 29}       &29     &29     &9      &-      &13     &8      &0      &0      &117    &102    &12.7\% &$10^{-06}$\\
RN-DM   &865    &{\bf 27}       &12     &16     &17     &15     &21     &5      &2      &4      &119    &110    &12.7\% &$10^{-05}$\\
RN-AT   &534    &10     &{\bf 19}       &13     &13     &14     &8      &7      &4      &3      &91     &82     &15.4\% &$10^{-04}$\\
CE-DM   &999    &{\bf 27}       &20     &12     &15     &19     &15     &7      &1      &1      &117    &113    &11.3\% &$10^{-03}$\\
MM-AT   &1233   &25     &{\bf 29}       &22     &19     &-      &11     &16     &4      &6      &132    &126    &10.2\% &$10^{-03}$ \\
MM-DM   &2108   &{\bf 40}       &26     &31     &31     &-      &19     &28     &7      &8      &190    &175    &8.3\%  &$0.02$ \\
RN-SC   &357    &{\bf 14}       &6      &7      &4      &-      &-      &5      &1      &1      &38     &38     &10.6\% &$0.03$ \\
SP-AT   &844    &11     &{\bf 18}       &14     &15     &-      &7      &7      &2      &3      &77     &69     &8.2\% &$0.05$ \\
CE-AT   &562    &{\bf 11}       &11     &5      &5      &7      &9      &7      &0      &0      &55     &53     &9.4\% &$0.4$ \\
AT-HS   &1864   &{\bf 17}       &9      &6      &6      &-      &-      &4      &3      &0      &45     &41     &2.2\%  &$0.5$ \\
RN-SP   &371    &{\bf 19}       &11     &11     &6      &10     &9      &7      &1      &1      &75     &65     &17.5\% &$0.8$ \\
SP-MM   &812    &{\bf 17}       &6      &10     &10     &-      &4      &6      &1      &1      &55     &49     &6.0\% &$1$ \\
SP-CE   &403    &15             &13     &17     &12   &{\bf 20} &8      &8      &2      &1      &96     &84     &20.8\% &$1$ \\
CE-SC   &362    &{\bf 9}        &5      &5      &9      &-      &-      &3      &1      &1      &33     &33     &9.1\% &$1$ \\
SC-MM   &745    &{\bf 14}       &13     &8      &8      &-      &-      &4      &1      &3      &51     &44     &5.9\% & $1$ \\
DM-AT   &1477   &{\bf 19}       &14     &16     &16     &-      &13     &9      &4      &3      &94     &92     &6.2\% &$1$ \\
CE-MM   &1136   &{\bf 19}       &5      &14     &9      &-      &11     &11     &1      &0      &70     &68     &6.0\% &$1$ \\
RN-CE   &726    &{\bf 25}       &11     &16     &16     &13     &15     &10     &0      &4      &110    &104    &14.3\% &$1$ \\
SC-AT   &977    &8              &5      &6      &7      &-      &-    &{\bf 9}  &7      &2      &44     &43     &4.4\%  &$1$ \\
\hline
total   &36992  &{\bf 859}      &523    &521    &504    &137    &200    &301    &72     &65     &3182   &2912   &7.87\% & $5\times 10^{-144}$ \\
\hline
    \end{tabular}
    \caption{\small {\bf Recovered orthologs:} Each row is one of 28 species pairs from BioGRID 3.4.164, using the 2-letter abbreviations of Table \ref{tab:BioGRID}.
    Orth.pairs=number of orthologs between the BioGRID network pair according to NCBI Homologene. 
    Columns EC through lg report the number of unique orthologs recovered in 100 alignments of SANA optimising said measure (abbreviations: grt=GREAT, ghst=GHOST, mag=MAGNA, gr=graphlets, lg=L-GRAAL); the best performing measure is in boldface. A dash '-' indicates the algorithm failed to produce  similarities for SANA to optimize. Objectives are sorted left-to-right by total recovery (last row, with the totals for GREAT and GHOST adjusted for failures). ``sum'' is total orthologs recovered across all 9 measures, including duplicates recovered by more than one measure; ``uniq''=sum with duplicates removed; rate=uniq/orth.pairs; $p$-value=probability of the boldfaced number of orthologs being properly paired at least once across 100 random alignments (see Methods, \S \ref{sec:Ortho-p-values}). Rows are sorted by $p$-value. Finally, all $p$-values in the last column have been increased by a factor of 9 to account for the multiple hypothesis testing across the 9 objective functions.
    }
    \label{tab:orthologs}
\end{table}



Interestingly, yeast-human has the top $p$-value, and is a very close second to mouse-human in raw number of recovered orthologs. Observing Table \ref{tab:BioGRID}, we see that mouse is about an order of magnitude less complete than human, and significantly less complete than yeast---and yet mouse-human does {\em better} than yeast-human at raw number of recovered orthologs. Likely this is because mouse and human are more closely related than yeast and human. We interpret these observations to bode extremely well for future network analyses: viz., if mouse-human does so well with such incomplete network coverage, and yeast-human does so well despite the vast taxonomic distance between them, then topology-driven network alignments are likely to become even more effective as the volume of high-quality network data continues to grow.

\section{Discussion}\label{sec:Discussion}
\subsection{Statistical significance}\label{sec:statSig}
To our knowledge, the $p$-values displayed in this paper are orders of magnitude more significant than published $p$-values derived from any other topology-only methods of PPI network analysis. We believe our results provide---by far---the strongest evidence to date demonstrating the relationship between global PPI network topology and functional similarity between regions of PPI networks, as well as for individual proteins. While we have several lines of evidence to soundly reject R0, we have demonstrated that the widespread failure of existing topological methods are likely a combination of all three of R1, R2, and R3. Namely,

\begin{itemize}
    \item Information theory suggests that most existing PPI networks have too low an edge density for topology-only methods to produce robust, well-defined global alignments. However, repeated sampling of high-scoring global alignments may expose local regions that are robustly aligned; our companion paper \cite{WangAtkinsonHayesGOpredict} demonstrates that robust GO-term predictions are possible even using decade-old BioGRID and GO releases.
    \item Though many topological objective functions have been proposed in the literature, the algorithms published to optimize those objectives almost universally fail to optimize those objectives to their maximum values. This failure to produce near-optimal solutions has hampered the ability of previous aligners to produce statistically significant topology-function relationships.
    \item When optimized more fully, the various objective functions show a wide disparity in their ability to align functionally similar proteins---at least on current data. (This may change as more data becomes available.)
    \item The most effective objective function on current data appears to be EC, which is able to align functionally similar regions of the denser PPI networks with $p$-values hundreds, or in some cases thousands, of orders of magnitude below published $p$-values in the literature. However, the dominance of EC may be temporary---an artefact of its ability to align networks of highly disparate of completeness; we expect more symmetric measures, such as $S^3$, to become more prominent as the densities of PPI networks become more comparable.
\end{itemize}


As we can see from Table \ref{tab:orthologs}, {\em all} of the topological measures tested are able to recover a statistically significant number of orthologs between species, especially when the network density is high enough in both networks. However, the EC measure recovers the greatest number of orthologs for almost all species pairs from BioGRID. Additionally, EC is usually the best measure for recovering common GO terms, and far outperforms all other measures on the most complete PPI networks. As we explain in section \ref{sec:EC-vs-S3}, we believe this is likely due simply to the wide disparity in network completeness across the BioGRID networks; if the 2018 IID networks \cite{kotlyar2018iid} are any indication of the amount of similar topology that {\em truly} exists, then more symmetric measures such as the $S^3$ score are likely to come to the fore as PPI networks become more complete.
At the moment, however, we note that using EC alone is not a good idea: the “uniq” column of Table \ref{tab:orthologs} shows that there is little overlap in the orthologs recovered by the various measures, suggesting that all of them have merit and that more information may be gleaned by using all of them rather than any one of them, including EC.

Figure \ref{fig:IIDrat-mouse} (as well as Supplementary Figure \ref{fig:degree-orthologs-recovered}) demonstrate that orthologs that were recovered in Table \ref{tab:orthologs} typically had a degree in their own networks of 3--5x the degree of orthologs that were not recovered. Not surprisingly, this clearly indicates that degree---essentially a surrogate for local edge density---plays a large role in the ability of network topology to recover biologically relevant alignments. We believe the discussion around Hypothesis R1 in section \ref{sec:infoTheory}, especially the number of bits discussed in relation to the size of the search space (Equation \ref{eq:searchSpace}), provides the beginnings of an understanding of {\em how much} topology is required before ``good'' topological network alignments can be expected to exist.

The number of BioGRID edges for human has been growing steadily by about 30\% per year over the past decade. This strong growth of PPI network data suggests that the information theory requirements will be met within the next few years, suggesting a bright future on the horizon for topology-based analysis of PPI networks.

\section{Methods}\label{sec:Methods}
\subsection{Information theory in the context of network alignment}\label{sec:InfoTheoryIntro}
Information theory, first introduced by Claud Shannon in 1948 \cite{shannon1948mathematical} is a very mature and well-developed field with many texts available (eg., \cite{kullback1997information,klir2005uncertainty}). One question it can help answer is, ``What is the bare minimum amount of information required to answer a well-posed question?'' Information is often quantified in bits. For example, consider the domain of a set with exactly two elements, $\{a,b\}$. Uniquely identifying one of the elements requires exactly 1 bit of information: the answer to the question ``is it $a$?'' (1 for yes, 0 for no, the latter directly implying the chosen member is $b$). If there are 4 items $\{a,b,c,d\}$, then two bits are required: the first bit splits the set in two (eg $\{a,b\}$ and $\{c,d\}$), and the second bit chooses between the first and second element of the appropriate pair. In general, if there are $n=2^k$ elements in a set, one absolutely requires at least $k$ bits to uniquely identify one element in that set. Taking the base-2 log of both sides, we see that $\log_2(n)=k$ bits are required to uniquely identify one element among $n$. Thus, from the base-2 log of Equation \ref{eq:searchSpace}, we can compute the absolute bare minimum amount of information, in bits, required to uniquely identify one alignment. If we would be satisfied---or are only able---to identify any member of a set of $N$ ``equivalent'' alignments, the information requirement is reduced by $log_2 N$ bits.

We emphasize that this bound is a fundamental limit. It does not depend on the objective function used, or the search algorithm. It's a statement that, if not enough information is on hand, then in general it will be impossible for {\em any} algorithm to isolate a relevant alignment.

Thus, the flip side of this question is, ``how much information, in bits, do we have at hand, given two networks we are trying to align?" If we can quantify the information content of the two networks, we can determine if they contain {\em enough} information to identify relevant alignments. The problem is that precisely measuring the amount of information in a ``message" (eg., a network) is very hard. This is why, for example, compression algorithms are constantly being improved: how much information, in bits, is in an English sentence? It’s clearly fewer than the bits required to represent all its characters, b/c fr xampl we cn remv most vwls and stll ndrstnd th sntnce. (Note: the spelling mistakes are {\em intentional}, to demonstrate our point.)

So, the information content of a network is hard to estimate---and can depend on context. For example, if we first specify the domain of networks to be cliques and nothing else, then we need only specify the number of nodes in the clique---even a 1,000-node clique requires only enough bits to specify the number 1,000 (10 bits), which is far less than the number of edges. However, we don't know enough about the structure of PPI networks to constrain them to any particular subset of networks (a PPI network may not even be connected). So for now, given that the number of `1' bits in the adjacency matrix of an undirected network is exactly the number of edges, we hypothesize that the number of edges provides an approximate lower bound on the amount of information contained in a network, and leave a better estimate to further work.

As Figure \ref{fig:IIDrat-mouse} demonstrates, the gap between our estimate of the information content of networks, and the number of bits required to uniquely specify an alignment is only a few tens of perent---though this still amounts to thousands of bits. Closing this gap further will require significant advances towards a theory of network alignment, and significant development in the formal topological information content of PPI networks. 

\subsection{Exactly computing the logarithm of large integers}\label{sec:base2log}
Equation \ref{eq:searchSpace} represents the size of the search space when aligning two networks; given that $n_1$ and $n_2$ are typically in the thousands, that number is enormous. However, we can easily compute its base-2 logarithm {\em exactly}, as follows:
\begin{align}
\log_2\left(\frac{n_2!}{(n_2-n_1)!}\right)
    &= \log_2 \left[n_2\times(n_2-1)\times(n_2-2)\times(n_2-3)\times\ldots \times(n_2-n_1+1)\right] \nonumber \\
    &= \log_2\prod_{k=n_2-n_1+1}^{n_2} k \nonumber \\
    &= \sum_{k=n_2-n_1+1}^{n_2} \log_2(k).\label{eq:searchSpaceBits}
\end{align}
Since $k$ has values ranging ``only'' into the thousands, we can easily compute all the required base-2 logarithms, and sum them. These are the values listed in the third column of Table \ref{tab:BioGRID}, for $n_1=n_2=n$ for an $n$-node network.

\subsection{Comparing SANA's alignments to those of competing aligners}\label{sec:WeBeatMethods}
In the cases of GRAAL, MAGNA++, OptNetAlign, L-GRAAL, HubAlign, ModuleAlign, WAVE, PROPER, and INDEX, we natively programmed their custom objective functions into SANA. In the case of GHOST, we modified GHOST to output its similarities for all node pairs, and then read that similarity file into SANA. In the case of GREAT, we used GREAT to generate required edge-based graphlet degree vectors (EGDVs) and their pairwise similarities, and then read in the edge-pair matrix that evaluates edge-to-edge similarity. (Note that this is a {\em large} matrix of size $m_1\times m_2$, where $m_i$ is the number of {\em edges} in network $G_i$.) Since we are interested in alignments based {\em only} on topology, we omit comparison against any algorithm that {\em requires} sequence---i.e., any algorithm that cannot perform topology-only network alignment.  For example, via personal communication with the authors, we have verified that NATALIE\cite{NATALIE}, NATALIE 2.0\cite{NATALIE2}, and PrimAlign\cite{kalecky2018primalign} all {\em require} the use of pairwise sequence similarities. Several published alignment papers \cite{Ulign,gligorijevic2015fuse,alberich2019alignet} explicitly state that they regard the use of sequence similarities as necessary, and in any case their topological objectives are usually some minor variant of EC and not worth considering independent of EC.

We ran each of these algorithms on all 28 combinations of the 8 largest PPI networks from BioGRID version 3.4.156 from January 2018. 

\subsection{Computing the $p$-value of recovered orthologs}\label{sec:Ortho-p-values}
Given a pair of networks $G_1,G_2$ with $n_1\le n_2$ nodes, assume $h$ is the number of known orthologs between them. To compute the $p$-value of having recovered $r$ orthologs, we need to know the expected number of aligned orthologs in an alignment chosen uniformly at random from the search space. (Note that this has nothing to do with running SANA; we are estimating the ortholog recovery rate in {\em random} alignments, not good ones.) We describe two methods for computing this expected value: first, a simpler method that works well when $h/n_1\ll 1$, and then a more accurate one that, although still approximate, appears to work well for any value of $h/n_1$.

Given a particular pair of orthologs $(u_1,u_2)$, the probability that $u_1$ is aligned to $u_2$ by chance in a random alignment is $1/n_2$; they are misaligned with probability $(1-1/n_2)$. Given two orthologous pairs $(u_1,u_2)$ and $(v_1,v_2)$, the probability that both are misaligned is well-approximated by $(1-1/n_2)^2$; this value is not exact because, since SANA produces only 1-to-1 alignments, $u_1$ and $v_1$ cannot simultaneously be aligned to (for example) $u_2$. This results in a very weak but difficult to disentangle dependence, which---in this first, simpler method---we shall ignore. Thus, the probability that all $h$ orthologs are misaligned can be approximated by $(1-1/n_2)^h\approx (1-h/n_2)$ since $1/n_2$ is small. Thus, we see that the probability that {\em at least one} of any of the $h$ orthologs is correctly aligned is $\approx 1-(1-h/n_2)$, which is simply $h/n_2$. This value makes intuitive sense if we consider that each of the $h$ orthologs contributes an equal ``chance'' to being aligned, each with probability $1/n_2$.

The above approximation can break down when $h/n_2$ is not small.
The second method involves using the Hypergeometric distribution, though still only approximately. We start by noting that when $n_1<n_2$, then only $n_1$ out of the $n_2$ nodes of $G_2$ are aligned. In a random alignment, we therefore expect that, on average, only $h'=\frac{n_1}{n_2}h$ of $G_2$'s orthologs are included among its aligned nodes. (For simplicity, assume the value of $h'$ is rounded to the nearest integer.) This in turn directly implies that there are only $h'$ orthologs {\em available} for alignment with $G_1$, even though $G_1$'s nodes include all $h$ of its orthologs. Recall the Hypergeometric distribution $\mathcal H(i,m,M,N)$ describes the probability of $i$ successes in $m$ draws from a finite population $N$ in which $M$ objects have the desired property. To estimate the probability of 1 or more aligned orthologs, we should compute $1-\mathcal H(0,n_1,h',n_1^2)$, since there are $n_1$ aligned pairs (the number of draws), and $h'$ available ``successes'' out of $n_1^2$ pairs---the first $n_1$ coming from $G_1$, the second from the subset of $G_1$'s nodes that are aligned. This would be exact except for the complication that, though the Hypergeometric distribution already accounts for non-replacement, we have the additional constraint imposed by the 1-to-1 property of our alignments: after choosing an aligned pair $(u,v)$ from the population of $N$, we must remove not only that pair, but {\em all other} pairs that contain {\em either} $u$ or $v$. We call this the {\it 1-to-1 constraint}, and it violates the model of the Hypergeometric distribution in which the population is {\em constant} and that only {\em one member} of the population is removed with each draw. To approximately account for this, we must estimate the size of a surrogate population that provides good estimates of the desired probabilities. We have found empirically that using $n_1$ throughout gives probabilities that are far too low, but that good results are obtained if we substitute $n_1$ with the approximate number of orthologs across ``valid'' populations---i.e., populations that satisfy the 1-to-1 constraint. Thus, our surrogate population, of constant size $N'$ designed to mimic the effect of a dynamically decreasing population, is
\begin{equation}
    N' = n_1^2 - \sum_{k=1}^{h''} k = n_1^2 - \frac{h''(h''+1)}{2},\label{eq:RandomOrtho-HyperSum}
\end{equation}
where $h''$ is the arithmetic mean of $h$ and $h'$. We then use $N'$ instead of $N$ as the constant population size for the Hypergeometric distribution above, arriving finally at the expression $\lambda_{n_1,n_2,h}=1-\mathcal H(0,n_1,h',N')$ which estimates the mean recovery rate of aligned orthologs in random alignments.

\begin{figure}[tb]
    \centering
    \includegraphics[width=0.49 \textwidth]{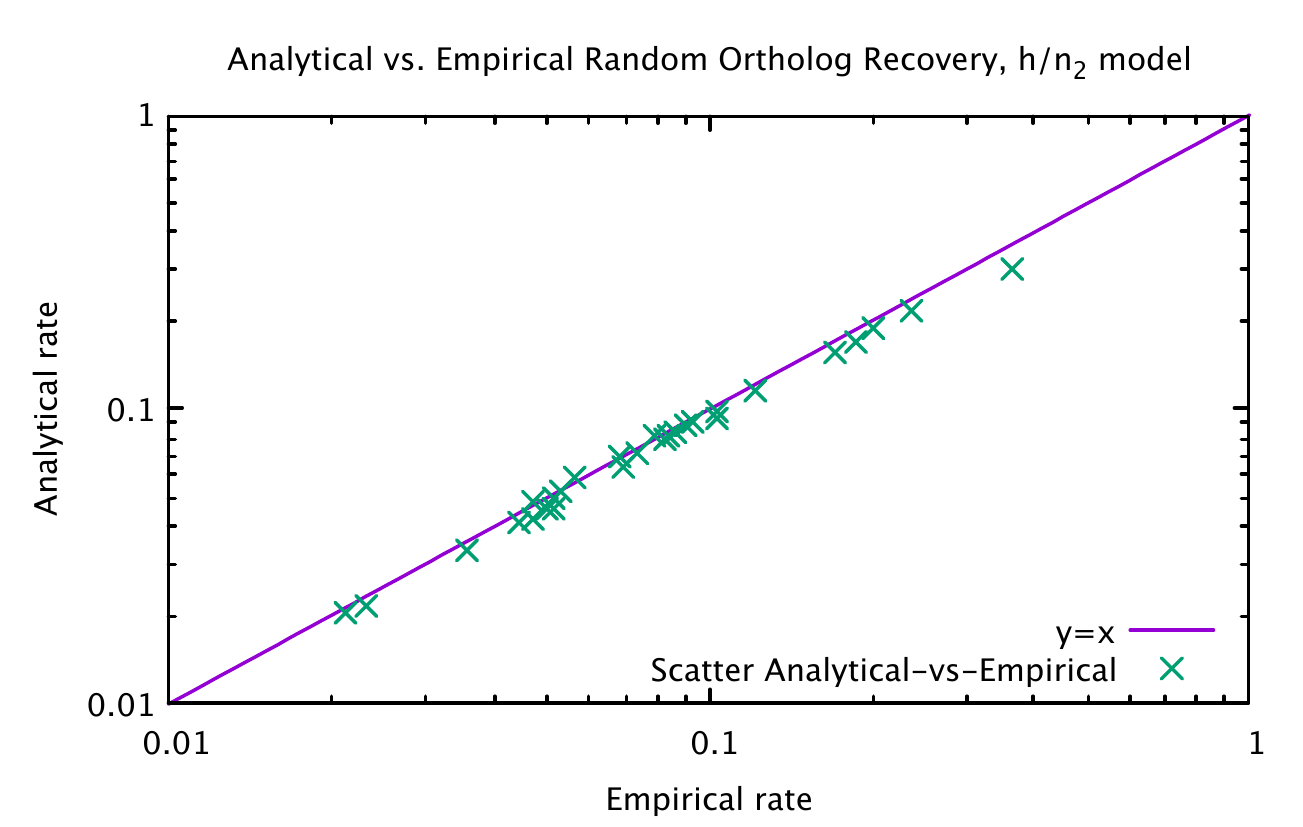}
    \includegraphics[width=0.49 \textwidth]{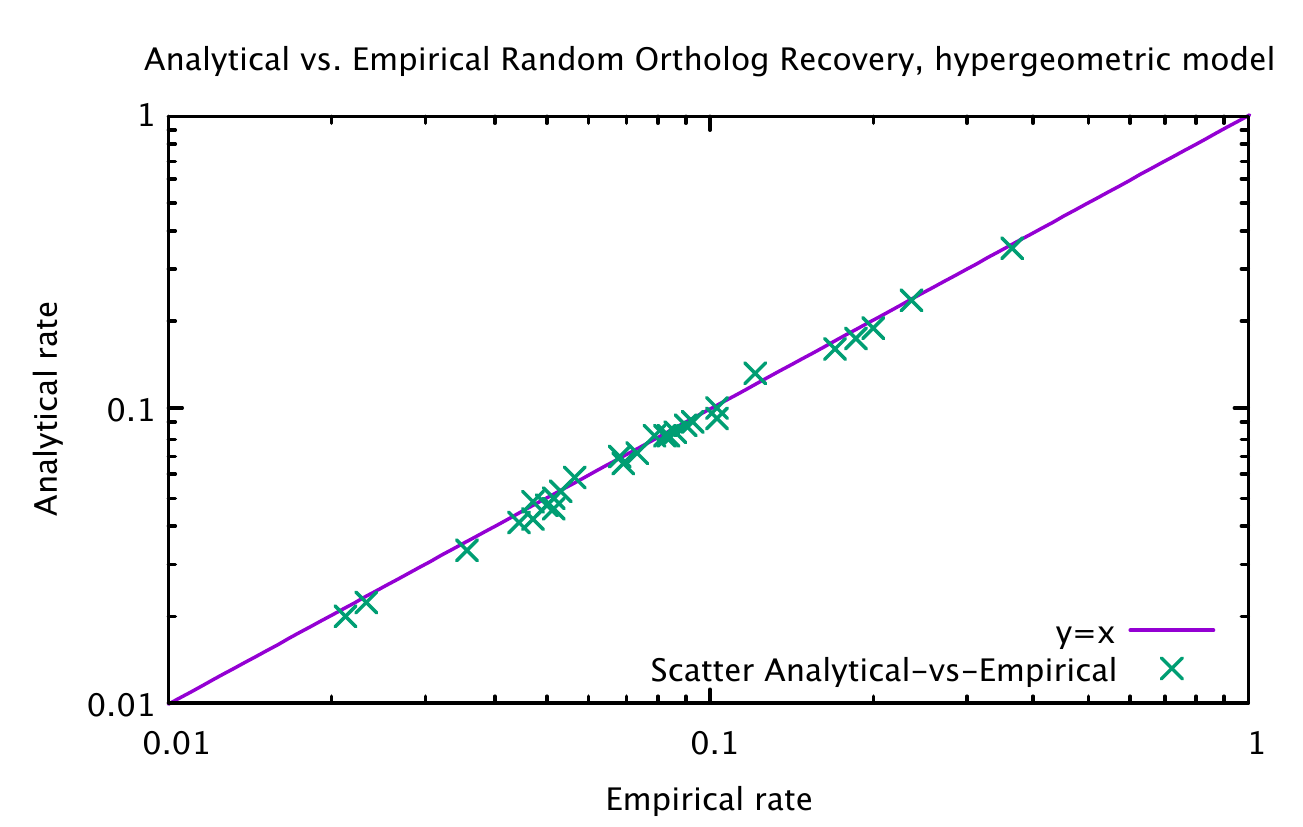}
    \caption{{\bf Estimating rate of ortholog pair recovery in {\em random} alignments:} We compare the empirically determined mean number of paired orthologs recovered in random alignments across 10,000 randomly generated alignments (horizontal axis) vs. the analytical version using the simpler model (left), and the more complex Hypergeometric model of Equation \ref{eq:RandomOrtho-HyperSum} (right). They diverge slightly towards the upper right, where the estimates from the simpler $h/n_2$ model are slightly less accurate than the Hypergeometric-based one.}
    \label{fig:RandomOrtho-vs-HyperSum}
\end{figure}

We generated 10,000 random alignments of each of the 28 species pairs, in order to empirically estimate the mean ortholog recovery rate of random alignments, and compared the result to both analytical estimates. The result, depicted in Figure \ref{fig:RandomOrtho-vs-HyperSum}, shows good agreement.

With the per-alignment rate of recovering ortholog pairs in a single random alignment, we now finally come to the task of estimating the $p$-values in Table \ref{tab:orthologs}. We interpret each random alignment as a Bernoulli trial with ``success'' defined as ``at least one pair of correctly aligned orthologs'', with probability computed via the second method above. The result is that the number of recovered orthologs per random alignment follows a Poisson distribution with expected recovery rate $\lambda_{n_1,n_2,h}$. Since each alignment is independent, and the expected total rate among $T$ independent Poisson distributions is the sum of their individual rates, we see that in $T$ random alignments, the expected number of recovered orthologs is $T\lambda_{n_1,n_2,h}$. The $p$-value listed in the last column of Table \ref{tab:orthologs} is computed taking $i$ to be the number of actual recovered orthologs for the ``winning'' measure (eg., $i=72$ for the EC measure in the SC-HS row), and then summing the tail of the Poisson distribution from $i$ upwards towards infinity until new terms are too small to change the floating-point sum.

Note that this analysis works even in the case of non-1-to-1 orthologs, so long as we count every possible pair among non-1-to-1 orthologs. So for example if $y_1,y_2,y_3,y_4$ are orthologs in yeast that are each orthologous to all of $h_1,h_2,h_3$ in human, then there are $4\times 3=12$ totals pairs. The number in the ``orthologs'' column of Table \ref{tab:orthologs} accounts for all such pairs.

\subsection{Computing the $p$-value of shared GO terms in an alignment}\label{sec:GOp-value}
We use GO terms downloaded from the Gene Ontology website in Sept. 2018, and eliminate all sequence-based GO terms. In particular, we only allow GO terms with evidence codes EXP, HDA, HEP, HGI, HMP, IC, IDA, IEP, IGI, IKR, IMR, IPI, IRD, NAS, ND, and TAS. Similar to the $p$-value computation of recovered orthologs detailed above, we use the hypergeometric distribution $\mathcal H(i,m,M,N)$, where $i$ is the observed number of protein pairs sharing GO term $g$, $m$ is the number of protein pairs in the alignment, $N$ is the number of protein pairs in the full Cartesian product of $V_1\times V_2$ (ie., all possibly protein pairings between $G_1$ and $G_2$), and $M$ is the number of such pairs that share $g$.

\subsection{On the importance of choosing the right measure of topological similarity}\label{sec:EC-vs-S3}

To demonstrate that no one topological measure is universally best, we will compare two commonly used measures of topological similarities: EC \cite{GRAAL} and $S^3$---the {\it Symmetric Substructure Score}\cite{MAGNA}. Both are a ratio of edge counts. In both cases, the numerator is the number of ``purple'' edges in the alignment (cf. Figure \ref{fig:NetAlign}). Clearly, more purple is better since purple edges highlight exactly what we mean by ``common topology'' between networks. The difference between EC and $S^3$ is in the denominator: EC simply uses the number of edges in $G_1$, since that is clearly an upper bound on the number of purple edges. However, as pointed out by \cite{MAGNA}, EC is asymmetric with respect to edge ``color'': if we swap $G_1$ and $G_2$, then the value of EC changes. Even more worrisome is that unaligned ``red'' edges are not penalized at all by EC: an alignment can have many unaligned red edges so long as blue edges are minimized. To rectify this, the {\em Symmetric Substructure Score} $S^3$ was created \cite{MAGNA}: the numerator is still the number of purple edges, but the denominator counts {\em all edges of all colors} in the induced subgraph on the purple nodes. Thus, if $G_1$ can be embedded with the same numerator in two different places in $G_2$, then $S^3$ will prefer the embedding that has fewer unaligned {\em red} edges, even though both embeddings have the same EC score.

The $S^3$ score clearly is a better measure when we know---or expect---that two networks genuinely contain regions of high similarity---meaning, there exist large regions with highly similar {\em induced} structure. In that case, a ``good'' alignment will exist in which there is a large number of purple edges, and very few red or blue unaligned edges. Such is the case, for example, with any alignment between the mammals of the 2018 release of the IID networks \cite{kotlyar2018iid}. The mammal PPI networks in IID-2018 were created by taking the experimentally determined edges for that mammal, and adding in all edges between proteins that had 1-to-1 orthologs with human. The result is that most of the edges in these networks are transferred directly from human experimental PPIs, which makes them all {\em highly} similar to the human PPI network---possibly overly so. For example, the next most complete mammal PPI network after human in BioGRID 3.4.164 is mouse, but it contains only about 20,000 edges on 7,000 nodes, compared to human's 280,000 edges on 17,000 nodes (cf. Table \ref{tab:BioGRID}). In other words, the mouse PPI network is likely less than 11\% complete (20,000/280,000$\approx$0.11)
Observing Table \ref{tab:IID_EC_vs_S3}, we see that when IID synthetically augments mammalian PPI networks using interologs from human, then using SANA to optimize $S^3$ for 1 hour between human and any of these networks recovers about 12,500 out of 16,000 orthologous proteins---essentially a 75\% rate of ``node correctness''  between species. Conversely, optimizing EC recovers only about 10,200 orthologs, a decrease of about 20\%. Clearly, $S^3$ is the better measure in cases like this where the PPI networks contain large regions that are almost identical.

Now consider a very different scenario: starting with some complete network $G_2$ that has more than enough edges to self-align according to Equation \ref{eq:Netbits}, and let $G_1$ be a sub-network where 95\% of $G_2$'s edges have been removed at random.
In the parlance of Figure \ref{fig:NetAlign}, the ``correct'' alignment would have an EC score of exactly 1 (since every blue edge from $G_1$ exists in $G_2$), but the $S^3$ score is only 0.05, since 95\% of red edges between purple nodes remain unaligned. In this case, $S^3$ may prefer to align the nodes of $G_1$ over the ``least dense'' regions of $G_2$, in order to minimize the number of unaligned red edges. This may result in sacrificing purple edges in the numerator in order to reduce red edges in the denominator, resulting in an alignment far from the ``correct'' one. Thus, we see that when edge densities are unbalanced, EC may be a more appropriate measure than $S^3$, precisely {\em because} it is an imbalanced measure, and this imbalance matches the imbalance in edge densities.

This latter case---of EC being a better measure when edge densities are grossly imbalanced---is exactly what we see when we use SANA to align virtually any pair of BioGRID networks: optimizing EC results in alignments that are much better able to align regions of high functional similarity as measured in our companion paper by ability to successfully predict GO terms \cite{WangAtkinsonHayesGOpredict}. This is presumably because the BioGRID networks have highly variable network coverage, so that given any two of them, one is likely to have far more edges than the other simply due to different levels of completeness.


\section{Data availability}
The BioGRID networks we used are available from TheBioGRID.org; only physical interactions between nodes in the stated species were included. SANA is available in source code form at \href{https://github.com/waynebhayes/SANA}{https://github.com/waynebhayes/SANA}.

\section*{Acknowledgements}
We thank Annie Raichev for running NAUTY \cite{NAUTY} on the BioGRID Human PPI network to count the number of automorphisms.

\section*{Author contributions statement}

ADK produced the initial SIF results that launched the project, while BJF updated the SIF results on a wider range of networks and aligners. PK and BAM performed all locking + saturation experiments. XC performed all H-GRAAL and most information theory experiments. SW+BJF together performed all runs and initial analyses for Table \ref{tab:orthologs}. WBH conceived and guided the entire project and performed some of the analyses related to information theory, frequency, GO terms and $p$-values, and created the final manuscript based on drafts of the respective sections written by each author. All authors were given the opportunity to review the final manuscript.
\bibliography{wayne-all}

\begin{thebibliography}{10}
\urlstyle{rm}
\expandafter\ifx\csname url\endcsname\relax
  \def\url#1{\texttt{#1}}\fi
\expandafter\ifx\csname urlprefix\endcsname\relax\def\urlprefix{URL }\fi
\expandafter\ifx\csname doiprefix\endcsname\relax\def\doiprefix{DOI: }\fi
\providecommand{\bibinfo}[2]{#2}
\providecommand{\eprint}[2][]{\url{#2}}

\bibitem{phillips2013physical}
\bibinfo{author}{Phillips, R.}, \bibinfo{author}{Kondev, J.},
  \bibinfo{author}{Theriot, J.} \& \bibinfo{author}{Garcia, H.~G.}
\newblock \emph{\bibinfo{title}{Physical Biology of the Cell}}
  (\bibinfo{publisher}{New York, NY: Garland Science. Available from
  http://www.ncbi.nlm.nih}, \bibinfo{year}{2013}).

\bibitem{alberts2015molecular}
\bibinfo{author}{Alberts, B.} \emph{et~al.}
\newblock \emph{\bibinfo{title}{Molecular Biology of the Cell. The
  Extracellular Matrix of Animals}} (\bibinfo{publisher}{New York, NY: Garland
  Science. Available from http://www.ncbi.nlm.nih}, \bibinfo{year}{2015}).

\bibitem{sequencing2005initial}
\bibinfo{author}{Waterson, R.~H.} \emph{et~al.}
\newblock \bibinfo{journal}{\bibinfo{title}{Initial sequence of the chimpanzee
  genome and comparison with the human genome}}.
\newblock {\emph{\JournalTitle{Nature}}} \textbf{\bibinfo{volume}{437}},
  \bibinfo{pages}{69} (\bibinfo{year}{2005}).

\bibitem{pennacchio2003insights}
\bibinfo{author}{Pennacchio, L.~A.}
\newblock \bibinfo{journal}{\bibinfo{title}{Insights from human/mouse genome
  comparisons}}.
\newblock {\emph{\JournalTitle{Mammalian genome}}}
  \textbf{\bibinfo{volume}{14}}, \bibinfo{pages}{429--436}
  (\bibinfo{year}{2003}).

\bibitem{pearson2013introduction}
\bibinfo{author}{Pearson, W.~R.}
\newblock \bibinfo{journal}{\bibinfo{title}{An introduction to sequence
  similarity (``homology'') searching}}.
\newblock {\emph{\JournalTitle{Current protocols in bioinformatics}}}
  \textbf{\bibinfo{volume}{42}}, \bibinfo{pages}{3--1} (\bibinfo{year}{2013}).

\bibitem{kulmanov2020deepgoplus}
\bibinfo{author}{Kulmanov, M.} \& \bibinfo{author}{Hoehndorf, R.}
\newblock \bibinfo{journal}{\bibinfo{title}{Deepgoplus: improved protein
  function prediction from sequence}}.
\newblock {\emph{\JournalTitle{Bioinformatics}}} \textbf{\bibinfo{volume}{36}},
  \bibinfo{pages}{422--429} (\bibinfo{year}{2020}).

\bibitem{zhang2017cofactor}
\bibinfo{author}{Zhang, C.}, \bibinfo{author}{Freddolino, P.~L.} \&
  \bibinfo{author}{Zhang, Y.}
\newblock \bibinfo{journal}{\bibinfo{title}{Cofactor: improved protein function
  prediction by combining structure, sequence and protein--protein interaction
  information}}.
\newblock {\emph{\JournalTitle{Nucleic acids research}}}
  \textbf{\bibinfo{volume}{45}}, \bibinfo{pages}{W291--W299}
  (\bibinfo{year}{2017}).

\bibitem{furuse1998claudin}
\bibinfo{author}{Furuse, M.}, \bibinfo{author}{Fujita, K.},
  \bibinfo{author}{Hiiragi, T.}, \bibinfo{author}{Fujimoto, K.} \&
  \bibinfo{author}{Tsukita, S.}
\newblock \bibinfo{journal}{\bibinfo{title}{Claudin-1 and-2: novel integral
  membrane proteins localizing at tight junctions with no sequence similarity
  to occludin}}.
\newblock {\emph{\JournalTitle{The Journal of cell biology}}}
  \textbf{\bibinfo{volume}{141}}, \bibinfo{pages}{1539--1550}
  (\bibinfo{year}{1998}).

\bibitem{schlicker2006new}
\bibinfo{author}{Schlicker, A.}, \bibinfo{author}{Domingues, F.~S.},
  \bibinfo{author}{Rahnenf{\"u}hrer, J.} \& \bibinfo{author}{Lengauer, T.}
\newblock \bibinfo{journal}{\bibinfo{title}{A new measure for functional
  similarity of gene products based on gene ontology}}.
\newblock {\emph{\JournalTitle{BMC bioinformatics}}}
  \textbf{\bibinfo{volume}{7}}, \bibinfo{pages}{302} (\bibinfo{year}{2006}).

\bibitem{kabsch1984use}
\bibinfo{author}{Kabsch, W.} \& \bibinfo{author}{Sander, C.}
\newblock \bibinfo{journal}{\bibinfo{title}{On the use of sequence homologies
  to predict protein structure: identical pentapeptides can have completely
  different conformations}}.
\newblock {\emph{\JournalTitle{Proceedings of the National Academy of
  Sciences}}} \textbf{\bibinfo{volume}{81}}, \bibinfo{pages}{1075--1078}
  (\bibinfo{year}{1984}).

\bibitem{morrone2011denatured}
\bibinfo{author}{Morrone, A.} \emph{et~al.}
\newblock \bibinfo{journal}{\bibinfo{title}{The denatured state dictates the
  topology of two proteins with almost identical sequence but different native
  structure and function}}.
\newblock {\emph{\JournalTitle{Journal of Biological Chemistry}}}
  \textbf{\bibinfo{volume}{286}}, \bibinfo{pages}{3863--3872}
  (\bibinfo{year}{2011}).

\bibitem{madsen1999psoriasis}
\bibinfo{author}{Madsen, P.} \emph{et~al.}
\newblock \bibinfo{journal}{\bibinfo{title}{Psoriasis upregulated phorbolin-1
  shares structural but not functional similarity to the mrna-editing protein
  apobec-1}}.
\newblock {\emph{\JournalTitle{Journal of Investigative Dermatology}}}
  \textbf{\bibinfo{volume}{113}}, \bibinfo{pages}{162--169}
  (\bibinfo{year}{1999}).

\bibitem{VidalPPI01}
\bibinfo{author}{Walhout, A.~J.} \& \bibinfo{author}{Vidal, M.}
\newblock \bibinfo{journal}{\bibinfo{title}{High-throughput yeast two-hybrid
  assays for large-scale protein interaction mapping}}.
\newblock {\emph{\JournalTitle{Methods}}} \textbf{\bibinfo{volume}{24}},
  \bibinfo{pages}{297--306} (\bibinfo{year}{2001}).

\bibitem{Milenkovic:2013:GNA:2506583.2508968}
\bibinfo{author}{Milenkovi\'{c}, T.}, \bibinfo{author}{Zhao, H.} \&
  \bibinfo{author}{Faisal, F.~E.}
\newblock \bibinfo{title}{Global network alignment in the context of aging}.
\newblock In \emph{\bibinfo{booktitle}{Proceedings of the International
  Conference on Bioinformatics, Computational Biology and Biomedical
  Informatics}}, BCB'13, \bibinfo{pages}{23:23--23:32},
  \doiprefix\url{10.1145/2506583.2508968} (\bibinfo{publisher}{ACM},
  \bibinfo{address}{New York, NY, USA}, \bibinfo{year}{2013}).

\bibitem{clark2014comparison}
\bibinfo{author}{Clark, C.} \& \bibinfo{author}{Kalita, J.}
\newblock \bibinfo{journal}{\bibinfo{title}{A comparison of algorithms for the
  pairwise alignment of biological networks}}.
\newblock {\emph{\JournalTitle{Bioinformatics}}} \textbf{\bibinfo{volume}{30}},
  \bibinfo{pages}{2351--2359} (\bibinfo{year}{2014}).

\bibitem{MamanoHayesSANA}
\bibinfo{author}{Mamano, N.} \& \bibinfo{author}{Hayes, W.~B.}
\newblock \bibinfo{journal}{\bibinfo{title}{{SANA}: Simulated annealing far
  outperforms many other search algorithms for biological network alignment.}}
\newblock {\emph{\JournalTitle{Bioinformatics (Oxford, England)}}}
  \textbf{\bibinfo{volume}{33}}, \bibinfo{pages}{2156--2164}
  (\bibinfo{year}{2017}).

\bibitem{GRAAL}
\bibinfo{author}{Kuchaiev, O.}, \bibinfo{author}{Milenkovi{\'c}, T.},
  \bibinfo{author}{Memi{\v s}evi{\'c}, V.}, \bibinfo{author}{Hayes, W.} \&
  \bibinfo{author}{Pr\v{z}ulj, N.}
\newblock \bibinfo{journal}{\bibinfo{title}{Topological network alignment
  uncovers biological function and phylogeny}}.
\newblock {\emph{\JournalTitle{Journal of The Royal Society Interface}}}
  \textbf{\bibinfo{volume}{7}}, \bibinfo{pages}{1341--1354},
  \doiprefix\url{10.1098/rsif.2010.0063} (\bibinfo{year}{2010}).

\bibitem{MIGRAAL}
\bibinfo{author}{Kuchaiev, O.} \& \bibinfo{author}{Pr\v{z}ulj, N.}
\newblock \bibinfo{journal}{\bibinfo{title}{Integrative network alignment
  reveals large regions of global network similarity in yeast and human}}.
\newblock {\emph{\JournalTitle{BIOINFORMATICS}}} \textbf{\bibinfo{volume}{27}},
  \bibinfo{pages}{1390--1396}, \doiprefix\url{bioinformatics/btr127}
  (\bibinfo{year}{2011}).

\bibitem{faisal2014global}
\bibinfo{author}{Faisal, F.~E.}, \bibinfo{author}{Zhao, H.} \&
  \bibinfo{author}{Milenkovi{\'c}, T.}
\newblock \bibinfo{journal}{\bibinfo{title}{Global network alignment in the
  context of aging}}.
\newblock {\emph{\JournalTitle{IEEE/ACM Transactions on Computational Biology
  and Bioinformatics}}} \textbf{\bibinfo{volume}{12}}, \bibinfo{pages}{40--52}
  (\bibinfo{year}{2014}).

\bibitem{DavisPrzulj2015TopoFunction}
\bibinfo{author}{Davis, D.}, \bibinfo{author}{Yavero{\u{g}}lu, {\"O}.~N.},
  \bibinfo{author}{Malod-Dognin, N.}, \bibinfo{author}{Stojmirovic, A.} \&
  \bibinfo{author}{Pr{\v{z}}ulj, N.}
\newblock \bibinfo{journal}{\bibinfo{title}{Topology-function conservation in
  protein--protein interaction networks}}.
\newblock {\emph{\JournalTitle{Bioinformatics}}} \textbf{\bibinfo{volume}{31}},
  \bibinfo{pages}{1632--1639}, \doiprefix\url{10.1093/bioinformatics/btv026}
  (\bibinfo{year}{2015}).

\bibitem{gaudelet2018higher}
\bibinfo{author}{Gaudelet, T.}, \bibinfo{author}{Malod-Dognin, N.} \&
  \bibinfo{author}{Pr{\v{z}}ulj, N.}
\newblock \bibinfo{journal}{\bibinfo{title}{Higher-order molecular organization
  as a source of biological function}}.
\newblock {\emph{\JournalTitle{Bioinformatics}}} \textbf{\bibinfo{volume}{34}},
  \bibinfo{pages}{i944--i953} (\bibinfo{year}{2018}).

\bibitem{malod2019functional}
\bibinfo{author}{Malod-Dognin, N.} \& \bibinfo{author}{Pr{\v{z}}ulj, N.}
\newblock \bibinfo{journal}{\bibinfo{title}{Functional geometry of protein
  interactomes}}.
\newblock {\emph{\JournalTitle{Bioinformatics}}}  (\bibinfo{year}{2019}).

\bibitem{Ulign}
\bibinfo{author}{Malod-Dognin, N.}, \bibinfo{author}{Ban, K.} \&
  \bibinfo{author}{Pr{\v{z}}ulj, N.}
\newblock \bibinfo{journal}{\bibinfo{title}{Unified alignment of
  protein-protein interaction networks}}.
\newblock {\emph{\JournalTitle{Scientific Reports}}}
  \textbf{\bibinfo{volume}{7}}, \bibinfo{pages}{953} (\bibinfo{year}{2017}).

\bibitem{gligorijevic2015fuse}
\bibinfo{author}{Gligorijevi{\'c}, V.}, \bibinfo{author}{Malod-Dognin, N.} \&
  \bibinfo{author}{Pr{\v{z}}ulj, N.}
\newblock \bibinfo{journal}{\bibinfo{title}{Fuse: multiple network alignment
  via data fusion}}.
\newblock {\emph{\JournalTitle{Bioinformatics}}} \bibinfo{pages}{btv731}
  (\bibinfo{year}{2015}).

\bibitem{NATALIE}
\bibinfo{author}{Klau, G.}
\newblock \bibinfo{journal}{\bibinfo{title}{A new graph-based method for
  pairwise global network alignment}}.
\newblock {\emph{\JournalTitle{BMC Bioinformatics}}}
  \textbf{\bibinfo{volume}{10}}, \bibinfo{pages}{S59},
  \doiprefix\url{10.1186/1471-2105-10-S1-S59} (\bibinfo{year}{2009}).

\bibitem{NATALIE2}
\bibinfo{author}{El-Kebir, M.}, \bibinfo{author}{Heringa, J.} \&
  \bibinfo{author}{Klau, G.~W.}
\newblock \bibinfo{title}{Lagrangian relaxation applied to sparse global
  network alignment}.
\newblock In \emph{\bibinfo{booktitle}{IAPR International Conference on Pattern
  Recognition in Bioinformatics}}, \bibinfo{pages}{225--236}
  (\bibinfo{organization}{Springer}, \bibinfo{year}{2011}).

\bibitem{kalecky2018primalign}
\bibinfo{author}{Kalecky, K.} \& \bibinfo{author}{Cho, Y.-R.}
\newblock \bibinfo{journal}{\bibinfo{title}{Primalign: Pagerank-inspired
  markovian alignment for large biological networks}}.
\newblock {\emph{\JournalTitle{Bioinformatics}}} \textbf{\bibinfo{volume}{34}},
  \bibinfo{pages}{i537--i546} (\bibinfo{year}{2018}).

\bibitem{alberich2019alignet}
\bibinfo{author}{Alberich, R.}, \bibinfo{author}{Alcala, A.},
  \bibinfo{author}{Llabres, M.}, \bibinfo{author}{Rossello, F.} \&
  \bibinfo{author}{Valiente, G.}
\newblock \bibinfo{journal}{\bibinfo{title}{Alignet: alignment of
  protein-protein interaction networks}}.
\newblock {\emph{\JournalTitle{arXiv preprint arXiv:1902.07107}}}
  (\bibinfo{year}{2019}).

\bibitem{H-GRAAL}
\bibinfo{author}{Milenkovi{\'c}, T.}, \bibinfo{author}{Ng, W.~L.},
  \bibinfo{author}{Hayes, W.} \& \bibinfo{author}{Pr{\v{z}}ulj, N.}
\newblock \bibinfo{journal}{\bibinfo{title}{Optimal network alignment with
  graphlet degree vectors}}.
\newblock {\emph{\JournalTitle{Cancer informatics}}}
  \textbf{\bibinfo{volume}{9}}, \bibinfo{pages}{121} (\bibinfo{year}{2010}).

\bibitem{GEDEVO}
\bibinfo{author}{Ibragimov, R.}, \bibinfo{author}{Malek, M.},
  \bibinfo{author}{Guo, J.} \& \bibinfo{author}{Baumbach, J.}
\newblock \bibinfo{title}{Gedevo: an evolutionary graph edit distance algorithm
  for biological network alignment}.
\newblock In \emph{\bibinfo{booktitle}{OASIcs-OpenAccess Series in
  Informatics}}, vol.~\bibinfo{volume}{34} (\bibinfo{organization}{Schloss
  Dagstuhl-Leibniz-Zentrum fuer Informatik}, \bibinfo{year}{2013}).

\bibitem{CytoGEDEVO}
\bibinfo{author}{Malek, M.}, \bibinfo{author}{Ibragimov, R.},
  \bibinfo{author}{Albrecht, M.} \& \bibinfo{author}{Baumbach, J.}
\newblock \bibinfo{journal}{\bibinfo{title}{{CytoGEDEVO}-global alignment of
  biological networks with cytoscape}}.
\newblock {\emph{\JournalTitle{Bioinformatics}}} \textbf{\bibinfo{volume}{32}},
  \bibinfo{pages}{1259--1261} (\bibinfo{year}{2016}).

\bibitem{zhu2017gmalign}
\bibinfo{author}{Zhu, Y.}, \bibinfo{author}{Li, Y.}, \bibinfo{author}{Liu, J.},
  \bibinfo{author}{Qin, L.} \& \bibinfo{author}{Yu, J.~X.}
\newblock \bibinfo{title}{Gmalign: A new network aligner for revealing large
  conserved functional components}.
\newblock In \emph{\bibinfo{booktitle}{2017 IEEE International Conference on
  Bioinformatics and Biomedicine (BIBM)}}, \bibinfo{pages}{120--127}
  (\bibinfo{organization}{IEEE}, \bibinfo{year}{2017}).

\bibitem{elmsallati2018index}
\bibinfo{author}{Elmsallati, A.}, \bibinfo{author}{Msalati, A.} \&
  \bibinfo{author}{Kalita, J.}
\newblock \bibinfo{journal}{\bibinfo{title}{Index-based network aligner of
  protein-protein interaction networks}}.
\newblock {\emph{\JournalTitle{IEEE/ACM Transactions on Computational Biology
  and Bioinformatics (TCBB)}}} \textbf{\bibinfo{volume}{15}},
  \bibinfo{pages}{330--336} (\bibinfo{year}{2018}).

\bibitem{xie2016adaptive}
\bibinfo{author}{Xie, J.} \emph{et~al.}
\newblock \bibinfo{journal}{\bibinfo{title}{An adaptive hybrid algorithm for
  global network alignment}}.
\newblock {\emph{\JournalTitle{IEEE/ACM Transactions on Computational Biology
  and Bioinformatics (TCBB)}}} \textbf{\bibinfo{volume}{13}},
  \bibinfo{pages}{483--493} (\bibinfo{year}{2016}).

\bibitem{PROPER}
\bibinfo{author}{Kazemi, E.}, \bibinfo{author}{Hassani, H.},
  \bibinfo{author}{Grossglauser, M.} \& \bibinfo{author}{Modarres, H.~P.}
\newblock \bibinfo{journal}{\bibinfo{title}{Proper: global protein interaction
  network alignment through percolation matching}}.
\newblock {\emph{\JournalTitle{BMC bioinformatics}}}
  \textbf{\bibinfo{volume}{17}}, \bibinfo{pages}{527} (\bibinfo{year}{2016}).

\bibitem{yasar2018iterative}
\bibinfo{author}{Yasar, A.} \& \bibinfo{author}{{\c{C}}ataly{\"u}rek,
  {\"U}.~V.}
\newblock \bibinfo{title}{An iterative global structure-assisted labeled
  network aligner}.
\newblock In \emph{\bibinfo{booktitle}{Proceedings of the 24th ACM SIGKDD
  International Conference on Knowledge Discovery \& Data Mining}},
  \bibinfo{pages}{2614--2623} (\bibinfo{organization}{ACM},
  \bibinfo{year}{2018}).

\bibitem{tame}
\bibinfo{author}{Mohammadi, S.}, \bibinfo{author}{Gleich, D.~F.},
  \bibinfo{author}{Kolda, T.~G.} \& \bibinfo{author}{Grama, A.}
\newblock \bibinfo{journal}{\bibinfo{title}{Triangular alignment (tame): A
  tensor-based approach for higher-order network alignment}}.
\newblock {\emph{\JournalTitle{IEEE/ACM transactions on computational biology
  and bioinformatics}}}  (\bibinfo{year}{2016}).

\bibitem{GHOST}
\bibinfo{author}{Patro, R.} \& \bibinfo{author}{Kingsford, C.}
\newblock \bibinfo{journal}{\bibinfo{title}{Global network alignment using
  multiscale spectral signatures}}.
\newblock {\emph{\JournalTitle{Bioinformatics}}} \textbf{\bibinfo{volume}{28}},
  \bibinfo{pages}{3105--3114}, \doiprefix\url{10.1093/bioinformatics/bts592}
  (\bibinfo{year}{2012}).
\newblock
  \eprint{http://bioinformatics.oxfordjournals.org/content/28/23/3105.full.pdf+html}.

\bibitem{alkan2015sipan}
\bibinfo{author}{Alkan, F.} \& \bibinfo{author}{Erten, C.}
\newblock \bibinfo{journal}{\bibinfo{title}{Sipan: simultaneous prediction and
  alignment of protein--protein interaction networks}}.
\newblock {\emph{\JournalTitle{Bioinformatics}}} \textbf{\bibinfo{volume}{31}},
  \bibinfo{pages}{2356--2363} (\bibinfo{year}{2015}).

\bibitem{HubAlign}
\bibinfo{author}{Hashemifar, S.} \& \bibinfo{author}{Xu, J.}
\newblock \bibinfo{journal}{\bibinfo{title}{{HubAlign: an accurate and
  efficient method for global alignment of protein-protein interaction
  networks}}}.
\newblock {\emph{\JournalTitle{Bioinformatics}}} \textbf{\bibinfo{volume}{30}},
  \bibinfo{pages}{i438--i444}, \doiprefix\url{10.1093/bioinformatics/btu450}
  (\bibinfo{year}{2014}).

\bibitem{LGRAAL}
\bibinfo{author}{Malod-Dognin, N.} \& \bibinfo{author}{Pr\v{z}ulj, N.}
\newblock \bibinfo{journal}{\bibinfo{title}{L-graal: Lagrangian graphlet-based
  network aligner}}.
\newblock {\emph{\JournalTitle{Bioinformatics}}}
  \doiprefix\url{10.1093/bioinformatics/btv130} (\bibinfo{year}{2015}).
\newblock
  \eprint{http://bioinformatics.oxfordjournals.org/content/early/2015/02/28/bioinformatics.btv130.full.pdf+html}.

\bibitem{Isorank}
\bibinfo{author}{Singh, R.}, \bibinfo{author}{Xu, J.} \&
  \bibinfo{author}{Berger, B.}
\newblock \bibinfo{journal}{\bibinfo{title}{Global alignment of multiple
  protein interaction networks with application to functional orthology
  detection}}.
\newblock {\emph{\JournalTitle{Proceedings of the National Academy of
  Sciences}}} \textbf{\bibinfo{volume}{105}}, \bibinfo{pages}{12763--12768},
  \doiprefix\url{10.1073/pnas.0806627105} (\bibinfo{year}{2008}).
\newblock \eprint{http://www.pnas.org/content/105/35/12763.full.pdf+html}.

\bibitem{mir2017index}
\bibinfo{author}{Mir, A.}, \bibinfo{author}{Naghibzadeh, M.} \&
  \bibinfo{author}{Saadati, N.}
\newblock \bibinfo{journal}{\bibinfo{title}{Index: Incremental depth extension
  approach for protein--protein interaction networks alignment}}.
\newblock {\emph{\JournalTitle{Biosystems}}} \textbf{\bibinfo{volume}{162}},
  \bibinfo{pages}{24--34} (\bibinfo{year}{2017}).

\bibitem{optnetalign}
\bibinfo{author}{Clark, C.} \& \bibinfo{author}{Kalita, J.}
\newblock \bibinfo{journal}{\bibinfo{title}{A multiobjective memetic algorithm
  for ppi network alignment}}.
\newblock {\emph{\JournalTitle{Bioinformatics}}} \textbf{\bibinfo{volume}{31}},
  \bibinfo{pages}{1988--1998}, \doiprefix\url{10.1093/bioinformatics/btv063}
  (\bibinfo{year}{2015}).
\newblock
  \eprint{http://bioinformatics.oxfordjournals.org/content/31/12/1988.full.pdf+html}.

\bibitem{PISwap}
\bibinfo{author}{Chindelevitch, L.}, \bibinfo{author}{Ma, C.-Y.},
  \bibinfo{author}{Liao, C.-S.} \& \bibinfo{author}{Berger, B.}
\newblock \bibinfo{journal}{\bibinfo{title}{Optimizing a global alignment of
  protein interaction networks}}.
\newblock {\emph{\JournalTitle{Bioinformatics}}} \textbf{\bibinfo{volume}{29}},
  \bibinfo{pages}{2765--2773}, \doiprefix\url{10.1093/bioinformatics/btt486}
  (\bibinfo{year}{2013}).
\newblock
  \eprint{http://bioinformatics.oxfordjournals.org/content/29/21/2765.full.pdf+html}.

\bibitem{modulealign}
\bibinfo{author}{Hashemifar, S.}, \bibinfo{author}{Ma, J.},
  \bibinfo{author}{Naveed, H.}, \bibinfo{author}{Canzar, S.} \&
  \bibinfo{author}{Xu, J.}
\newblock \bibinfo{journal}{\bibinfo{title}{Modulealign: module-based global
  alignment of protein--protein interaction networks}}.
\newblock {\emph{\JournalTitle{Bioinformatics}}} \textbf{\bibinfo{volume}{32}},
  \bibinfo{pages}{i658--i664} (\bibinfo{year}{2016}).

\bibitem{NETAL}
\bibinfo{author}{Neyshabur, B.}, \bibinfo{author}{Khadem, A.},
  \bibinfo{author}{Hashemifar, S.} \& \bibinfo{author}{Arab, S.~S.}
\newblock \bibinfo{journal}{\bibinfo{title}{Netal: a new graph-based method for
  global alignment of protein-protein interaction networks}}.
\newblock {\emph{\JournalTitle{Bioinformatics}}} \textbf{\bibinfo{volume}{29}},
  \bibinfo{pages}{1654--1662}, \doiprefix\url{10.1093/bioinformatics/btt202}
  (\bibinfo{year}{2013}).
\newblock
  \eprint{http://bioinformatics.oxfordjournals.org/content/29/13/1654.full.pdf+html}.

\bibitem{WAVE}
\bibinfo{author}{Sun, Y.}, \bibinfo{author}{Crawford, J.},
  \bibinfo{author}{Tang, J.} \& \bibinfo{author}{Milenkovi\`{c}, T.}
\newblock \bibinfo{title}{Simultaneous optimization of both node and edge
  conservation in network alignment via {WAVE}}.
\newblock In \bibinfo{editor}{Pop, M.} \& \bibinfo{editor}{Touzet, H.} (eds.)
  \emph{\bibinfo{booktitle}{Algorithms in Bioinformatics}}, vol.
  \bibinfo{volume}{9289} of \emph{\bibinfo{series}{Lecture Notes in Computer
  Science}}, \bibinfo{pages}{16--39},
  \doiprefix\url{10.1007/978-3-662-48221-6\_2} (\bibinfo{publisher}{Springer
  Berlin Heidelberg}, \bibinfo{address}{Germany}, \bibinfo{year}{2015}).

\bibitem{MAGNA}
\bibinfo{author}{Saraph, V.} \& \bibinfo{author}{Milenkovi{\'c}, T.}
\newblock \bibinfo{journal}{\bibinfo{title}{{MAGNA}: maximizing accuracy in
  global network alignment}}.
\newblock {\emph{\JournalTitle{Bioinformatics}}} \textbf{\bibinfo{volume}{30}},
  \bibinfo{pages}{2931--2940} (\bibinfo{year}{2014}).

\bibitem{MAGNA++}
\bibinfo{author}{Vijayan, V.}, \bibinfo{author}{Saraph, V.} \&
  \bibinfo{author}{Milenkovi{\'c}, T.}
\newblock \bibinfo{journal}{\bibinfo{title}{Magna++: Maximizing accuracy in
  global network alignment via both node and edge conservation}}.
\newblock {\emph{\JournalTitle{Bioinformatics}}} \textbf{\bibinfo{volume}{31}},
  \bibinfo{pages}{2409--2411} (\bibinfo{year}{2015}).

\bibitem{GREAT}
\bibinfo{author}{Crawford, J.} \& \bibinfo{author}{Milenkovi{\'c}, T.}
\newblock \bibinfo{title}{Great: graphlet edge-based network alignment}.
\newblock In \emph{\bibinfo{booktitle}{Bioinformatics and Biomedicine (BIBM),
  2015 IEEE International Conference on}}, \bibinfo{pages}{220--227}
  (\bibinfo{organization}{IEEE}, \bibinfo{year}{2015}).

\bibitem{gong2015global}
\bibinfo{author}{Gong, M.}, \bibinfo{author}{Peng, Z.}, \bibinfo{author}{Ma,
  L.} \& \bibinfo{author}{Huang, J.}
\newblock \bibinfo{journal}{\bibinfo{title}{Global biological network alignment
  by using efficient memetic algorithm}}.
\newblock {\emph{\JournalTitle{IEEE/ACM transactions on computational biology
  and bioinformatics}}} \textbf{\bibinfo{volume}{13}},
  \bibinfo{pages}{1117--1129} (\bibinfo{year}{2015}).

\bibitem{SPINAL}
\bibinfo{author}{Alada\u{g}, A.~E.} \& \bibinfo{author}{Erten, C.}
\newblock \bibinfo{journal}{\bibinfo{title}{Spinal: scalable protein
  interaction network alignment}}.
\newblock {\emph{\JournalTitle{Bioinformatics}}} \textbf{\bibinfo{volume}{29}},
  \bibinfo{pages}{917--924}, \doiprefix\url{10.1093/bioinformatics/btt071}
  (\bibinfo{year}{2013}).
\newblock
  \eprint{http://bioinformatics.oxfordjournals.org/content/29/7/917.full.pdf+html}.

\bibitem{ideker2012differential}
\bibinfo{author}{Ideker, T.} \& \bibinfo{author}{Krogan, N.~J.}
\newblock \bibinfo{journal}{\bibinfo{title}{Differential network biology}}.
\newblock {\emph{\JournalTitle{Molecular systems biology}}}
  \textbf{\bibinfo{volume}{8}}, \bibinfo{pages}{565} (\bibinfo{year}{2012}).

\bibitem{rolland2014proteome}
\bibinfo{author}{Rolland, T.} \emph{et~al.}
\newblock \bibinfo{journal}{\bibinfo{title}{A proteome-scale map of the human
  interactome network}}.
\newblock {\emph{\JournalTitle{Cell}}} \textbf{\bibinfo{volume}{159}},
  \bibinfo{pages}{1212--1226} (\bibinfo{year}{2014}).

\bibitem{thomas2012use}
\bibinfo{author}{Thomas, P.~D.} \emph{et~al.}
\newblock \bibinfo{journal}{\bibinfo{title}{On the use of gene ontology
  annotations to assess functional similarity among orthologs and paralogs: a
  short report}}.
\newblock {\emph{\JournalTitle{PLoS computational biology}}}
  \textbf{\bibinfo{volume}{8}} (\bibinfo{year}{2012}).

\bibitem{lockhart1994recovering}
\bibinfo{author}{Lockhart, P.~J.}, \bibinfo{author}{Steel, M.~A.},
  \bibinfo{author}{Hendy, M.~D.} \& \bibinfo{author}{Penny, D.}
\newblock \bibinfo{journal}{\bibinfo{title}{Recovering evolutionary trees under
  a more realistic model of sequence evolution.}}
\newblock {\emph{\JournalTitle{Molecular biology and evolution}}}
  \textbf{\bibinfo{volume}{11}}, \bibinfo{pages}{605--612}
  (\bibinfo{year}{1994}).

\bibitem{kachroo2015systematic}
\bibinfo{author}{Kachroo, A.~H.} \emph{et~al.}
\newblock \bibinfo{journal}{\bibinfo{title}{Systematic humanization of yeast
  genes reveals conserved functions and genetic modularity}}.
\newblock {\emph{\JournalTitle{Science}}} \textbf{\bibinfo{volume}{348}},
  \bibinfo{pages}{921--925} (\bibinfo{year}{2015}).

\bibitem{kotlyar2018iid}
\bibinfo{author}{Kotlyar, M.}, \bibinfo{author}{Pastrello, C.},
  \bibinfo{author}{Malik, Z.} \& \bibinfo{author}{Jurisica, I.}
\newblock \bibinfo{journal}{\bibinfo{title}{{IID} 2018 update: context-specific
  physical protein--protein interactions in human, model organisms and
  domesticated species}}.
\newblock {\emph{\JournalTitle{Nucleic acids research}}}
  \textbf{\bibinfo{volume}{47}}, \bibinfo{pages}{D581--D589}
  (\bibinfo{year}{2018}).

\bibitem{milenkovic2008uncovering}
\bibinfo{author}{Milenkovi{\'c}, T.} \& \bibinfo{author}{Pr{\v{z}}ulj, N.}
\newblock \bibinfo{journal}{\bibinfo{title}{Uncovering biological network
  function via graphlet degree signatures}}.
\newblock {\emph{\JournalTitle{Cancer informatics}}}
  \textbf{\bibinfo{volume}{6}}, \bibinfo{pages}{CIN--S680}
  (\bibinfo{year}{2008}).

\bibitem{page1999pagerank}
\bibinfo{author}{Page, L.}, \bibinfo{author}{Brin, S.},
  \bibinfo{author}{Motwani, R.} \& \bibinfo{author}{Winograd, T.}
\newblock \bibinfo{title}{The pagerank citation ranking: Bringing order to the
  web.}
\newblock \bibinfo{type}{Tech. Rep.}, \bibinfo{institution}{Stanford InfoLab}
  (\bibinfo{year}{1999}).

\bibitem{kirkpatrick1983optimization}
\bibinfo{author}{Kirkpatrick, S.}, \bibinfo{author}{Gelatt, C.~D.} \&
  \bibinfo{author}{Vecchi, M.~P.}
\newblock \bibinfo{journal}{\bibinfo{title}{Optimization by simulated
  annealing}}.
\newblock {\emph{\JournalTitle{science}}} \textbf{\bibinfo{volume}{220}},
  \bibinfo{pages}{671--680} (\bibinfo{year}{1983}).

\bibitem{mitra1985convergence}
\bibinfo{author}{Mitra, D.}, \bibinfo{author}{Romeo, F.} \&
  \bibinfo{author}{Sangiovanni-Vincentelli, A.}
\newblock \bibinfo{title}{Convergence and finite-time behavior of simulated
  annealing}.
\newblock In \emph{\bibinfo{booktitle}{Decision and Control, 1985 24th IEEE
  Conference on}}, \bibinfo{pages}{761--767} (\bibinfo{organization}{IEEE},
  \bibinfo{year}{1985}).

\bibitem{romeo1986efficient}
\bibinfo{author}{Romeo, F.}, \bibinfo{author}{Sangiovanni, V.~A.} \&
  \bibinfo{author}{Huang, M.}
\newblock \bibinfo{title}{An efficient general cooling schedule for simulated
  annealing} (\bibinfo{organization}{PROCEEDING OF IEEE INTERNATIONAL
  CONFERENCE ON COMPUTER AIDED DESIGN}, \bibinfo{year}{1986}).

\bibitem{autosa}
\bibinfo{author}{Park, M.-W.} \& \bibinfo{author}{Kim, Y.-D.}
\newblock \bibinfo{journal}{\bibinfo{title}{A systematic procedure for setting
  parameters in simulated annealing algorithms}}.
\newblock {\emph{\JournalTitle{Computers and Operations Research}}}
  \textbf{\bibinfo{volume}{25}}, \bibinfo{pages}{207 -- 217},
  \doiprefix\url{http://dx.doi.org/10.1016/S0305-0548(97)00054-3}
  (\bibinfo{year}{1998}).

\bibitem{szu1987FSA}
\bibinfo{author}{Szu, H.} \& \bibinfo{author}{Hartley, R.}
\newblock \bibinfo{journal}{\bibinfo{title}{Fast simulated annealing}}.
\newblock {\emph{\JournalTitle{Physics letters A}}}
  \textbf{\bibinfo{volume}{122}}, \bibinfo{pages}{157--162}
  (\bibinfo{year}{1987}).

\bibitem{meise1998convergence}
\bibinfo{author}{Meise, C.}
\newblock \bibinfo{journal}{\bibinfo{title}{On the convergence of parallel
  simulated annealing}}.
\newblock {\emph{\JournalTitle{Stochastic processes and their applications}}}
  \textbf{\bibinfo{volume}{76}}, \bibinfo{pages}{99--115}
  (\bibinfo{year}{1998}).

\bibitem{Strens2003}
\bibinfo{author}{Strens, M. J.~A.}
\newblock \bibinfo{title}{{Evolutionary MCMC Sampling and Optimization in
  Discrete Spaces}} (\bibinfo{year}{2003}).

\bibitem{suman2006survey}
\bibinfo{author}{Suman, B.} \& \bibinfo{author}{Kumar, P.}
\newblock \bibinfo{journal}{\bibinfo{title}{A survey of simulated annealing as
  a tool for single and multiobjective optimization}}.
\newblock {\emph{\JournalTitle{Journal of the operational research society}}}
  \textbf{\bibinfo{volume}{57}}, \bibinfo{pages}{1143--1160}
  (\bibinfo{year}{2006}).

\bibitem{dowsland2012simulated}
\bibinfo{author}{Dowsland, K.~A.} \& \bibinfo{author}{Thompson, J.~M.}
\newblock \bibinfo{title}{Simulated annealing}.
\newblock In \emph{\bibinfo{booktitle}{Handbook of natural computing}},
  \bibinfo{pages}{1623--1655} (\bibinfo{publisher}{Springer},
  \bibinfo{year}{2012}).

\bibitem{AguiareOliveiraJunior2012}
\bibinfo{author}{Aguiar~e Oliveira~Junior, H.}, \bibinfo{author}{Ingber, L.},
  \bibinfo{author}{Petraglia, A.}, \bibinfo{author}{Rembold~Petraglia, M.} \&
  \bibinfo{author}{Augusta Soares~Machado, M.}
\newblock \emph{\bibinfo{title}{Adaptive Simulated Annealing}},
  \bibinfo{pages}{33--62} (\bibinfo{publisher}{Springer Berlin Heidelberg},
  \bibinfo{address}{Berlin, Heidelberg}, \bibinfo{year}{2012}).

\bibitem{zhan2016list}
\bibinfo{author}{Zhan, S.-h.}, \bibinfo{author}{Lin, J.},
  \bibinfo{author}{Zhang, Z.-j.} \& \bibinfo{author}{Zhong, Y.-w.}
\newblock \bibinfo{journal}{\bibinfo{title}{List-based simulated annealing
  algorithm for traveling salesman problem}}.
\newblock {\emph{\JournalTitle{Computational intelligence and neuroscience}}}
  \textbf{\bibinfo{volume}{2016}}, \bibinfo{pages}{8} (\bibinfo{year}{2016}).

\bibitem{SANAtemperature}
\bibinfo{author}{Hayes, W.} \& \bibinfo{author}{{Mamano}, N.}
\newblock \bibinfo{journal}{\bibinfo{title}{Automatic temperature endpoints for
  simulated annealing}}.
\newblock {\emph{\JournalTitle{Submitted}}}  (\bibinfo{year}{2020}).

\bibitem{shannon1948mathematical}
\bibinfo{author}{Shannon, C.~E.}
\newblock \bibinfo{journal}{\bibinfo{title}{A mathematical theory of
  communication}}.
\newblock {\emph{\JournalTitle{Bell system technical journal}}}
  \textbf{\bibinfo{volume}{27}}, \bibinfo{pages}{379--423}
  (\bibinfo{year}{1948}).

\bibitem{johnson2003introduction}
\bibinfo{author}{Johnson~Jr, P.~D.}, \bibinfo{author}{Harris, G.~A.} \&
  \bibinfo{author}{Hankerson, D.}
\newblock \emph{\bibinfo{title}{Introduction to information theory and data
  compression}} (\bibinfo{publisher}{Chapman and Hall/CRC},
  \bibinfo{year}{2003}).

\bibitem{adamek2011foundations}
\bibinfo{author}{Adamek, J.}
\newblock \emph{\bibinfo{title}{Foundations of coding: Theory and applications
  of error-correcting codes with an introduction to cryptography and
  information theory}} (\bibinfo{publisher}{John Wiley \& Sons},
  \bibinfo{year}{2011}).

\bibitem{sakrison1968communication}
\bibinfo{author}{Sakrison, D.~J.}
\newblock \emph{\bibinfo{title}{Communication Theory: Transmission of Waveforms
  and Digital Information}}, vol. \bibinfo{volume}{968}
  (\bibinfo{publisher}{Wiley New York}, \bibinfo{year}{1968}).

\bibitem{dhulipala2016compressing}
\bibinfo{author}{Dhulipala, L.} \emph{et~al.}
\newblock \bibinfo{title}{Compressing graphs and indexes with recursive graph
  bisection}.
\newblock In \emph{\bibinfo{booktitle}{Proceedings of the 22nd ACM SIGKDD
  International Conference on Knowledge Discovery and Data Mining}},
  \bibinfo{pages}{1535--1544} (\bibinfo{year}{2016}).

\bibitem{Milenkovic2008}
\bibinfo{author}{Milenkovi\'{c}, T.} \& \bibinfo{author}{Pr\v{z}ulj, N.}
\newblock \bibinfo{journal}{\bibinfo{title}{Uncovering biological network
  function via graphlet degree signatures}}.
\newblock {\emph{\JournalTitle{Cancer Informatics}}}
  \textbf{\bibinfo{volume}{6}}, \bibinfo{pages}{257--273}
  (\bibinfo{year}{2008}).

\bibitem{Krogan2006}
\bibinfo{author}{Krogan, N.~J.} \emph{et~al.}
\newblock \bibinfo{journal}{\bibinfo{title}{Global landscape of protein
  complexes in the yeast {S}accharomyces cerevisiae}}.
\newblock {\emph{\JournalTitle{Nature}}} \textbf{\bibinfo{volume}{440}},
  \bibinfo{pages}{637--643} (\bibinfo{year}{2006}).

\bibitem{Collins2007yeast2}
\bibinfo{author}{Collins, S.~R.} \emph{et~al.}
\newblock \bibinfo{journal}{\bibinfo{title}{Toward a comprehensive atlas of the
  physical interactome of saccharomyces cerevisiae}}.
\newblock {\emph{\JournalTitle{Molecular and Cellular Proteomics}}}
  \textbf{\bibinfo{volume}{6}}, \bibinfo{pages}{439--450},
  \doiprefix\url{10.1074/mcp.M600381-MCP200} (\bibinfo{year}{2007}).
\newblock \eprint{http://www.mcponline.org/content/6/3/439.full.pdf+html}.

\bibitem{Radivojac2008}
\bibinfo{author}{Radivojac, P.} \emph{et~al.}
\newblock \bibinfo{journal}{\bibinfo{title}{An integrated approach to inferring
  gene-disease associations in humans.}}
\newblock {\emph{\JournalTitle{Proteins}}} \textbf{\bibinfo{volume}{72}},
  \bibinfo{pages}{1030--1037} (\bibinfo{year}{2008}).

\bibitem{ingber1989VFSA}
\bibinfo{author}{Ingber, L.}
\newblock \bibinfo{journal}{\bibinfo{title}{Very fast simulated re-annealing}}.
\newblock {\emph{\JournalTitle{Mathematical and computer modelling}}}
  \textbf{\bibinfo{volume}{12}}, \bibinfo{pages}{967--973}
  (\bibinfo{year}{1989}).

\bibitem{GareyJohnson79}
\bibinfo{author}{Garey, M.~R.} \& \bibinfo{author}{Johnson, D.~S.}
\newblock \emph{\bibinfo{title}{Computers and Intractability--A Guide to the
  Theory of NP-Completeness}} (\bibinfo{publisher}{W. H. Freeman And Company},
  \bibinfo{address}{New York}, \bibinfo{year}{1979}).

\bibitem{GO}
\bibinfo{author}{{The Gene Ontology Consortium}}.
\newblock \bibinfo{journal}{\bibinfo{title}{The gene ontology project in
  2008}}.
\newblock {\emph{\JournalTitle{Nucleic Acids Research}}}
  \textbf{\bibinfo{volume}{36}}, \bibinfo{pages}{D440--D444},
  \doiprefix\url{10.1093/nar/gkm883} (\bibinfo{year}{2008}).
\newblock
  \eprint{http://nar.oxfordjournals.org/content/36/suppl_1/D440.full.pdf+html}.

\bibitem{poole2016combining}
\bibinfo{author}{Poole, W.}, \bibinfo{author}{Gibbs, D.~L.},
  \bibinfo{author}{Shmulevich, I.}, \bibinfo{author}{Bernard, B.} \&
  \bibinfo{author}{Knijnenburg, T.~A.}
\newblock \bibinfo{journal}{\bibinfo{title}{Combining dependent p-values with
  an empirical adaptation of brown's method}}.
\newblock {\emph{\JournalTitle{Bioinformatics}}} \textbf{\bibinfo{volume}{32}},
  \bibinfo{pages}{i430--i436} (\bibinfo{year}{2016}).

\bibitem{WangAtkinsonHayesGOpredict}
\bibinfo{author}{Wang, S.}, \bibinfo{author}{Atkinson, G. R.~S.} \&
  \bibinfo{author}{Hayes, W.~B.}
\newblock \bibinfo{journal}{\bibinfo{title}{{SANA: Cross-Species Prediction of
  Gene Ontology GO Annotations via Topological Network Alignment}}}.
\newblock {\emph{\JournalTitle{npj Systems Biology (accepted)}}}
  (\bibinfo{year}{2022}).

\bibitem{kullback1997information}
\bibinfo{author}{Kullback, S.}
\newblock \emph{\bibinfo{title}{Information theory and statistics}}
  (\bibinfo{publisher}{Courier Corporation}, \bibinfo{year}{1997}).

\bibitem{klir2005uncertainty}
\bibinfo{author}{Klir, G.~J.}
\newblock \emph{\bibinfo{title}{Uncertainty and Information: Foundations of
  Generalized Information Theory}} (\bibinfo{publisher}{John Wiley \& Sons},
  \bibinfo{year}{2005}).

\bibitem{NAUTY}
\bibinfo{author}{Mckay, B.~D.}
\newblock \bibinfo{title}{Nauty} (\bibinfo{year}{2010}).

\bibitem{walhout2000protein}
\bibinfo{author}{Walhout, A.~J.} \emph{et~al.}
\newblock \bibinfo{journal}{\bibinfo{title}{Protein interaction mapping in c.
  elegans using proteins involved in vulval development}}.
\newblock {\emph{\JournalTitle{Science}}} \textbf{\bibinfo{volume}{287}},
  \bibinfo{pages}{116--122} (\bibinfo{year}{2000}).

\bibitem{krogan2006yeast1}
\bibinfo{author}{Krogan, N.~J.} \emph{et~al.}
\newblock \bibinfo{journal}{\bibinfo{title}{Global landscape of protein
  complexes in the yeast {S}accharomyces cerevisiae}}.
\newblock {\emph{\JournalTitle{Nature}}} \textbf{\bibinfo{volume}{440}},
  \bibinfo{pages}{637--643} (\bibinfo{year}{2006}).

\end{thebibliography}
\newpage

\section*{Supplementary}
\begin{table}[hbt]
    \centering\small
    \begin{tabular}{|ll|rrrrrr|}
    \hline
    species 1 & species 2 &  common orths $n$ & $m_1$ & $m_2$ & $N$ & interologs $i$ & $p$-value \\
    \hline
MMusculus & HSapiens & 5920 & 16140 & 120342 & 17520240 & 4222  & $10^{ -5172}$ \\
DMelanogaster & HSapiens & 2567 & 10194 & 30498 & 3293461 & 2364        & $10^{ -2480}$ \\
SCerevisiae & HSapiens & 1040 & 10160 & 7502 & 540280 & 1619    & $10^{ -1195}$ \\
RNorvegicus & HSapiens & 2013 & 2965 & 28672 & 2025078 & 743    & $10^{ -667}$ \\
SPombe & SCerevisiae & 654 & 992 & 6595 & 213531 & 445  & $10^{ -390}$ \\
SPombe & HSapiens & 483 & 792 & 3710 & 116403 & 382     & $10^{ -348}$ \\
SCerevisiae & DMelanogaster & 482 & 3458 & 1386 & 115921 & 441  & $10^{ -320}$ \\
RNorvegicus & MMusculus & 1134 & 1794 & 3030 & 642411 & 255     & $10^{ -283}$ \\
SPombe & DMelanogaster & 235 & 357 & 565 & 27495 & 153  & $10^{ -164}$ \\
AThaliana & HSapiens & 273 & 326 & 1143 & 37128 & 161   & $10^{ -153}$ \\
MMusculus & DMelanogaster & 622 & 911 & 1855 & 193131 & 161     & $10^{ -147}$ \\
CElegans & HSapiens & 573 & 701 & 3606 & 163878 & 141   & $10^{ -88}$ \\
CElegans & DMelanogaster & 287 & 306 & 593 & 41041 & 85 & $10^{ -82}$ \\
DMelanogaster & AThaliana & 117 & 144 & 123 & 6786 & 54 & $10^{ -59}$ \\
SCerevisiae & AThaliana & 87 & 179 & 80 & 3741 & 40     & $10^{ -31}$ \\
SCerevisiae & MMusculus & 103 & 478 & 81 & 5253 & 44    & $10^{ -24}$ \\
CElegans & SCerevisiae & 72 & 55 & 306 & 2556 & 36      & $10^{ -20}$ \\
CElegans & MMusculus & 45 & 35 & 31 & 990 & 16  & $10^{ -16}$ \\
SPombe & AThaliana & 38 & 40 & 29 & 703 & 17    & $10^{ -14}$ \\
MMusculus & AThaliana & 102 & 119 & 97 & 5151 & 21      & $10^{ -14}$ \\
RNorvegicus & DMelanogaster & 320 & 404 & 867 & 51040 & 34      & $10^{ -13}$ \\
SPombe & CElegans & 44 & 61 & 36 & 946 & 17     & $10^{ -11}$ \\
SPombe & MMusculus & 53 & 49 & 44 & 1378 & 15   & $10^{ -11}$ \\
CElegans & AThaliana & 11 & 6 & 6 & 55 & 6      & $10^{ -6}$ \\
RNorvegicus & SCerevisiae & 35 & 37 & 114 & 595 & 18    & $10^{ -4}$ \\
RNorvegicus & SPombe & 23 & 27 & 34 & 253 & 9   & $10^{ -1}$ \\
RNorvegicus & CElegans & 8 & 4 & 4 & 28 & 2     & $10^{ 0}$ \\
RNorvegicus & AThaliana & 13 & 12 & 8 & 78 & 1  & $10^{ 0}$ \\
    \hline
    \end{tabular}
    \caption{We test hypothesis R0 using the hypergeometric test on BioGRID 3.4.164:
    ``common orths $n$'' is the number of common 1-to-1 orthologs, according to NCBI Homologene, that exist in both BioGRID networks of species 1 and species 2, and thus the number of nodes $n$ in both the networks $H_1,H_2$ that are induced from the networks in BioGRID;
    $m_1$ is the number of edges between the nodes of $H_1$, and similarly for $m_2$ and $H_2$;
    $N$ is ${n \choose 2}$;
    interologs is the number of observed edges that co-occur between two pairs of orthologs.
    For each edge $e_1$ of $H_1$, we take its endpoint nodes $(u_1,v_1)$, and look at the orthologous nodes $(u_2,v_2)$ in $H_2$. If R0 is true, then the existence of $e_1$ should tell us little or nothing beyond chance about the existence of an edge between $u_2$ and $v_2$, so that edge should exist at random with probability equal to the edge density of $H_2$. This hypothesis can be tested using the hypergeometric distribution with $m_1$ being the number of draws, and $m_2$ being the number of successes that exist among the $N={n \choose 2}$ node pairs in $H_2$. The last column---on which the table is sorted---is the $p$-value of the actual number of observed interologs $i$ between $H_1$ and $H_2$.
    As can be seen, the number of interologs soundly rejects hypothesis R0, at least for BioGRID networks that have enough nodes and edges to have significant overlap with known orthologs. Mouse and Human, in particular, have enough interologs for a $p$-value less than {\em 10 to the power of negative five thousand} [sic].
    (The hypergeometric distribution boils down to ratios of factorials, each of whose base-10 logarithm can be computed exactly by summing logarithms similar to what is done in Equation \ref{eq:Netbits}. The result, rounded to the nearest integer, becomes the exponent in the $p$-value of the last column.)
    }
    \label{tab:interologs}
\end{table}
Hypothesis R0, if correct, would substantially disrupt much of modern molecular biology. The evidence against it is overwhelming: the quotes that start this article are {\em textbook} quotes, both literally and figuratively; the assumption underlies much of the success of modern evidence-based medicine\cite{thomas2012use} (eg., animal testing would be pointless unless animals shared a common biochemistry with humans); and R0 contradicts much of our understanding of molecular evolution\cite{lockhart1994recovering}. As just one recent dramatic {\it in vivo} example of how orthology implies both functional and PPI network conservation across a vast taxonomic distance, Kachroo {\it et al.}  \cite{kachroo2015systematic} found that out of 414 {\em essential} yeast genes with human orthologs, 47\% of them could be wholesale {\em replaced} by the human orthologs without destroying the viability of the yeast organism. As these authors state, their experiment demonstrates that ``critical ancestral functions of many essential genes are thus retained {\bf in a pathway-specific manner} [our emphasis], resilient to drift in sequences, splicing, and protein interfaces''. The authors further confirmed in several cases that the proteins expressed by the human genes were the human proteins, further validating the hypothesis that orthologous {\em proteins} (not just their genes) perform virtually identical functions across vast taxonomic distances. The latter requires physically compatible interfaces on the surface of the proteins, in order for the {\em human} proteins to interact with their appropriate {\em yeast} partners, which obviously implies conservation of PPI network topology at the local scale, extending at least partially to the global scale when such a large number of orthologous proteins retain their function across more than a billion years of evolution.

To additionally test R0 ourselves, we performed the following novel test. If R0 were true, then orthologous proteins across species would rarely perform similar function. Since the function of a protein is defined by its interaction partners and its resulting placement in the larger fabric of the network, this means that there would be little statistical significance to the common interactions between {\em pairs} of orthologous proteins across species---such a common interaction is called an {\it interolog} \cite{walhout2000protein}. This hypothesis can be tested via the {\it hypergeometric distribution} $\mathcal H(i,m,M,N)$, which describes the probability of $i$ success in $m$ draws from a finite population $N$ in which $M$ objects have the desired property. In our case, assume we are given two BioGRID networks $G_1$ and $G_2$, and a list of $l$ orthologous protein pairs between them. Taking the induced subgraphs of $G_1$ and $G_2$ on their common $n$ orthologs ($n$ can be less than $l$ if either of $G_1,G_2$ do not contain some of the orthologs) gives two graphs $H_1,H_2$ with the same number of nodes, $n$. Imposing the network alignment implied by the $n$ ortholog pairs, we can test whether the number of resulting interologs $i$ is consistent with being random, as R0 would suggest. In particular, assume there are $m_1$ edges in $H_1$, $m_2$ edges in $H_2$, and $i$ observed interologs (ie., common interactions between orthologs). Then there are $N={n \choose 2}$ total node pairs in the population, $m_1$ represents the number of draws of node pairs from one network, and $m_2$ represents the number of potential successes from the other. Then the $p$-value of observing $i$ interologs is described by the tail of $\mathcal H(i,m_1,m_2,N)$. Table \ref{tab:interologs} demonstrates that---at least for BioGRID networks having sufficient edge density for a meaningful measurement, and barring the existence of {\em extremely} potent and as yet unknown systemic biases in the data---there is little room for debate on the existence of interologs.


\subsection{Examples of objective function saturation}\label{sec:saturationSupp}
\begin{figure}
    \centering
    \includegraphics[scale=0.49]{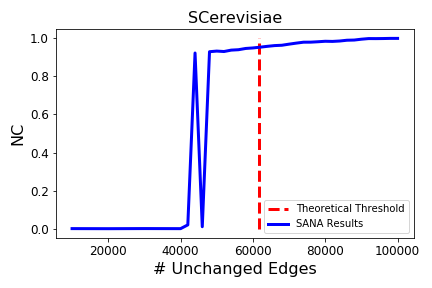}
    \includegraphics[scale=0.49]{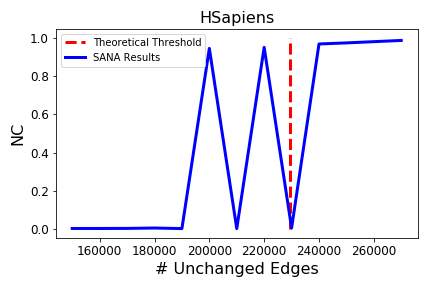}
    \caption{\textbf{Correctly aligning a large graph to itself even in the presence of noise}
    We align a noisy subgraph $G_1$ consisting of 75\% of the nodes of the BioGRID networks (v. 3.4.164) of {\it S. cerevisiae} (left) and {\it H. sapiens} (right) PPI networks to the full version $G_2$ of the same network, optimizing $S^3$. At the far right of each figure, $G_1$ has all of its edges, induced from $G_2$. (Although not depicted, the $S^3$ score was exactly 1 for all runs of SANA that took longer than 12 minutes, with at most a handful of degree-1 ``leaf'' nodes misaligned due to their being topologically indistinguishable from sibling leaf nodes.)
    Moving left in each figure, we progressively rewire more and more edges of $G_1$, effectively removing information useful for alignment. (``Rewiring'' means deleting a true edge, and replacing it by an edge between two randomly chosen nodes.) The horizontal axis depicts the number of ``clean'' edges (ie., not rewired), which approximates the information content remaining after rewiring. The vertical dotted red line is the information bound as computed by Equation \ref{eq:Netbits}.
    The vertical axis (blue curve) is {\it node correctness} or $NC$, the fraction of $G_1$'s nodes that are correctly self-aligned in a 2 hour run of SANA optimizing $S^3$ (cf. Figure \ref{fig:NetAlign}).
    SANA is able to produce correct alignments down to a number of edges significantly below that dictated by Equation \ref{eq:Netbits}.
    The ragged nature of the blue curve (jumping several times from near zero to near 1) occurs because, once near the theoretical threshold, SANA's random search has a non-negligible probability of failing to find the best solution in a 2-hour run. Once significantly below the threshold, the probability of recovering the correct alignment drops to zero as the noise of rewired edges overwhelms any signal remaining in the ``clean'' edges.
    }
    \label{fig:SC-HS-self}
\end{figure}

In the case of aligning a network to itself, the fraction of nodes that are correctly aligned to themselves---commonly called {\it node correctness} or NC---acts as a surrogate to ``functional similarity''. For topological measures, we will use EC and $S^3$, which are both equal to 1 in a self-alignment; if one network has a superset of edges over the other, then the perfect alignment scores 1 in EC but less than 1 in $S^3$.  Figure \ref{fig:SC-HS-self} provides an example both of (1) how SANA easily performs near-perfect self-alignments when the number of edges is above our hypothesized information-theoretic bound, and (2) that when the number of edges drops significantly below this bound---which is obviously approximate---the NC score drops precipitously to zero.

\begin{figure}
\centering
\includegraphics[scale=0.51]{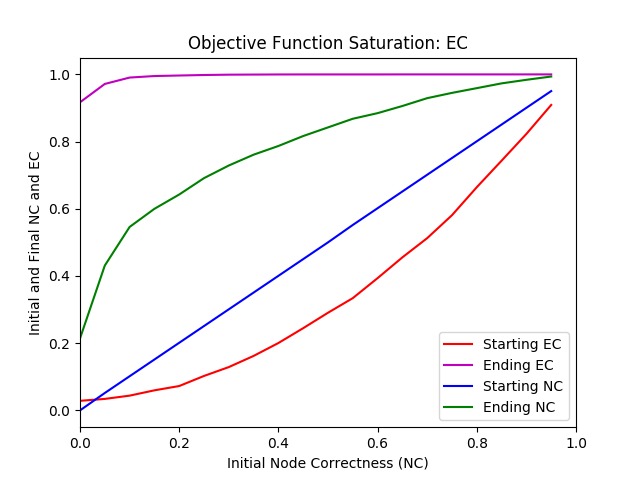}
\includegraphics[width=0.47\linewidth]{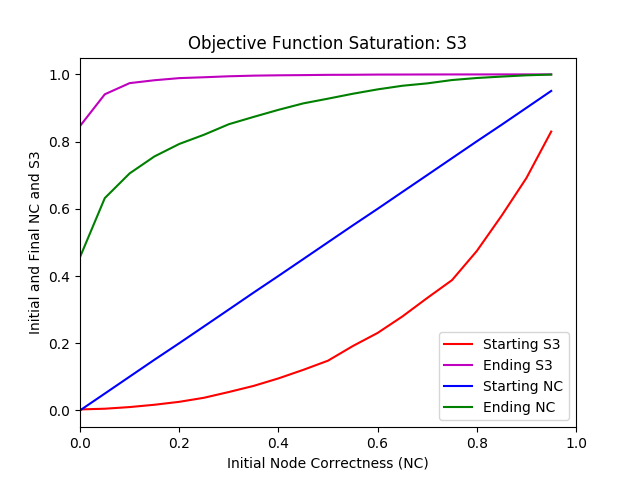}
\caption{\small{\bf Objective Function Saturation and the multitudes of near-optimal alignments}: We show two examples of objective function saturation: on the left we align the highest-confidence, largest connected component of the 2006 Krogan {et al.} ``clean'' yeast network\cite{krogan2006yeast1} to a later and larger yeast network by Collins\cite{Collins2007yeast2}; on the right is the latter network being aligned to itself. Since both networks are the same species, we know the correct mapping.
We start with some fraction NC of nodes {\em locked} to their correct positions (horizontal axis, and the blue line $y=x$), while the remaining nodes are initially scattered at random. The lowest parabolic (red) curves are the starting EC (left) and $S^3$ (right). (They are parabolic because if each node has probability $p$ of being correctly aligned, then the probability of a random edge being aligned is $p^2$, since both its endpoints must be correctly aligned.) We then run SANA optimizing EC (left) or $S^3$ (right), subject to the constraint that locked nodes cannot move. We see that SANA is able to push the values of EC and $S^3$ to their optimal values of 1 (purple curves) even though this results in ending NC values (green curves) well below 1. All curves show the mean score across 1,000 one-minute runs of SANA.
}
\label{fig:sat_graph}
\end{figure}

Let $a^*$ be an optimal alignment under some topological objective function $f$, so that $f(a^*)\ge f(a) \forall a$.
It has been shown\cite{GRAAL,H-GRAAL} that $a^*$ is often not unique.
Even if there exists some idea of a ``correct'' alignment between $G_1$ and $G_2$, and even if that correct alignment has the optimal value of the topological objective, the non-uniqueness of topologically optimal alignments may mean that topology alone may be unable to distinguish between alignments of high and low functional similarity. We refer to this phenomenon as {\it objective function saturation}, because the objective ``saturates'' to its maximum value before isolating alignments of high functional relevance.

As an example of saturation in the first case, we take two yeast networks from 2006\cite{krogan2006yeast1} and 2007\cite{Collins2007yeast2}, which we will call $G_1$ and $G_2$, respectively; they have 1004 and 2390 nodes, respectively. Equation \ref{eq:Netbits} tells us that aligning $G_1$ to itself requires $\approx 8500$ bits of information, while $G_1$ has 8323 edges; aligning $G_1$ to $G_2$ requires 10,910 bits, which is far above $G_1$'s edge count; and finally, with 2390 nodes, $G_2$ aligned to itself requires 23381 bits while it has only 16,127 edges. As we can see, all of these combinations have edge counts that are either very close to, or well below, the bit requirement of Equation \ref{eq:Netbits}. Thus, we {\em expect} that saturation will occur, due to lack of information. Here we study how bad the ``incorrect'' alignments can be even when the topological score is near-optimal (ie., saturated).

Consider Figure \ref{fig:sat_graph}. We create alignments with a pre-determined correctness by ``locking'' some fraction of the nodes to their correct positions. This bounds the correctness of the alignment from below. We then view how ``badly'' the rest of the nodes can be aligned once the objective reaches saturation. We refer to the fraction of nodes that are correctly aligned as the {\em node correctness} (NC).
We initialize the alignment with a fraction NC of $G_1$'s nodes locked to themselves in $G_2$, and then
scatter $G_1$'s remaining nodes to random locations in $G_2$. In this initial configuration, a randomly chosen edge will lie ``on top of'' another edge if (a) both of its endpoints are correctly aligned, or (b) by chance alone to another edge if both endpoints are misaligned. The latter can be shown to have extremely low probability compared to $NC^2$. Thus, edges are initially aligned correctly with probability proportional to $NC^2$ and so, when applied to an initial alignment, an edge-based measure of common topology will have a score proportional to $NC^2$. These are the red curves in Figure \ref{fig:sat_graph}.

Starting with an alignment along the red curves of Figure \ref{fig:sat_graph}, we then run SANA for 1 minute, optimizing the stated objective (EC on the left, $S^3$ on the right). The resulting score of the topological objective is plotted in purple. As we can see, the topological objectives easily get very close to 1 (which is the optimal score). However, the final NC value (green curves) remain far below 1 except when the {\em initial} NC is close to 1. These curves demonstrate that there are a large number of different alignments that have optimal topological scores (purple curves), even though most have ``correctness'' which is far from optimal (green curves).

The above demonstrates the existence of objective function saturation when we are quite certain that the networks have too little information for {\em any} topological measure to produce ``good'' alignments. In contrast, consider the networks of the {\it Integrated Interaction Database}\cite{kotlyar2018iid}, or IID. These networks take it as given that 1-to-1 orthologous proteins across mammals have near-identical functions with near-identical interaction partners, and uses this assumption to transfer interologs from the more dense PPI networks (mostly human and a bit of mouse) to the less dense PPI networks of other mammals. The result is that the PPI networks of all mammal species in the IID are highly similar---possibly overly so, compared to the real (but unknown) PPI networks. These networks all have approximately 15,000-20,000 nodes and about 300,000 edges---well above the 257,000 bits required of Equation \ref{eq:Netbits} for aligning networks of 20,000 nodes. As a result, we expect to achieve relatively robust alignments. This expectation is corroborated by Table \ref{tab:IID_EC_vs_S3}, where we show the results of applying SANA optimizing either EC or $S^3$ for 1 hour to these networks. We see that EC is able to recover about 10,218 orthologs on average (far right column), while $S^3$ is able to recover almost 22\% more, at 12,441. Note that these results are {\em robust} and {\em repeatable}: each 1-hour run of SANA on any pair of these networks gives virtually the same set of recovered orthologs. (The ones that are recovered tend to have much higher degree compared to those that are not---cf. Figures \ref{fig:IIDrat-mouse} and Supplementary Figure \ref{fig:degree-orthologs-recovered}.) However, even with these robust results, we see that $S^3$ {\em always} recovers more orthologs than EC. This makes sense, since the symmetry of $S^3$ means that it will tend to do well in aligning regions that are virtually identical, and in this case it does better than EC since the networks {\em do} have large regions that are almost identical, by construction.

\begin{table}[hbt]
    \centering
    \begin{tabular}{|c|rrrrrrrrr|l|}
    \hline
    measure &rabbit &mouse  &sheep  &cat    &horse  &dog    &pig    &cow    &rat    &mean \\
    \hline
    $|$human orthologs$|$     &14339  &15996  &15601  &15500  &15552  &15616  &15744  &15641  &15206  &15466.1 \\
    $S^3$   &11676  &12627  &12579  &12541  &12678  &12525  &12670  &12499  &12174  &12441 \\
    EC      &10483  &9128   &10995  &10459  &11016  &9049   &11406  &11485  &7944   &10218 \\
    \hline
    \end{tabular}
    \caption{Recovery of correct 1-to-1 orthologs when aligning mammals in 2018 IID networks \cite{kotlyar2018iid} to the human one running SANA {\em just once} optimizing $S^3$ and EC for 1 hour---no multiple runs are required, in contrast to those associated with Table \ref{tab:orthologs}. Since all networks except human are largely synthetic and based on orthology with human, the PPI networks are very similar to each other. Thus, using $S^3$ to measure topological similarity produces alignments that correctly recover on average about 20\% more orthologs than EC; compare to BioGRID in Table \ref{tab:orthologs}, where EC does significantly better than $S^3$; a potential explanation of the discrepancy is offered in section \ref{sec:EC-vs-S3}. (Note: The human IID network has 17373 nodes.)
    }
    \label{tab:IID_EC_vs_S3}
\end{table}

The takeaway from this section is that there are degrees of saturation: for any given topological objective function, more data tends to allow that objective to achieve better ``correctness'' at saturation. However, at a given level of data completeness, different topological measures can still have different abilities to recover the ``correct'' alignment. Here we saw that $S^3$ was better than EC at recovering orthologs in the IID networks. However, as we will see below, the same cannot be said when aligning the BioGRID networks.




\begin{figure}[tb]
    \centering
\includegraphics[width=0.24 \textwidth]{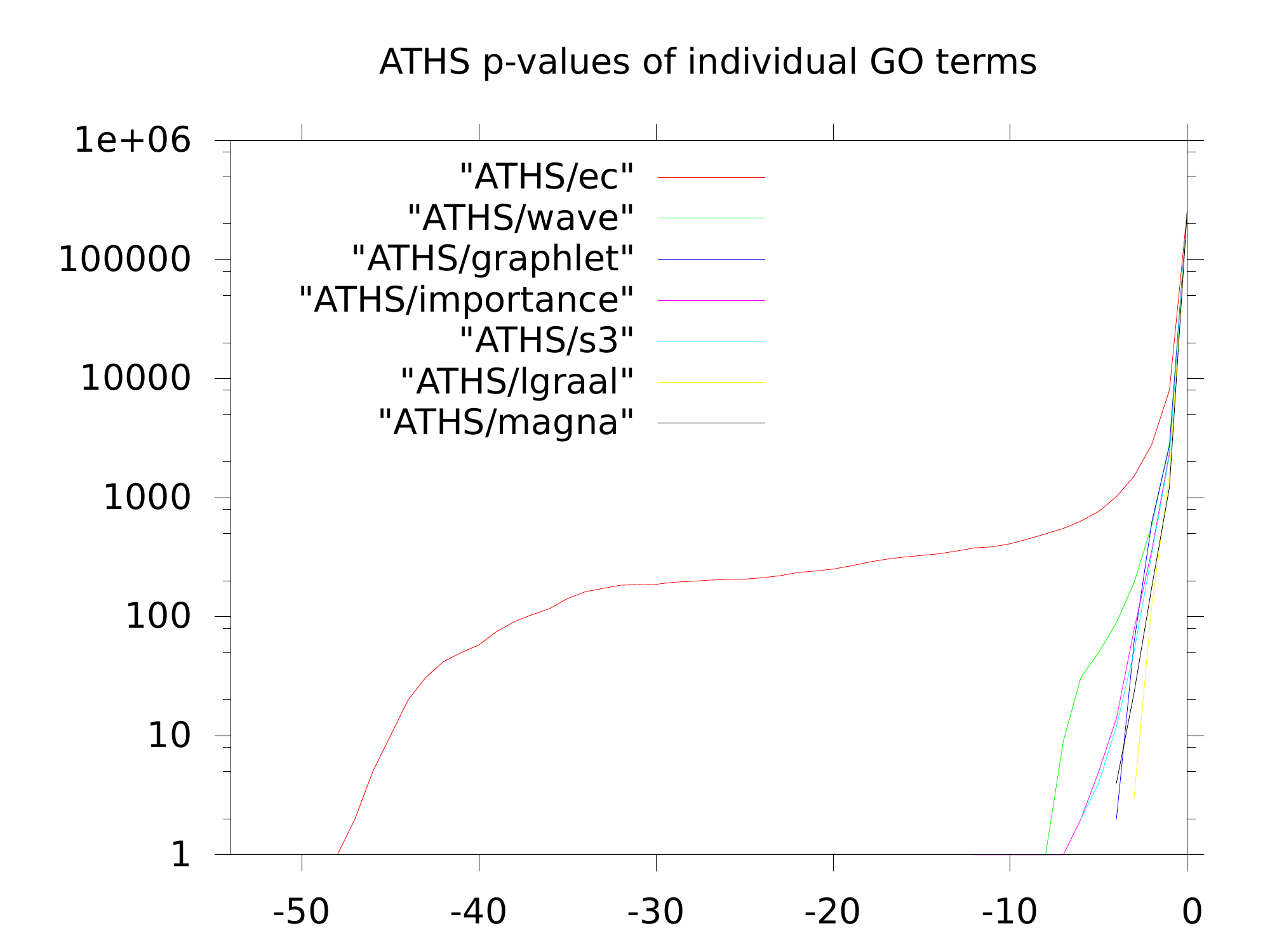}
\includegraphics[width=0.24 \textwidth]{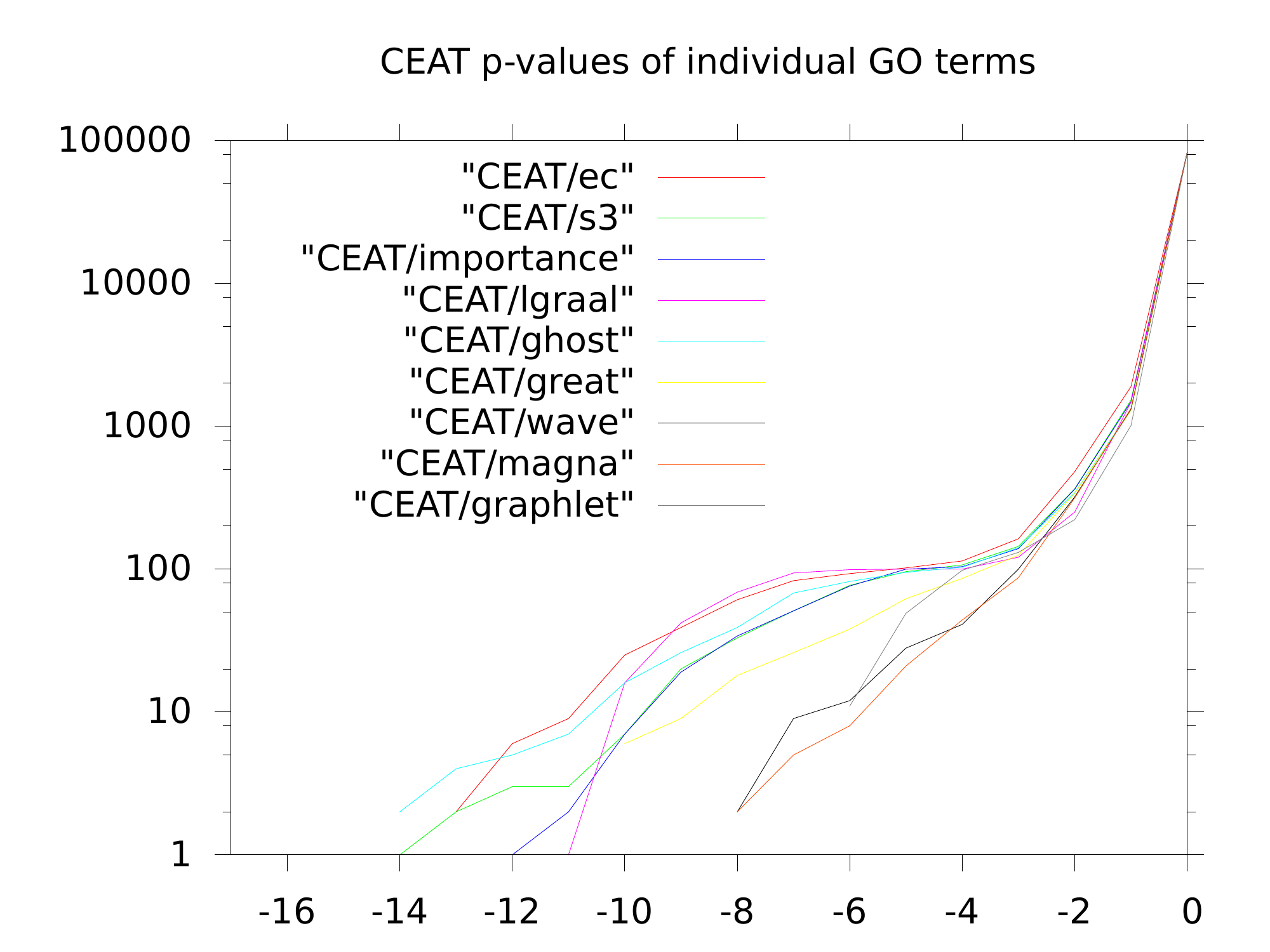}
\includegraphics[width=0.24 \textwidth]{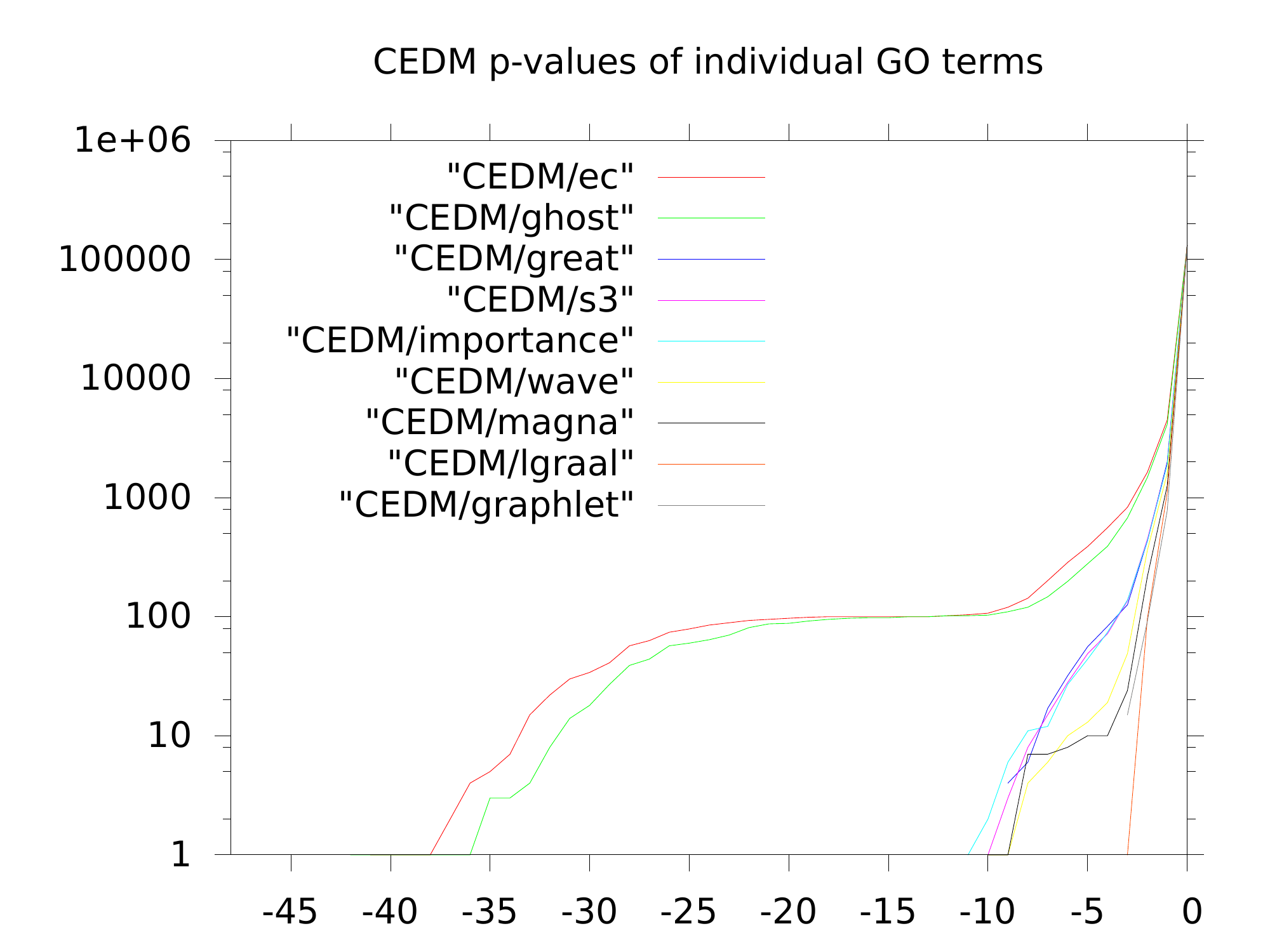}
\includegraphics[width=0.24 \textwidth]{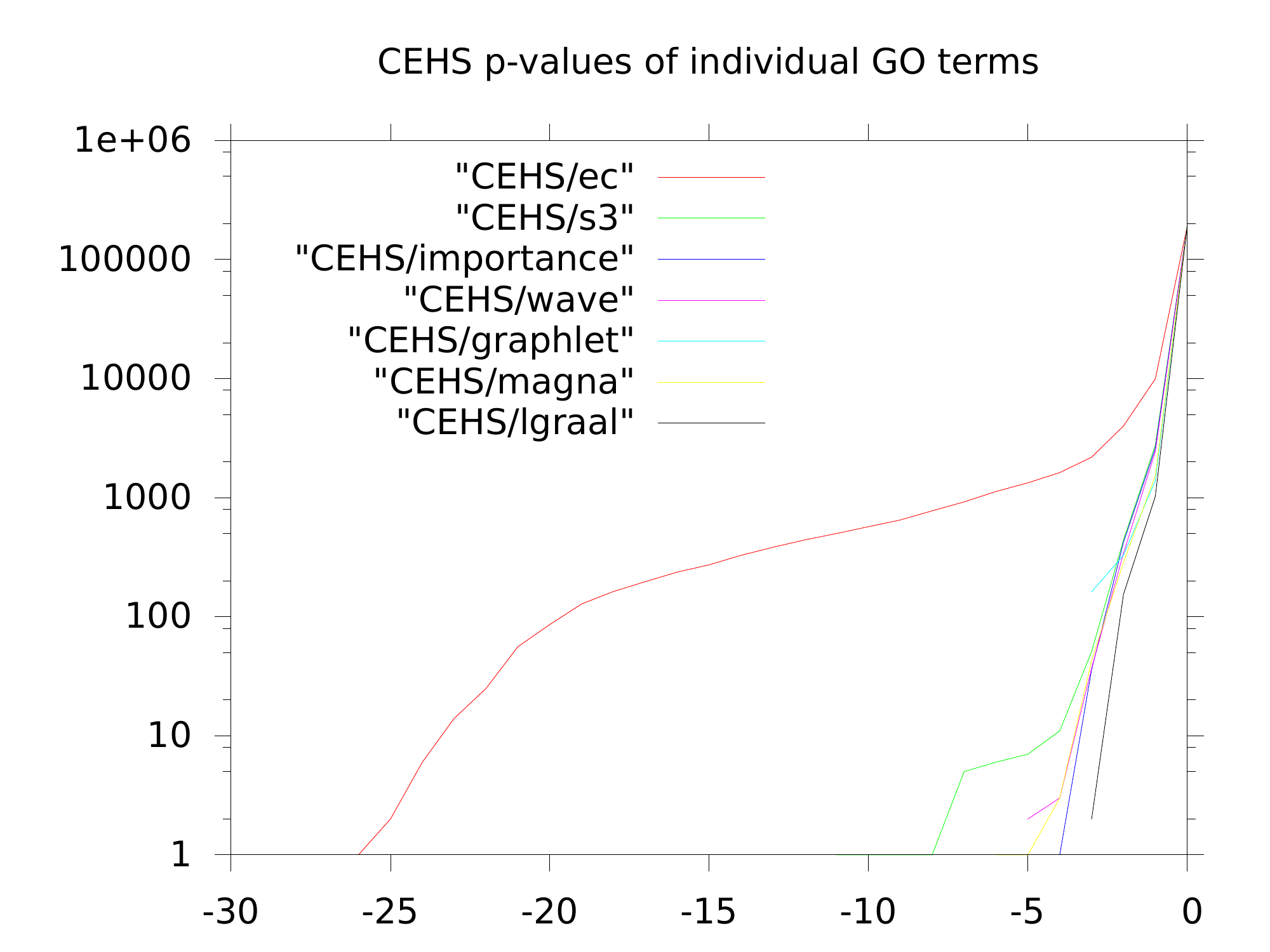}
\includegraphics[width=0.24 \textwidth]{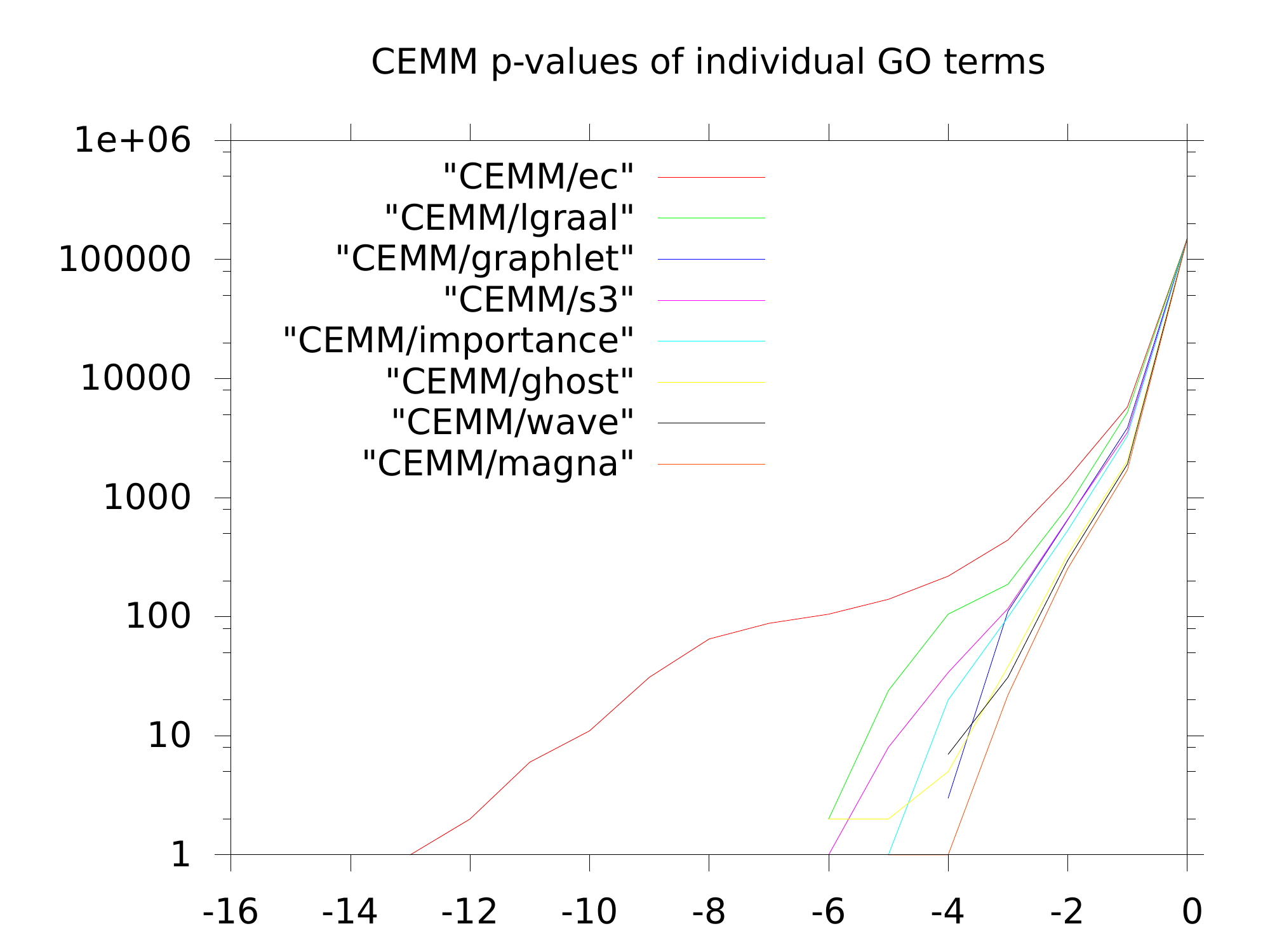}
\includegraphics[width=0.24 \textwidth]{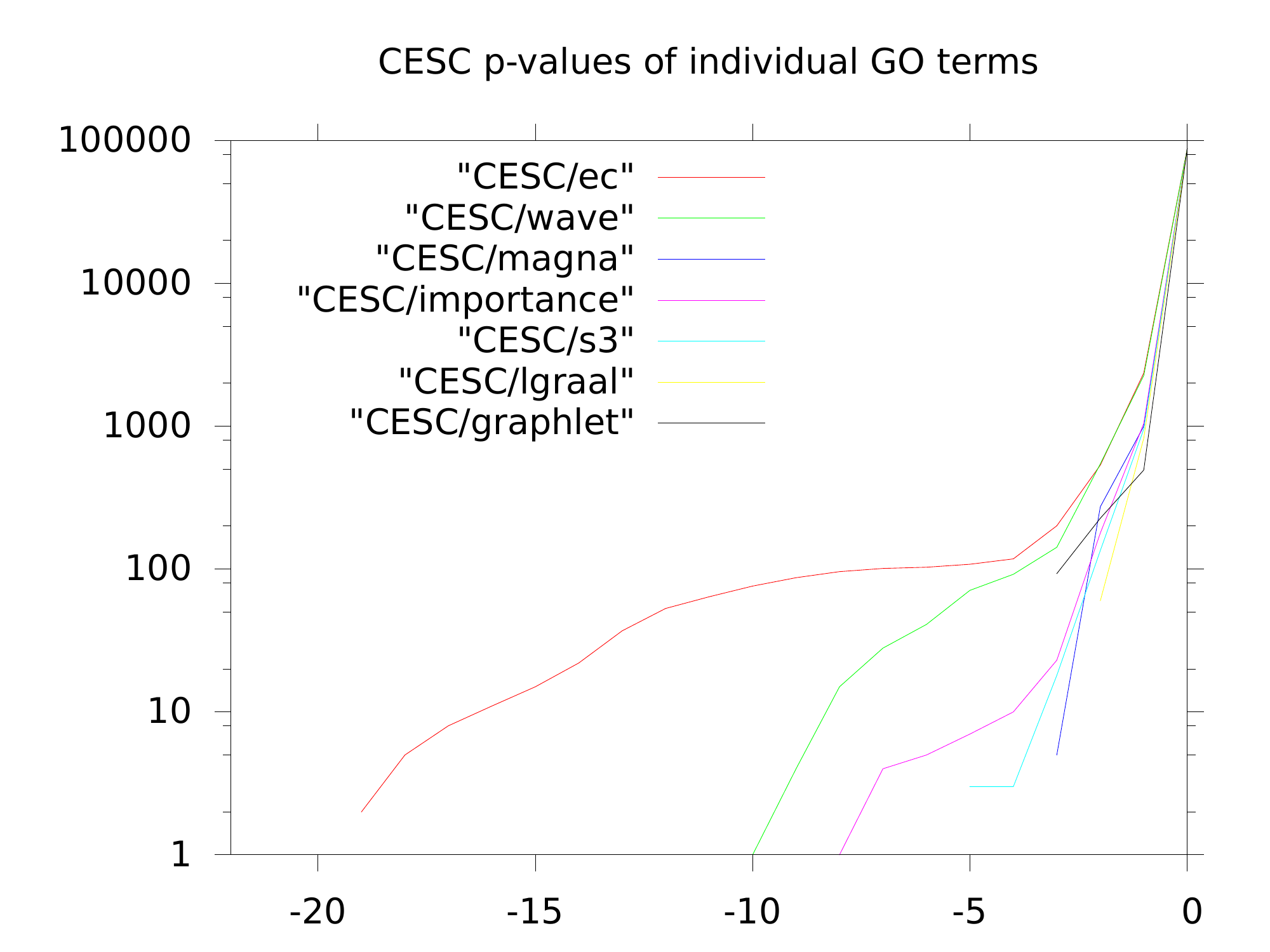}
\includegraphics[width=0.24 \textwidth]{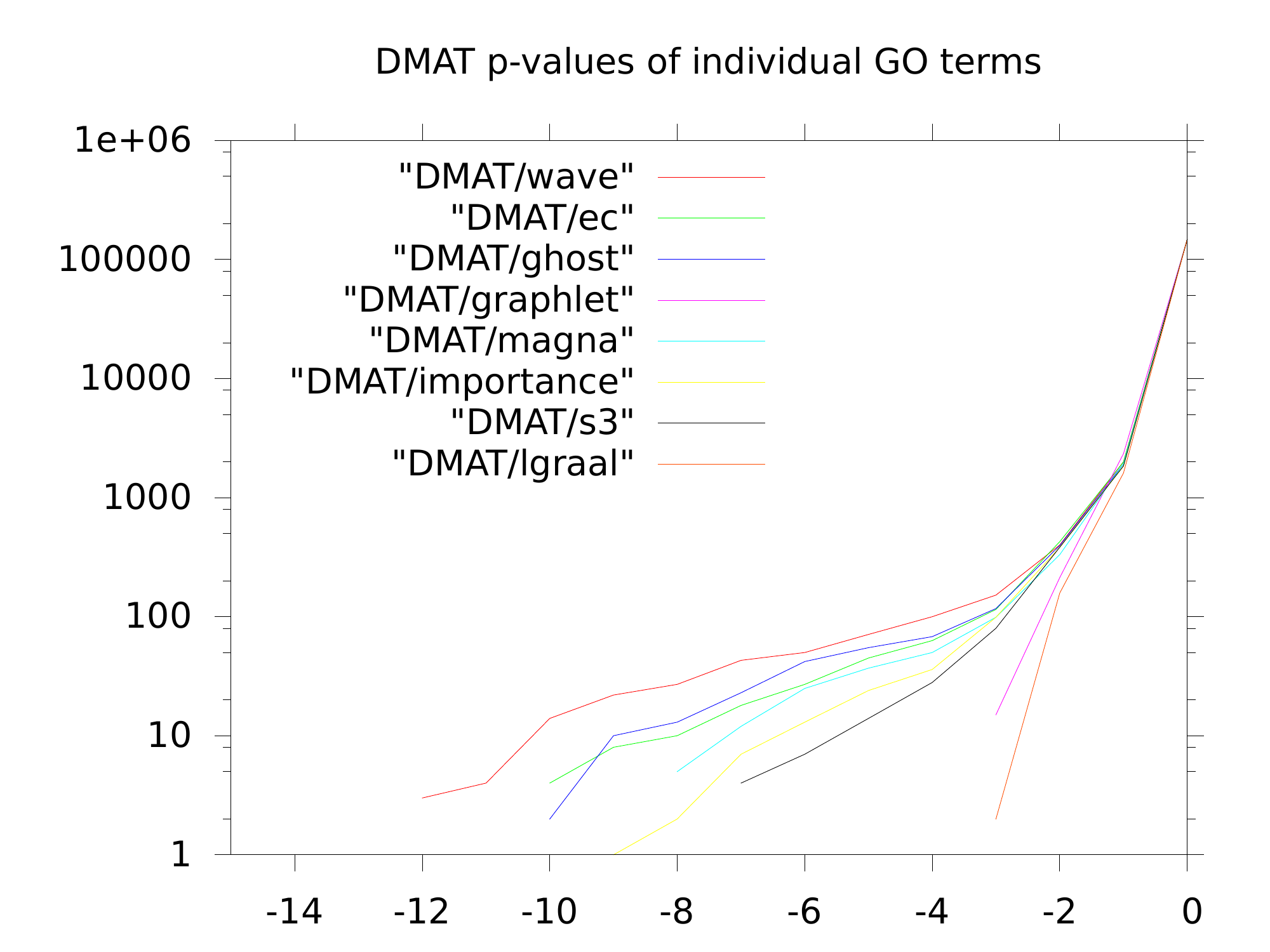}
\includegraphics[width=0.24 \textwidth]{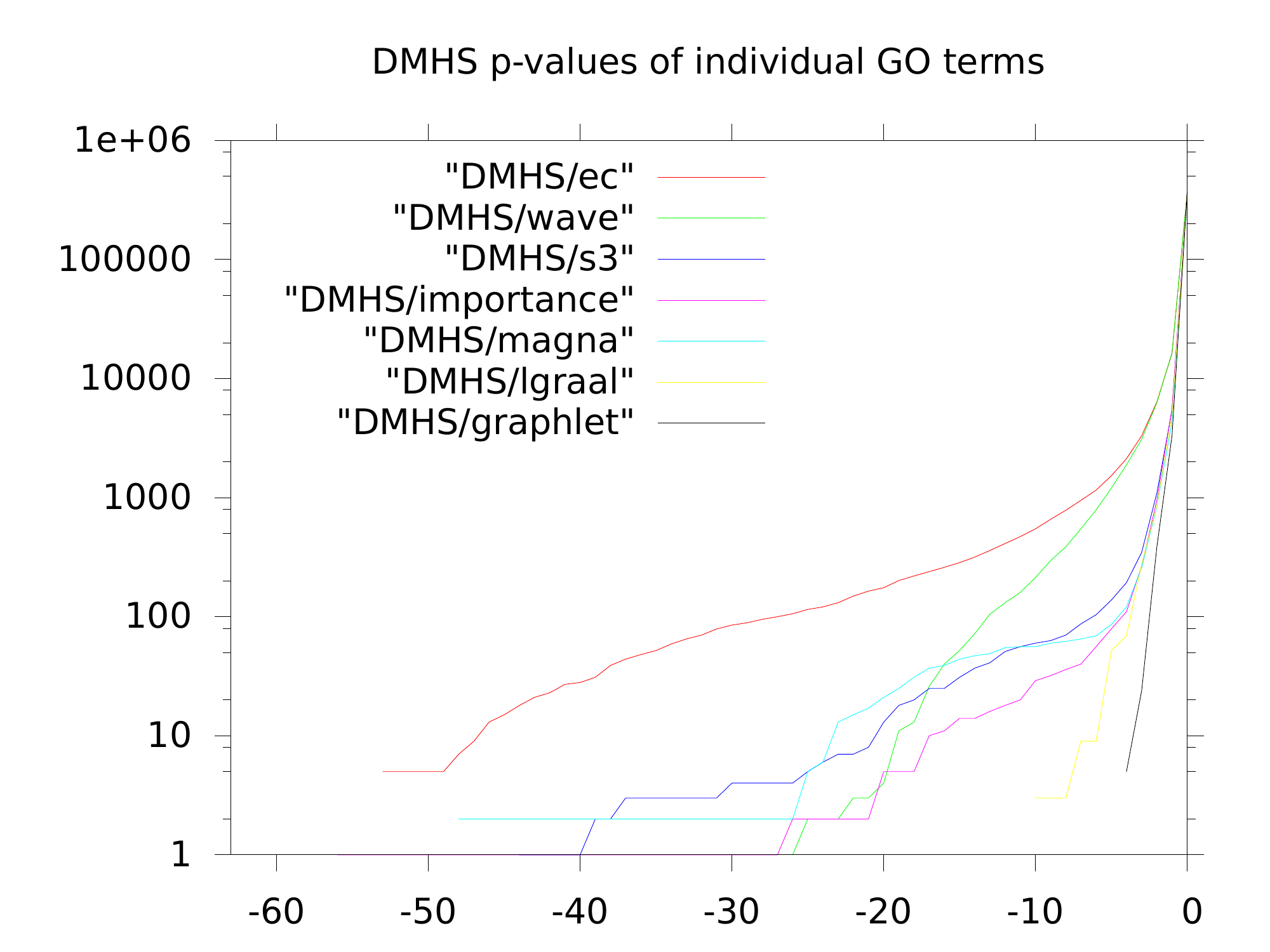}
\includegraphics[width=0.24 \textwidth]{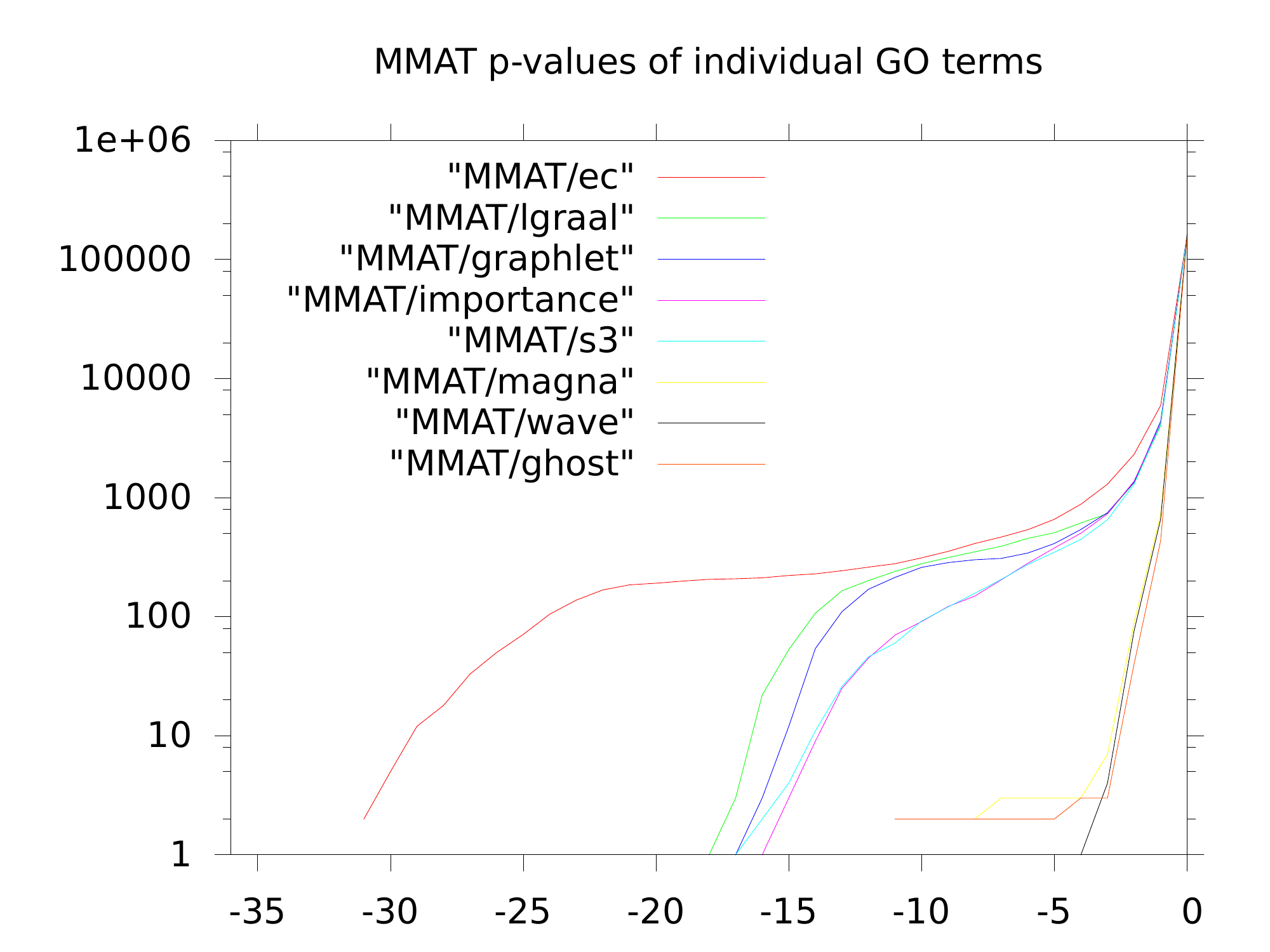}
\includegraphics[width=0.24 \textwidth]{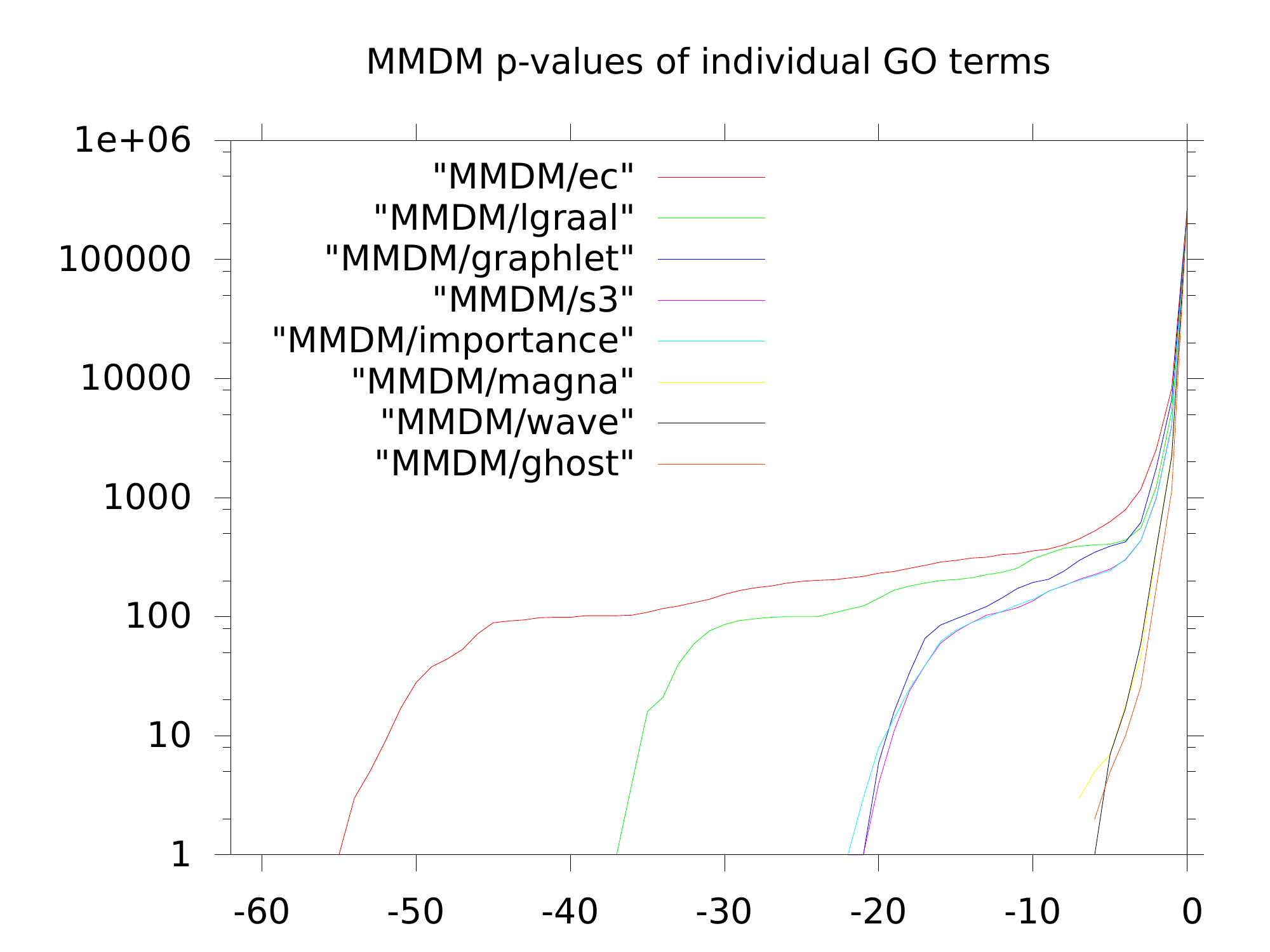}
\includegraphics[width=0.24 \textwidth]{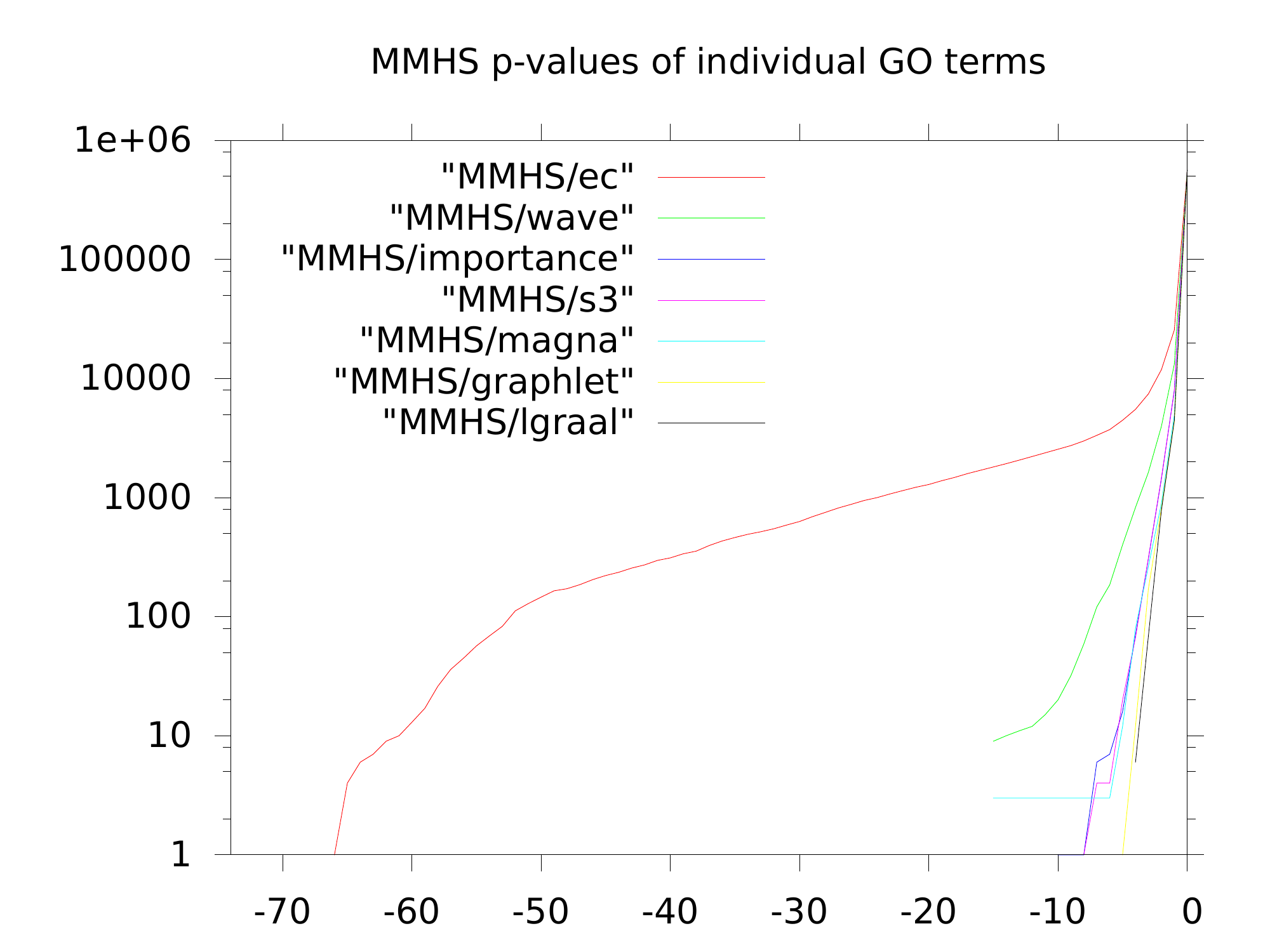}
\includegraphics[width=0.24 \textwidth]{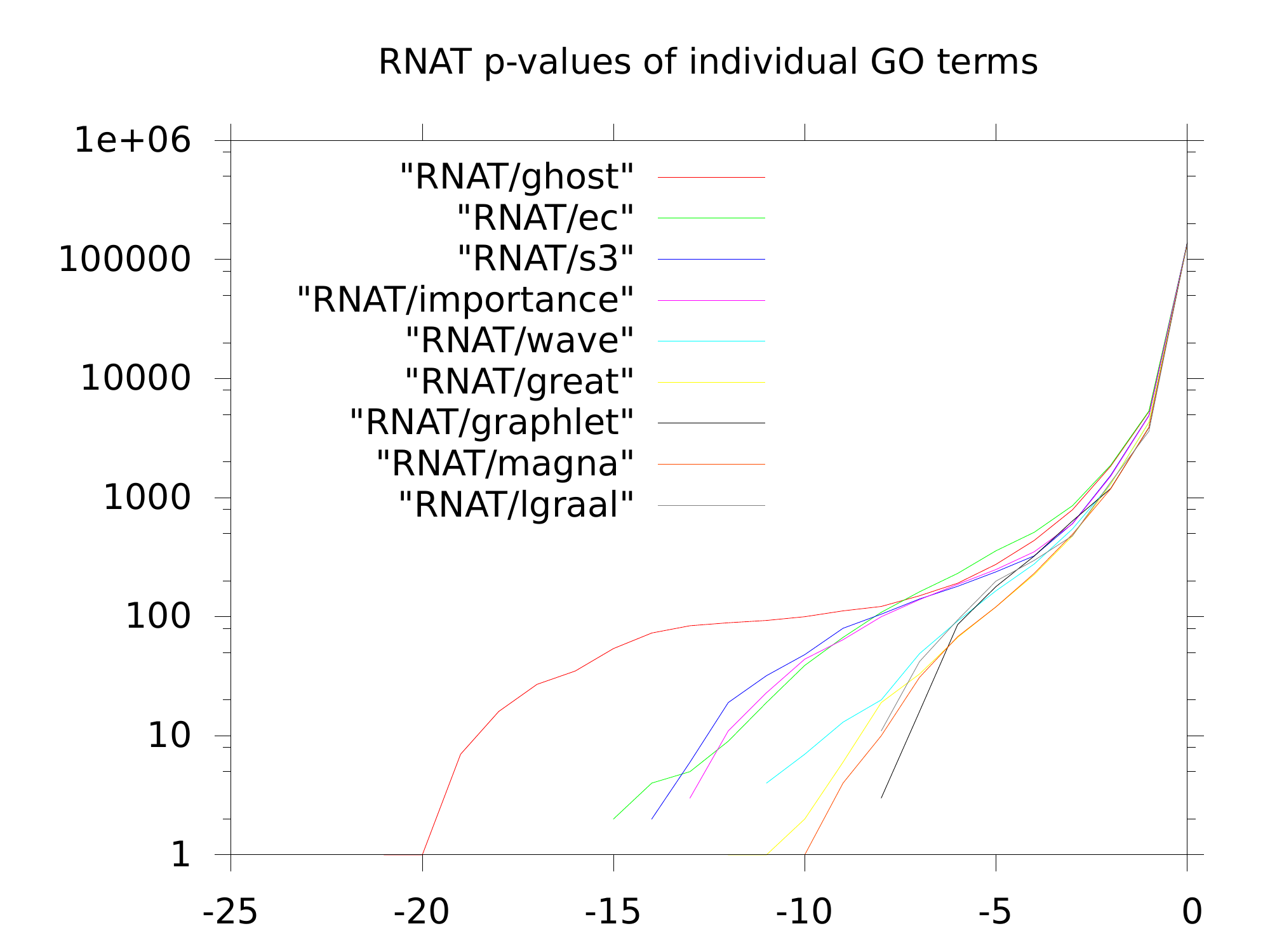}
\includegraphics[width=0.24 \textwidth]{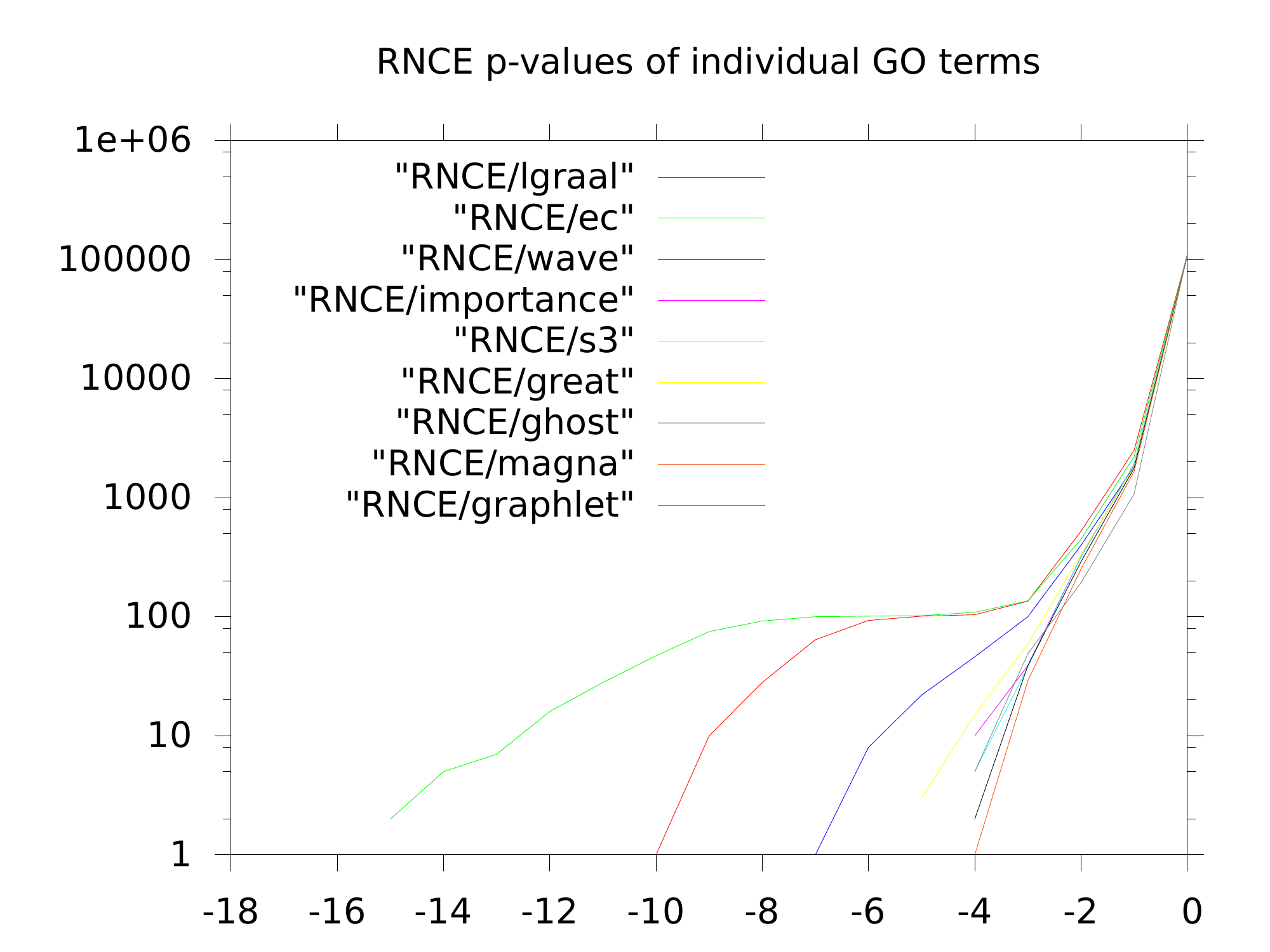}
\includegraphics[width=0.24 \textwidth]{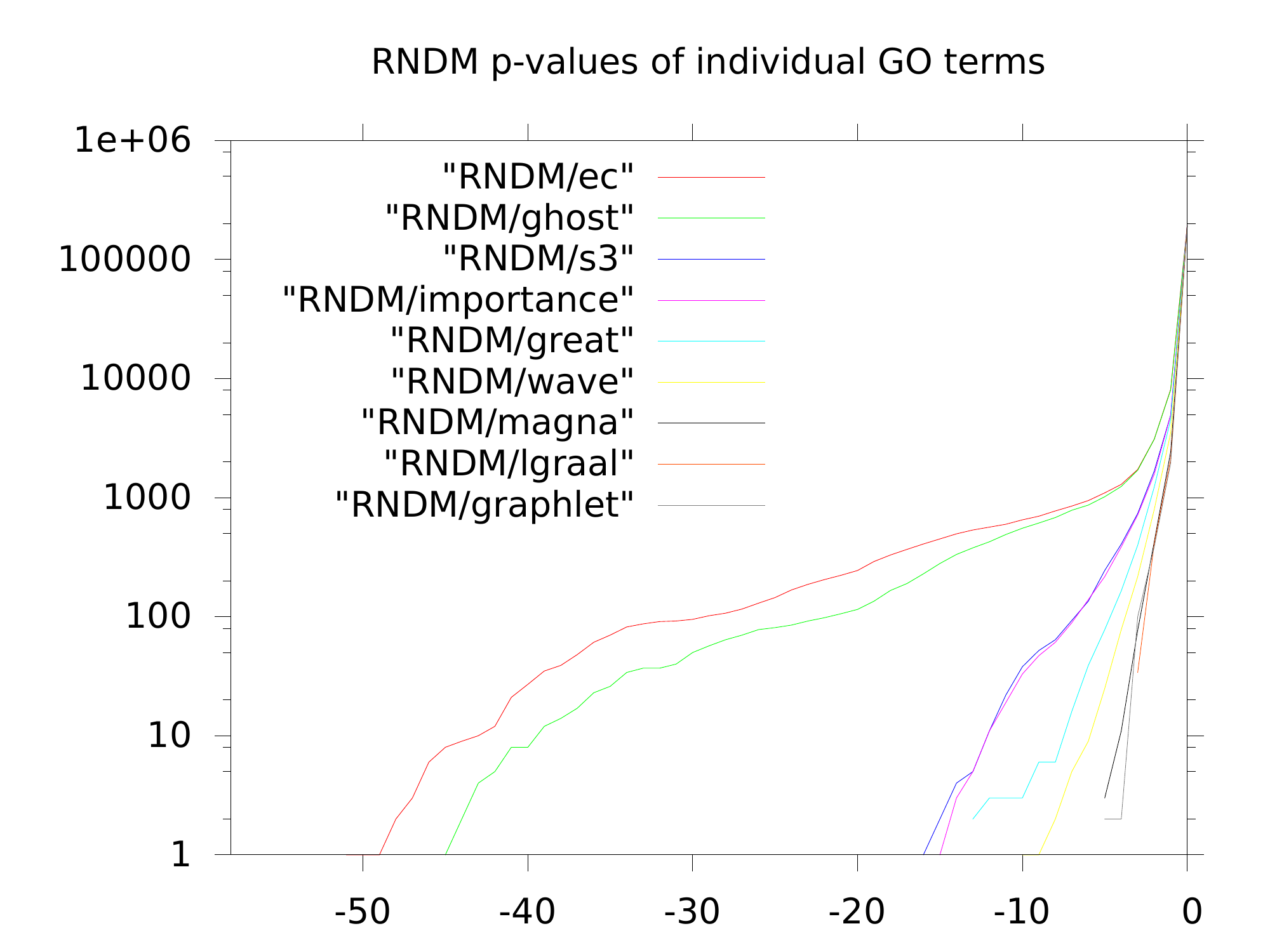}
\includegraphics[width=0.24 \textwidth]{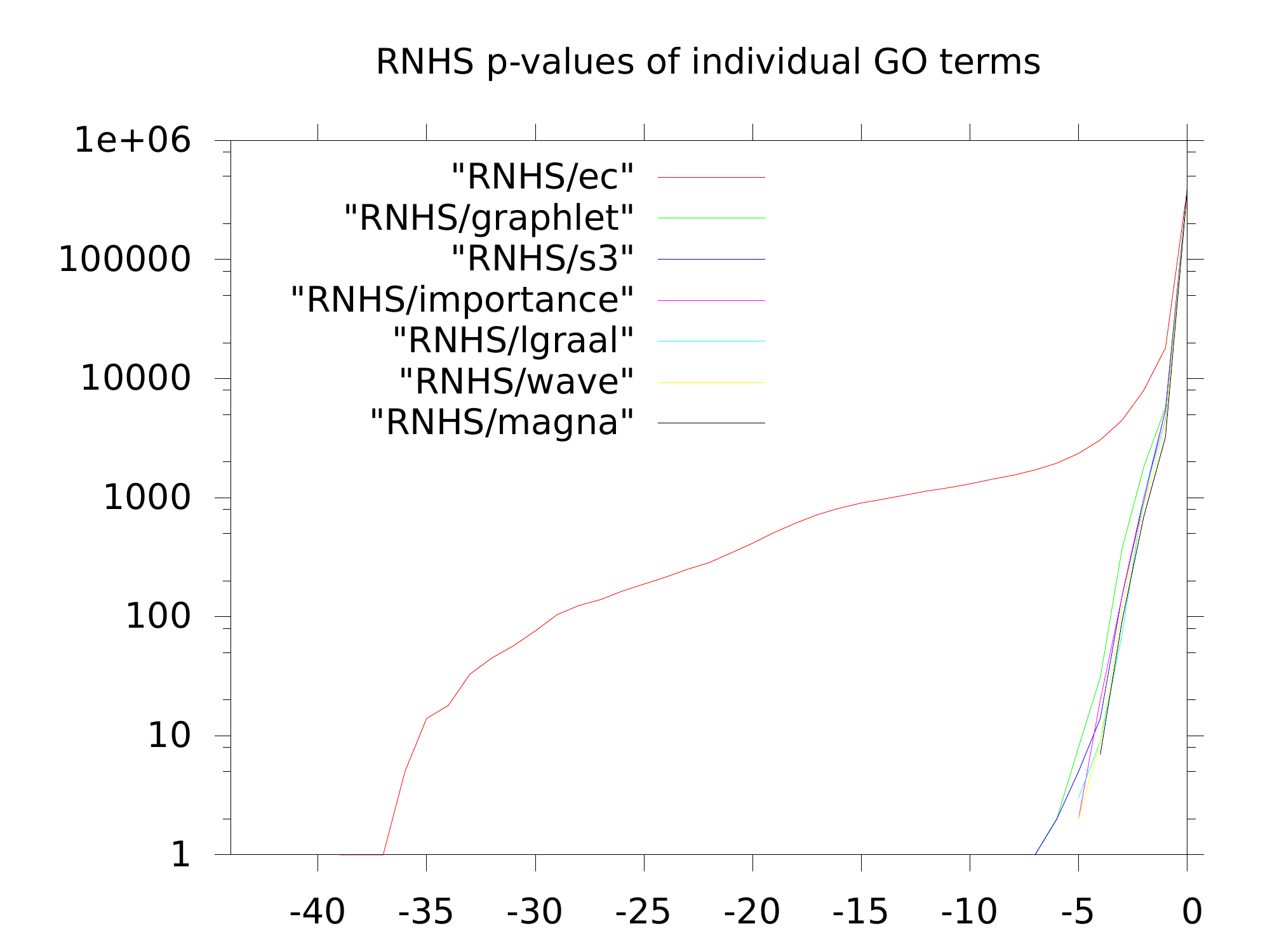}
\includegraphics[width=0.24 \textwidth]{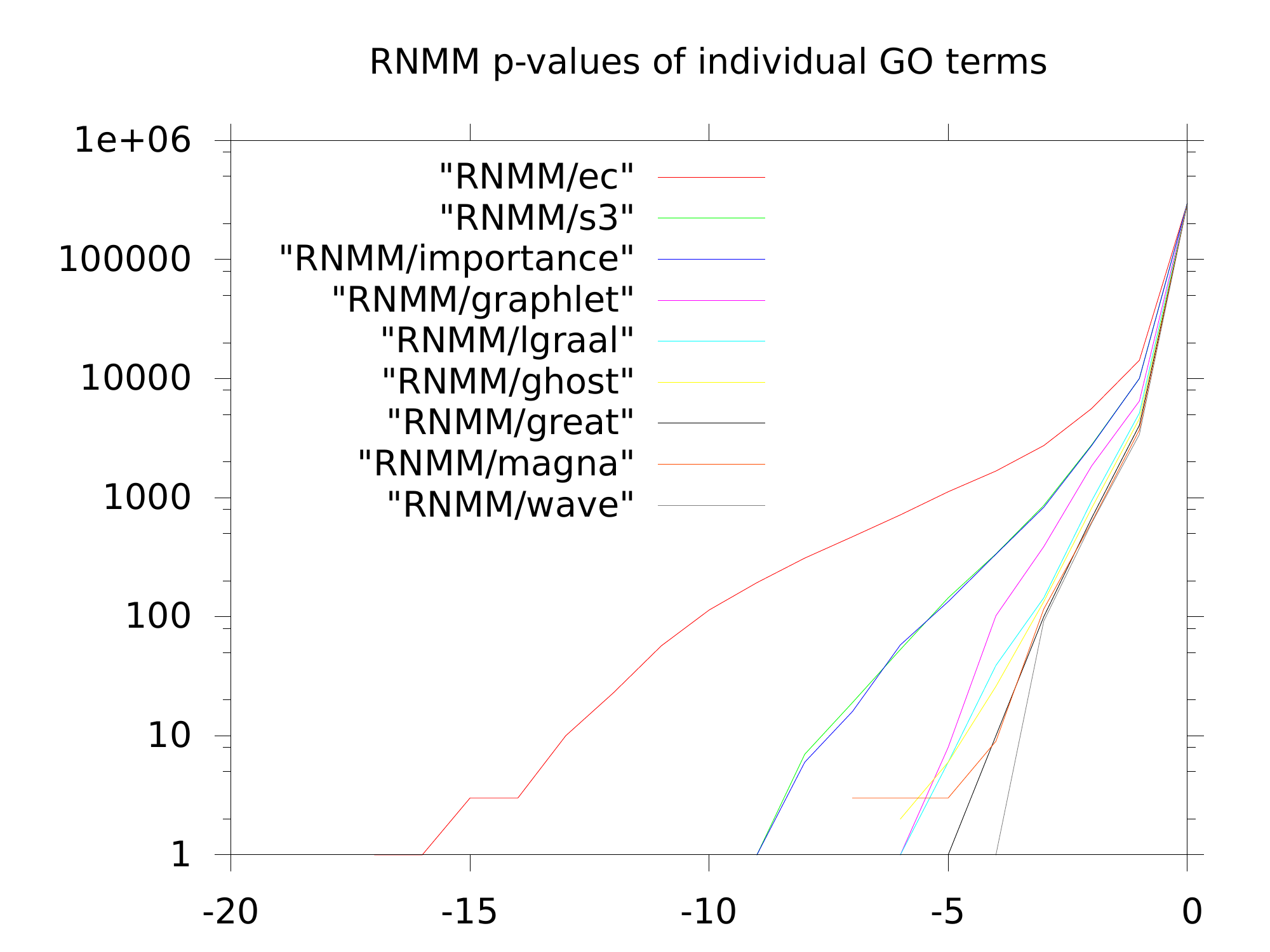}
\includegraphics[width=0.24 \textwidth]{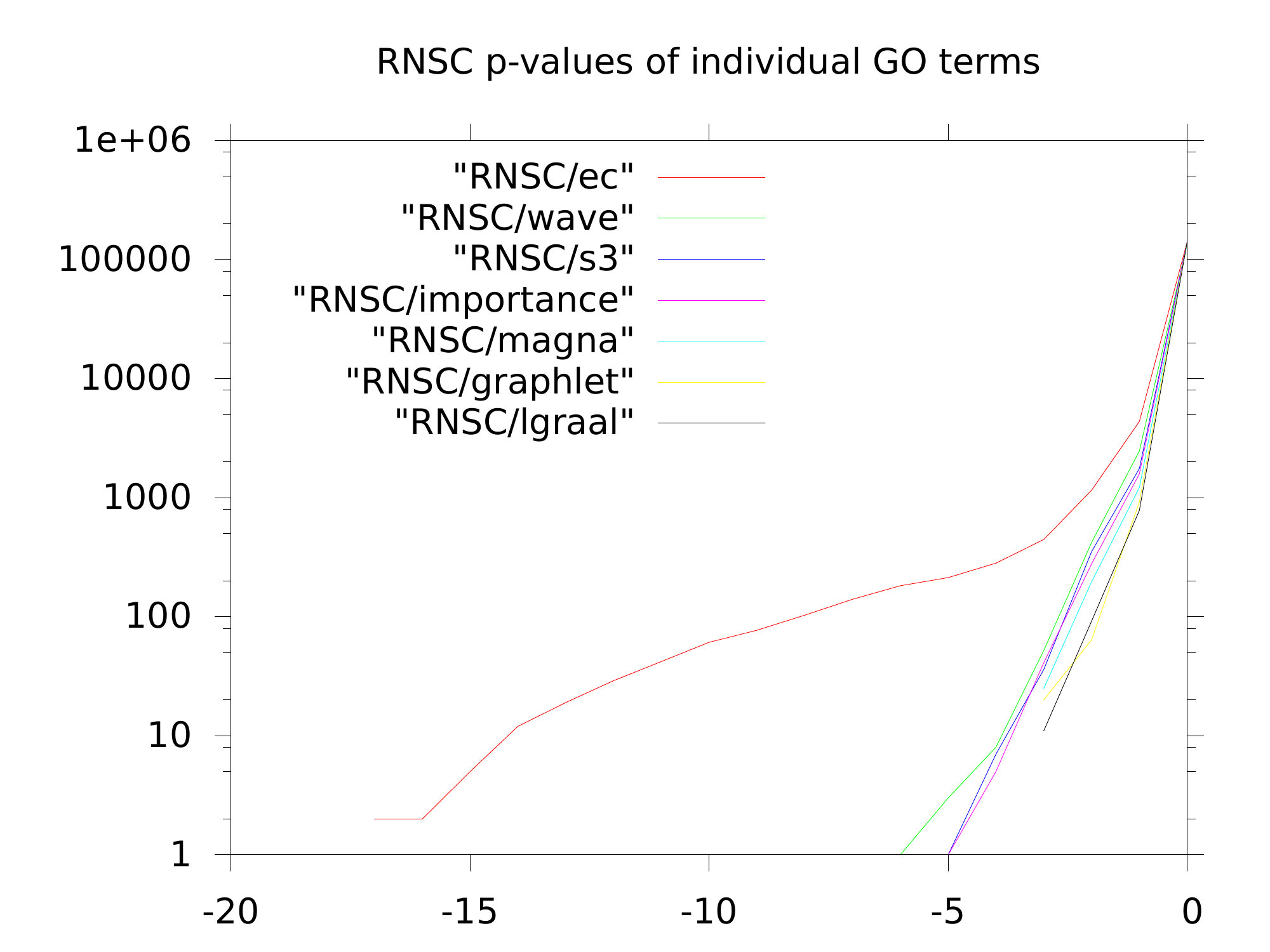}
\includegraphics[width=0.24 \textwidth]{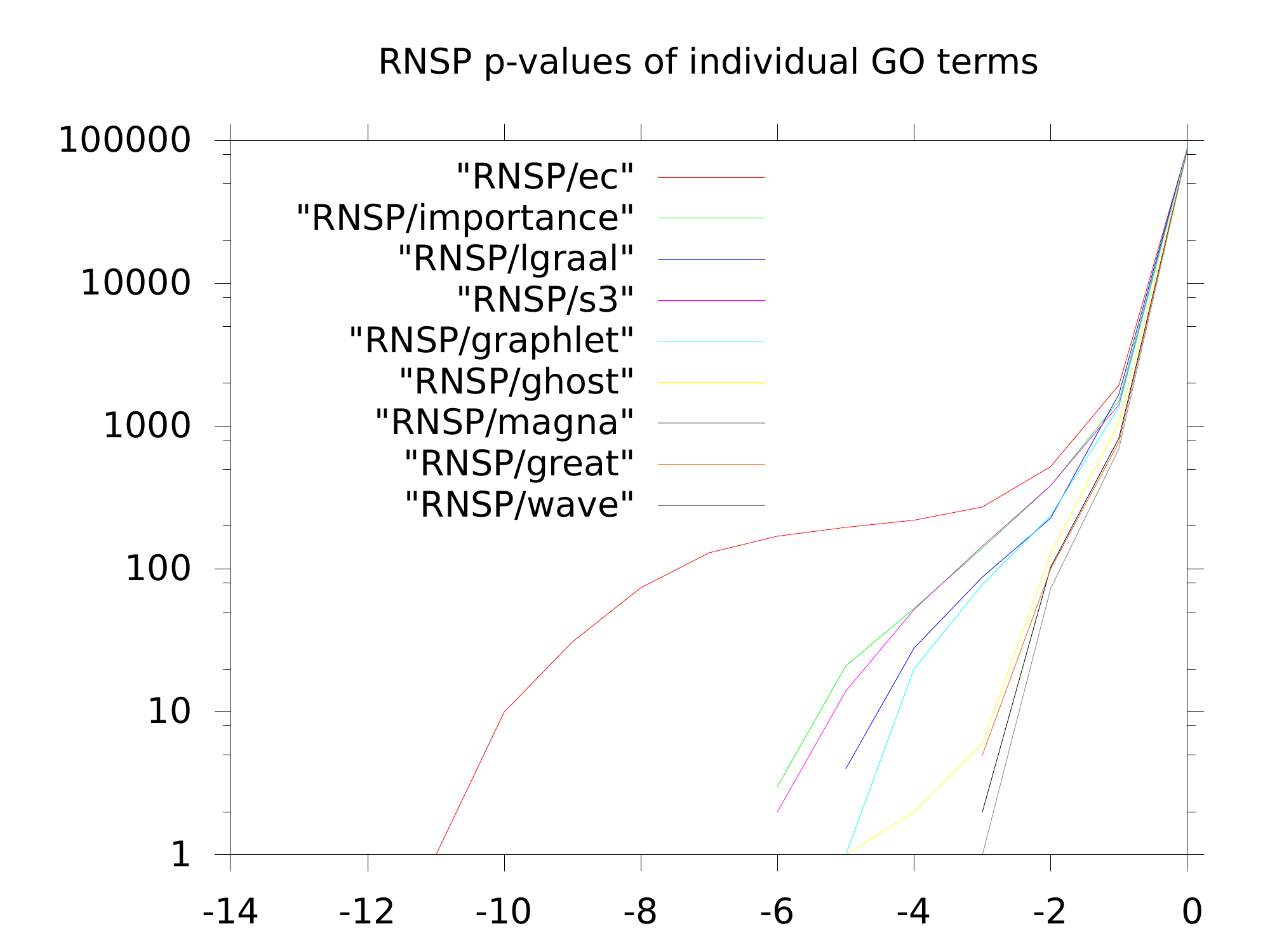}
\includegraphics[width=0.24 \textwidth]{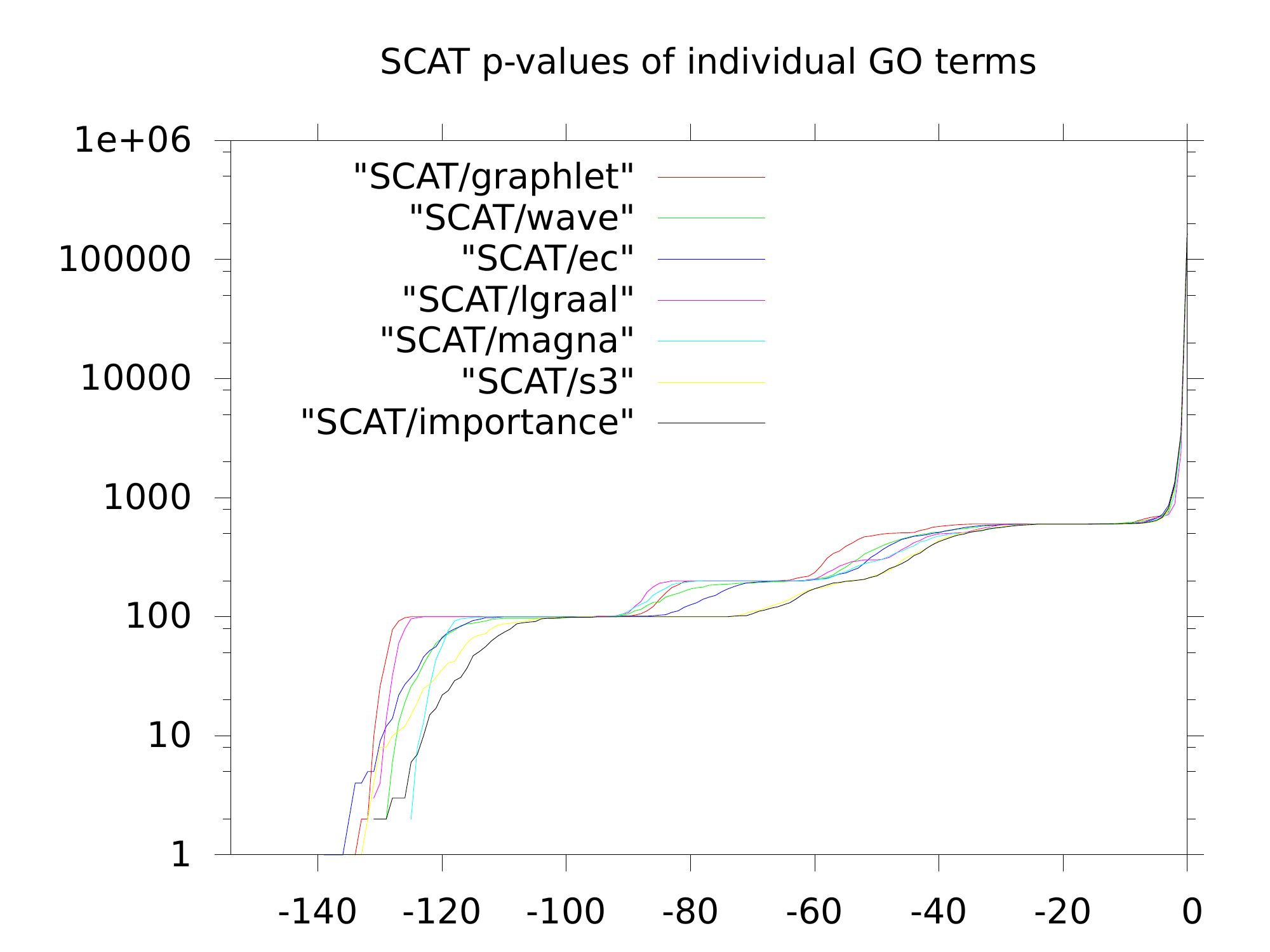}
\includegraphics[width=0.24 \textwidth]{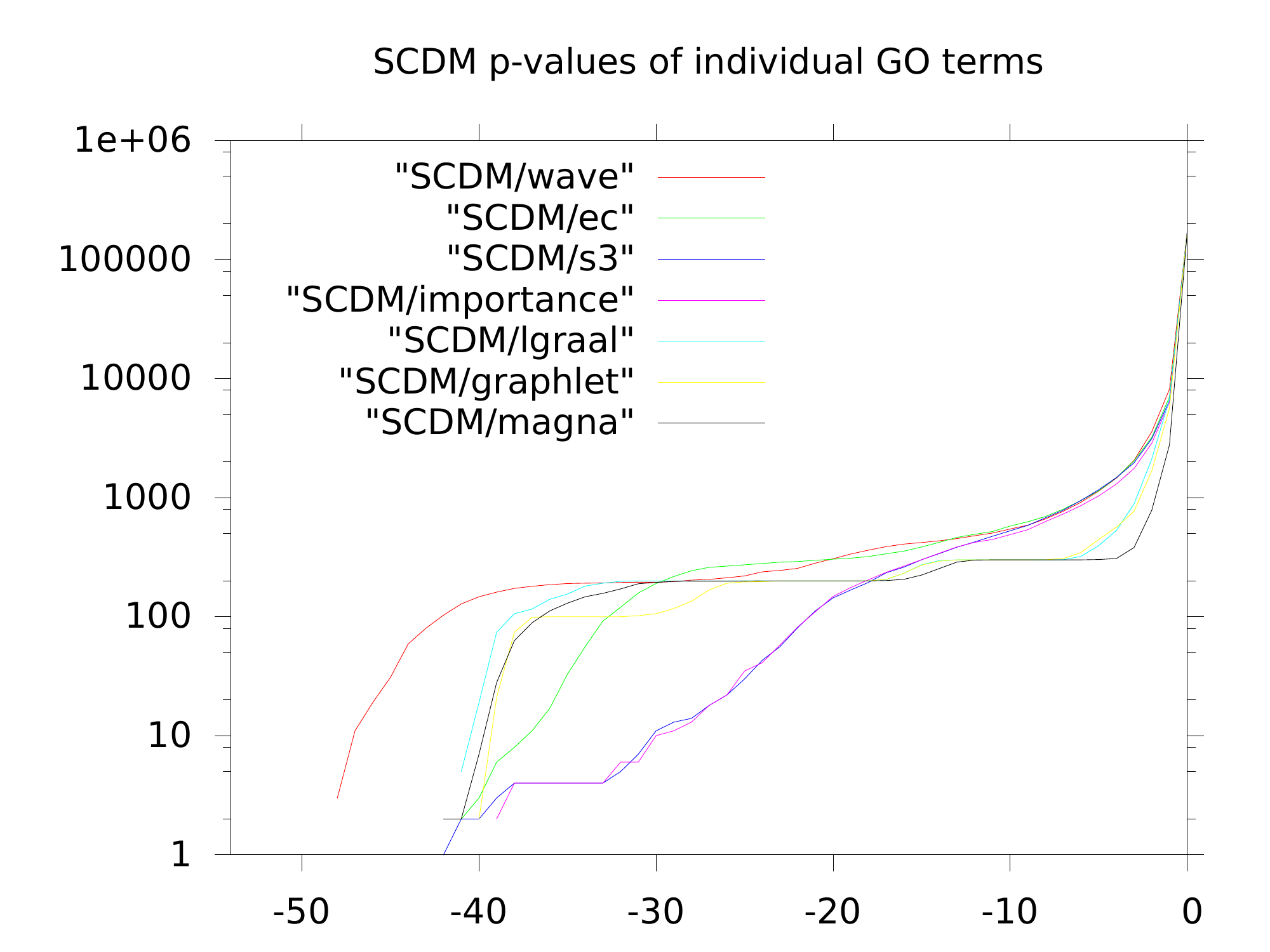}
\includegraphics[width=0.24 \textwidth]{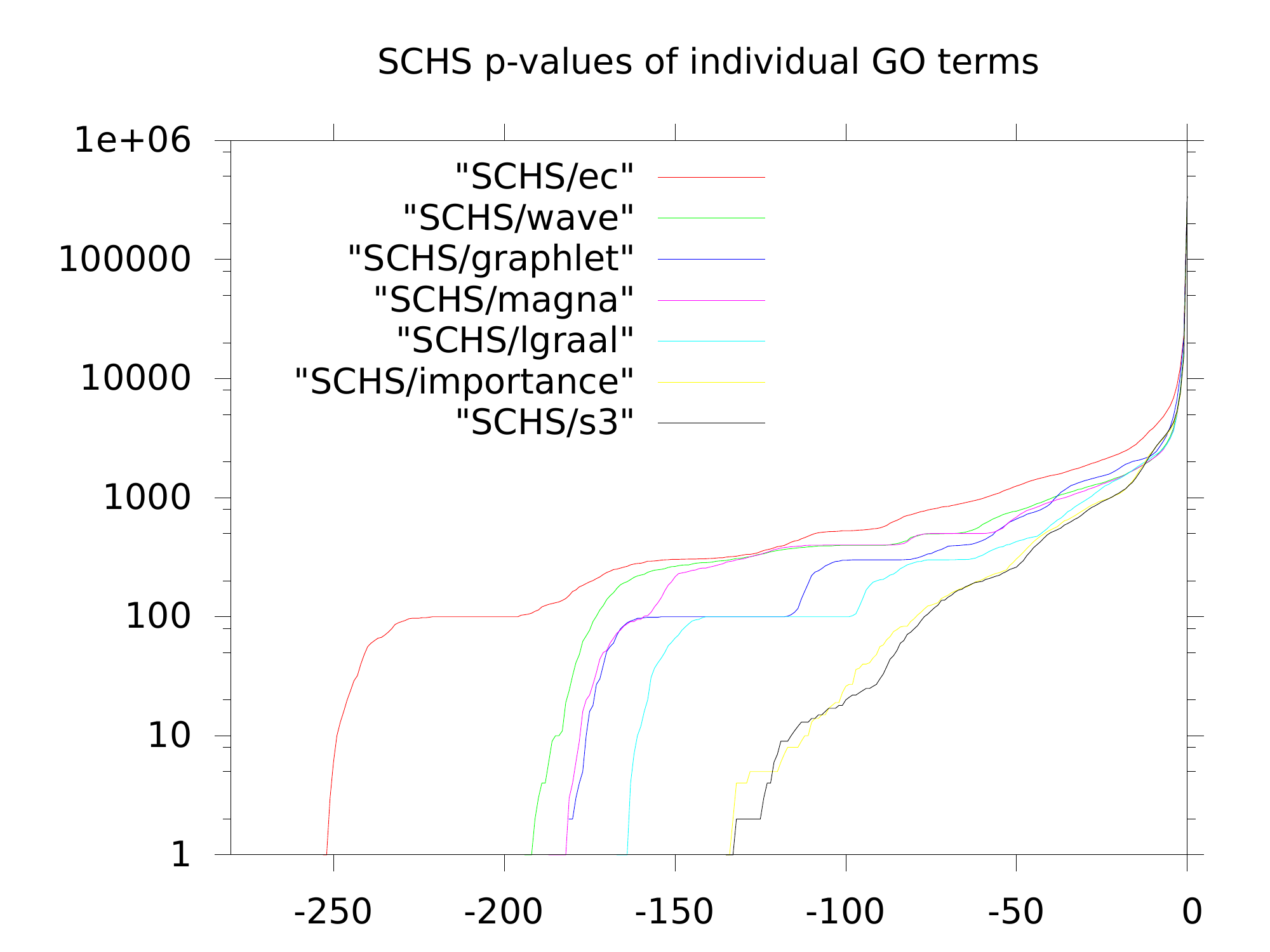}
\includegraphics[width=0.24 \textwidth]{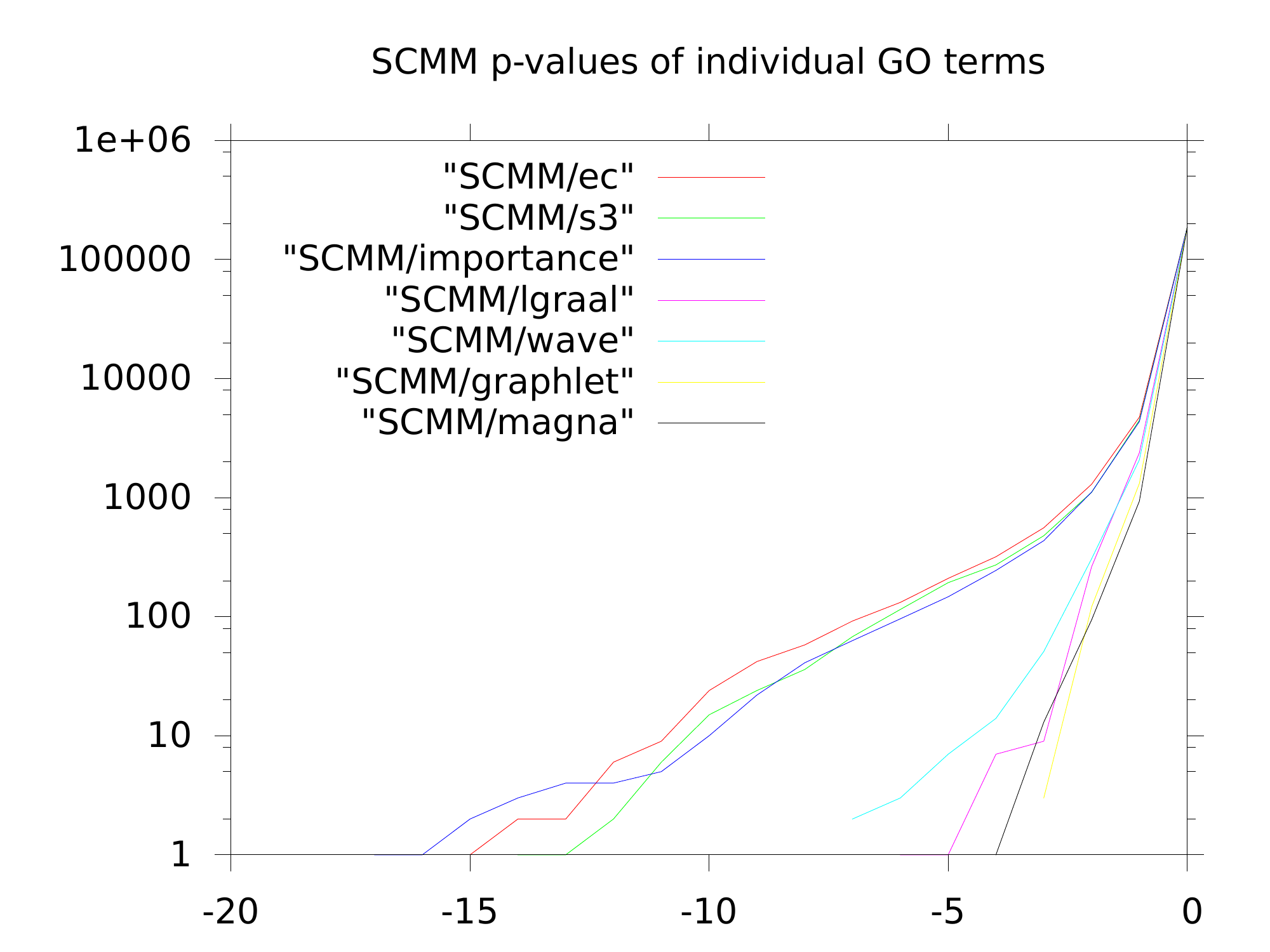}
\includegraphics[width=0.24 \textwidth]{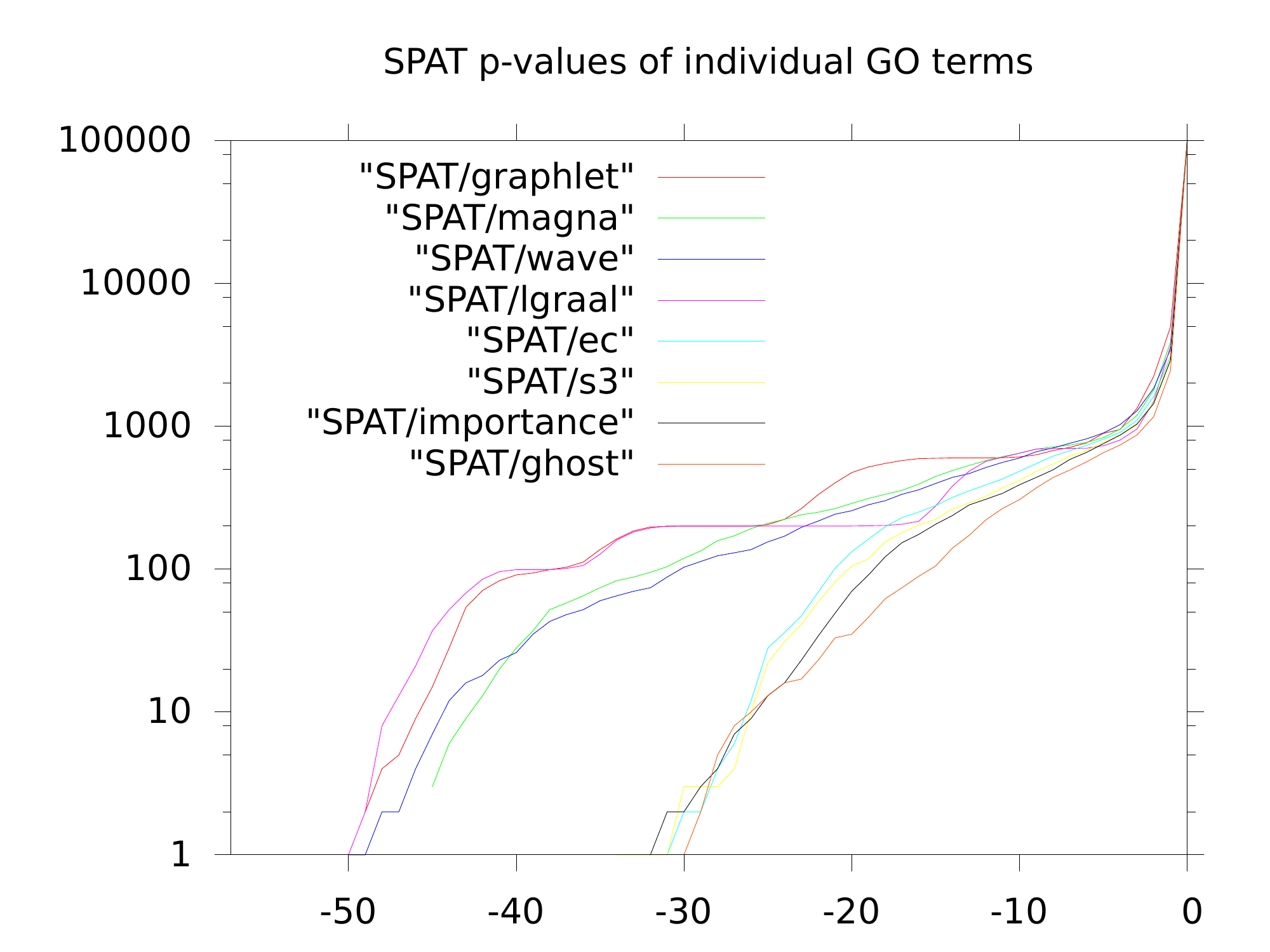}
\includegraphics[width=0.24 \textwidth]{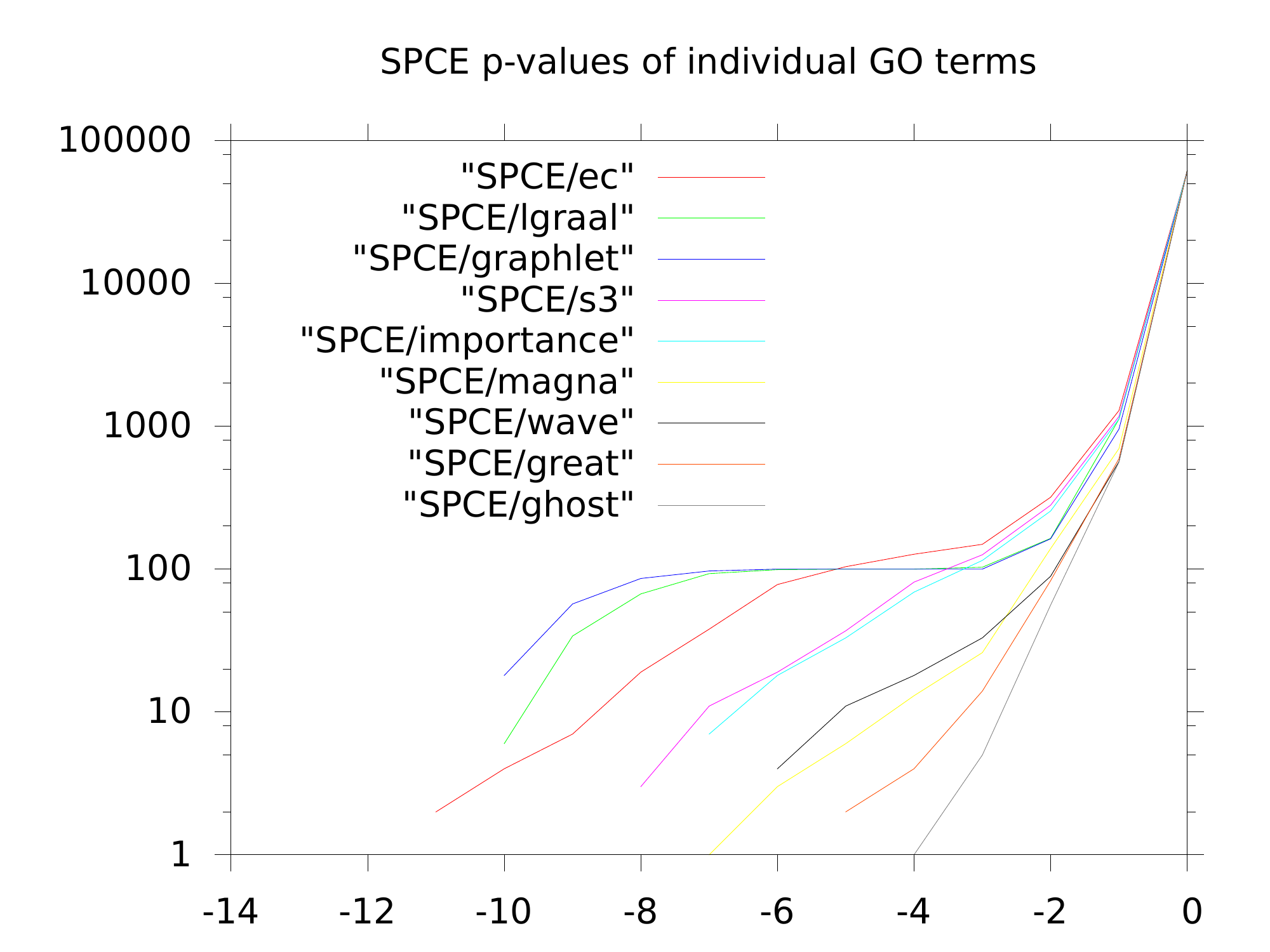}
\includegraphics[width=0.24 \textwidth]{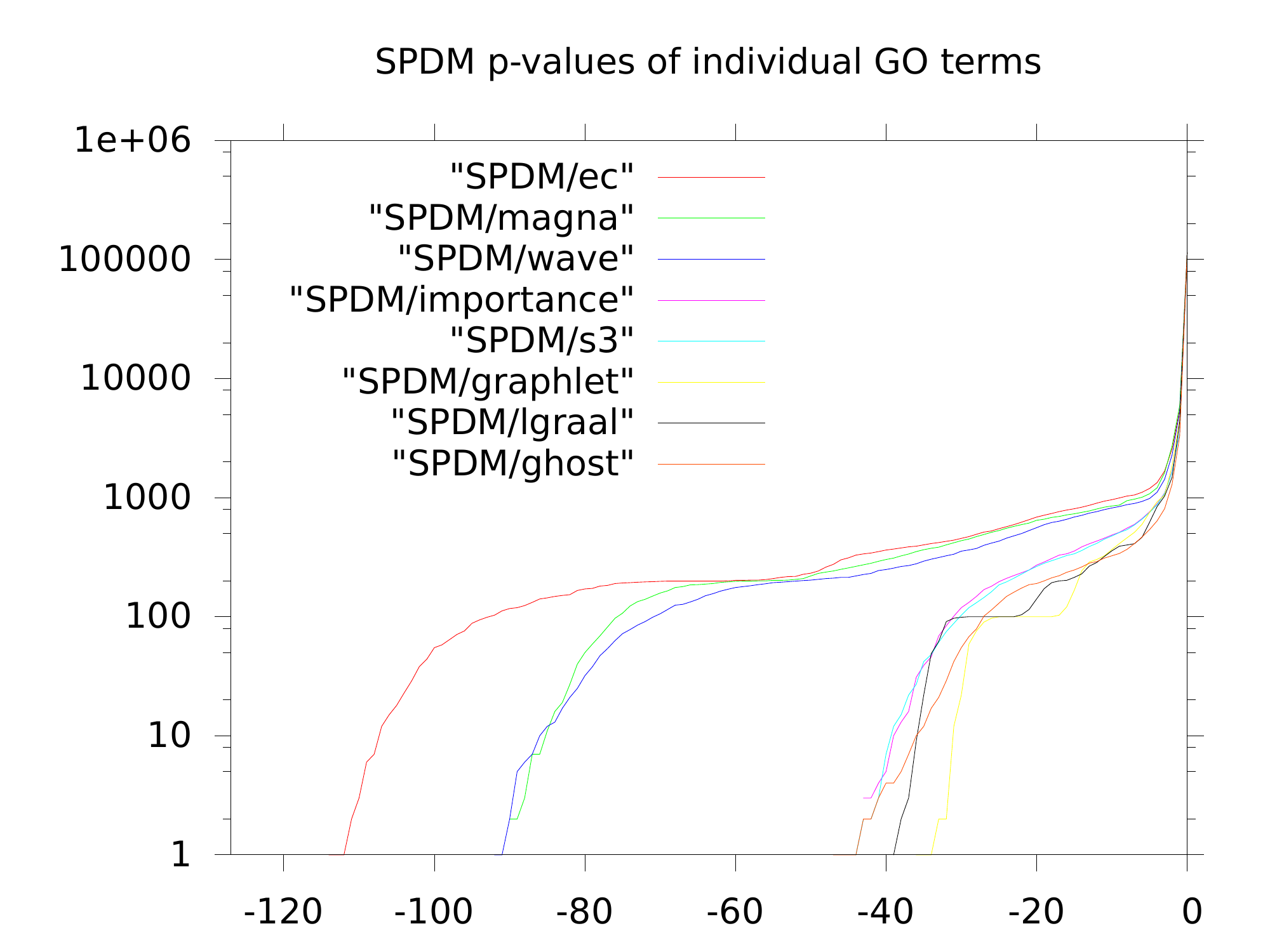}
\includegraphics[width=0.24 \textwidth]{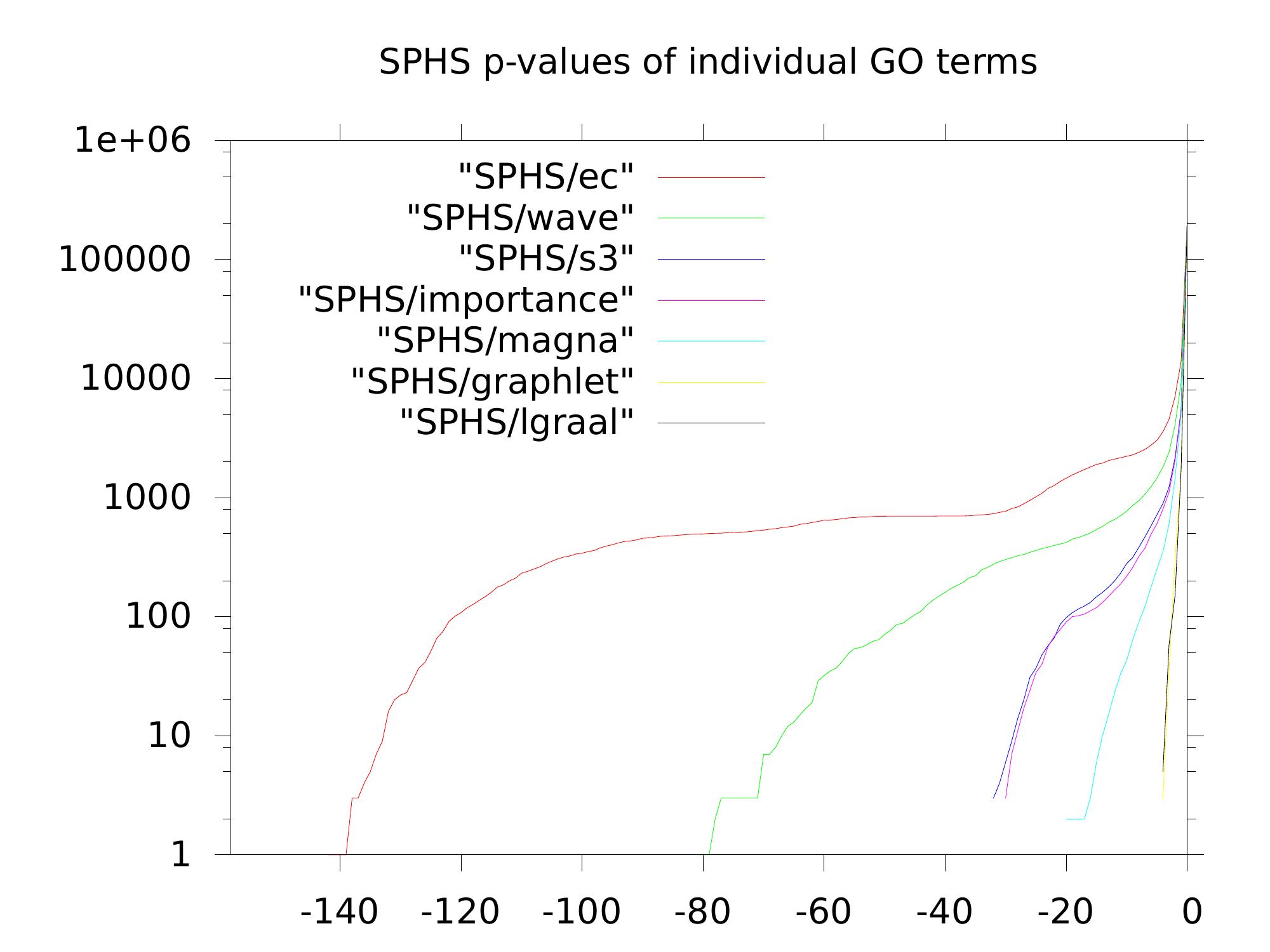}
\includegraphics[width=0.24 \textwidth]{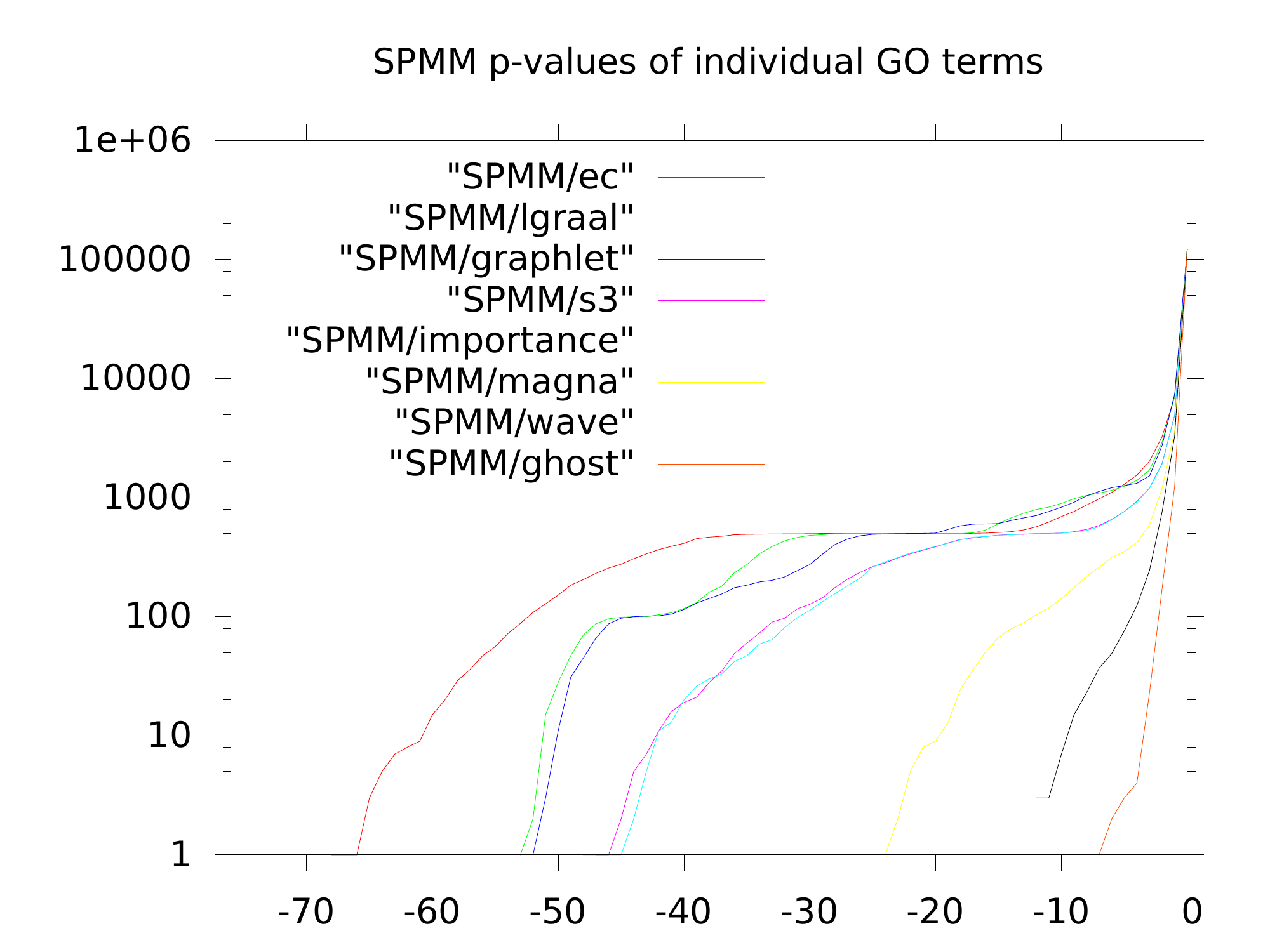}
\includegraphics[width=0.24 \textwidth]{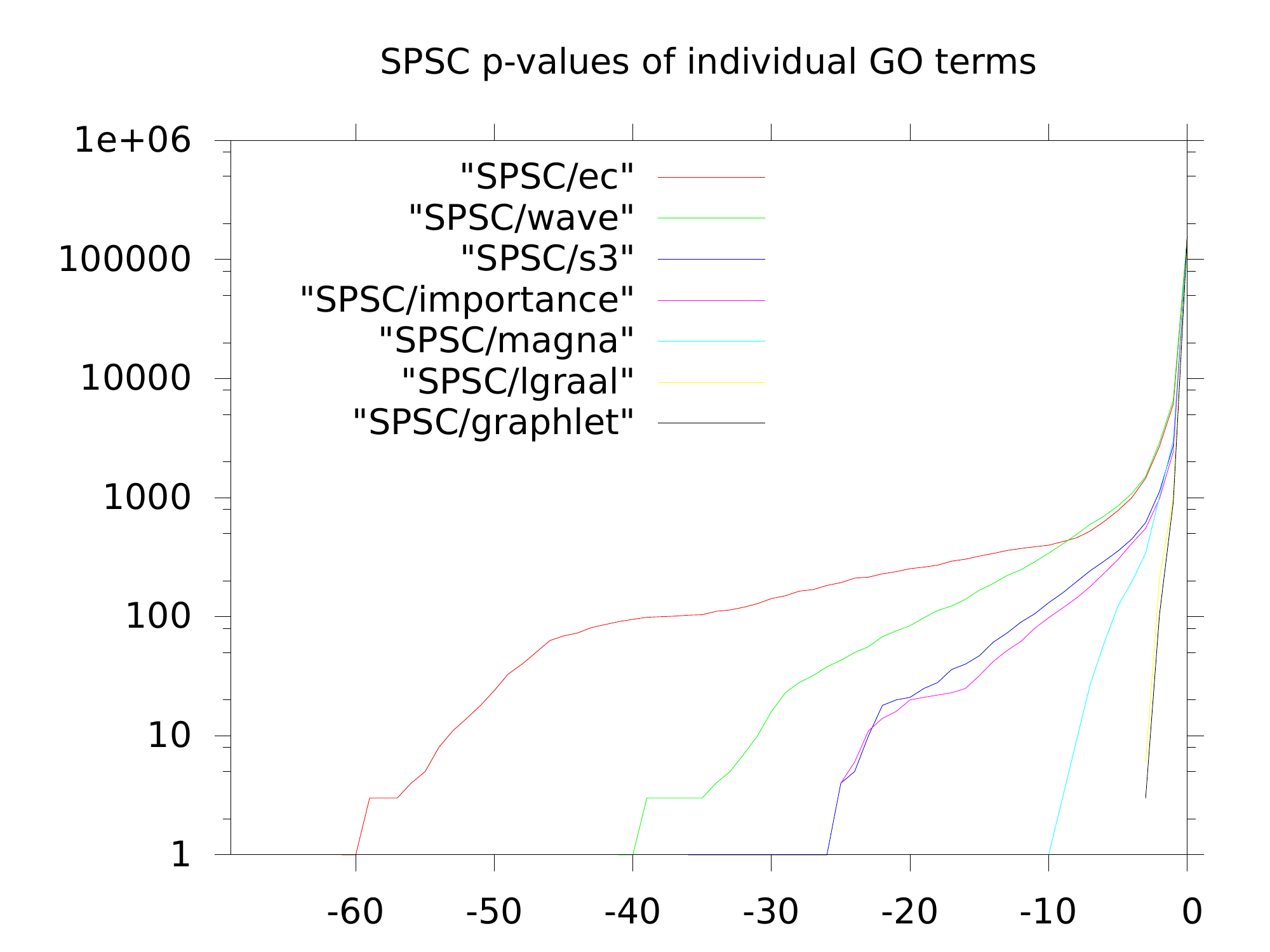}
\caption{Similar to Figure \ref{fig:GOhist}, but for all network pairs (alphabetical order).
Note the horizontal axis is different for each pair.}
    \label{fig:GOhistALL}
\end{figure}

\begin{figure}[htb]
    \centering
\includegraphics[width=0.24 \textwidth]{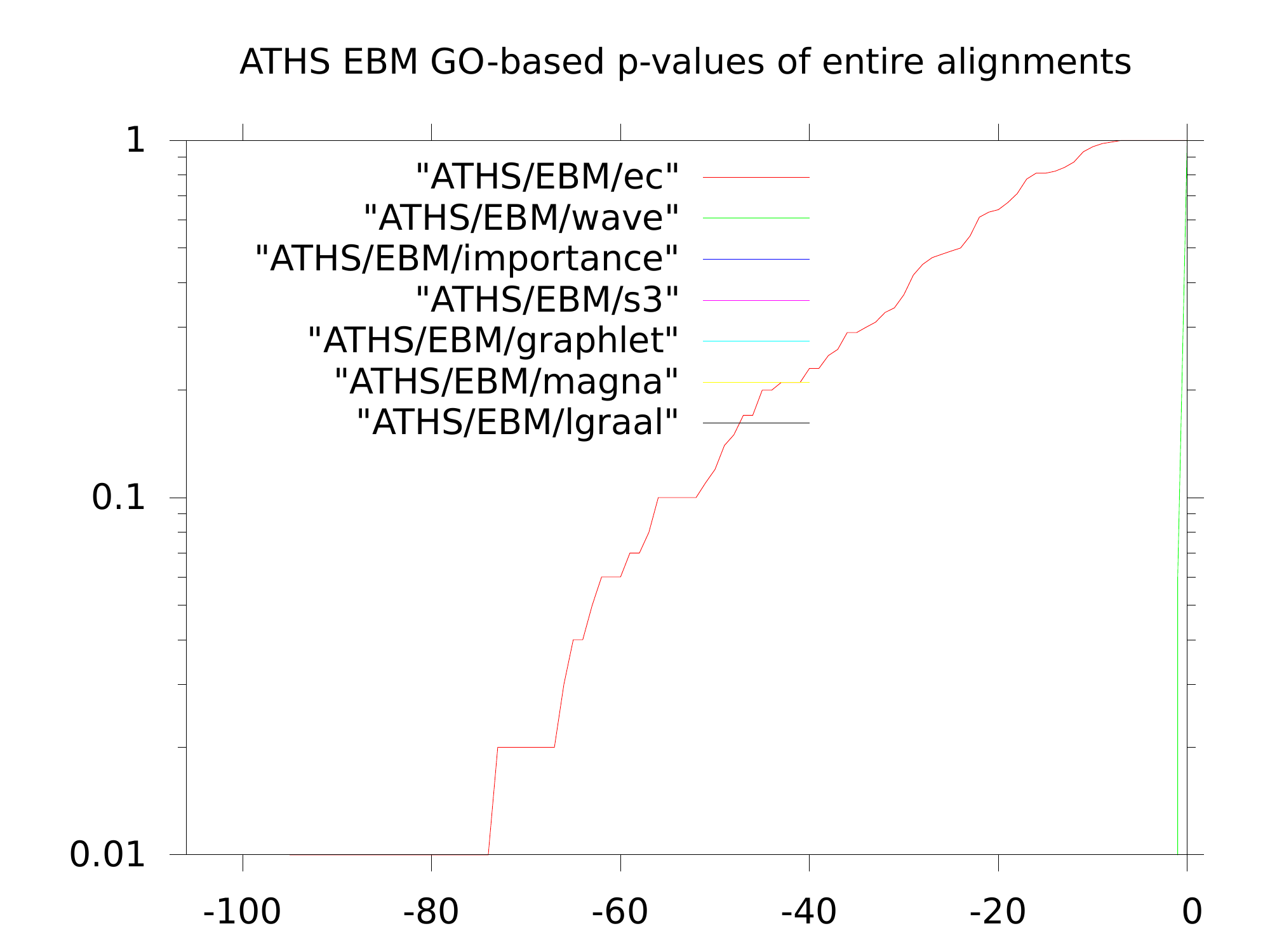}
\includegraphics[width=0.24 \textwidth]{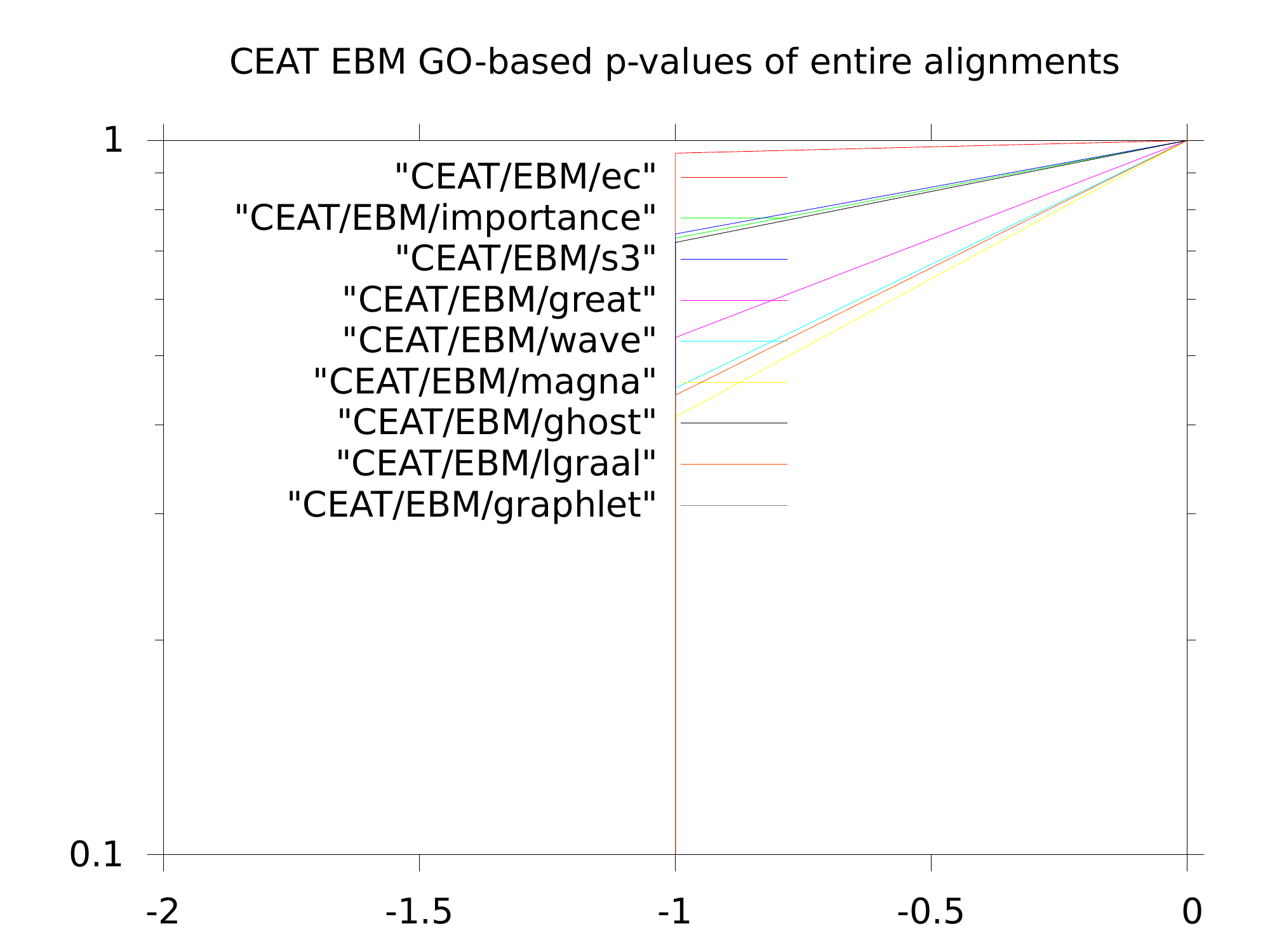}
\includegraphics[width=0.24 \textwidth]{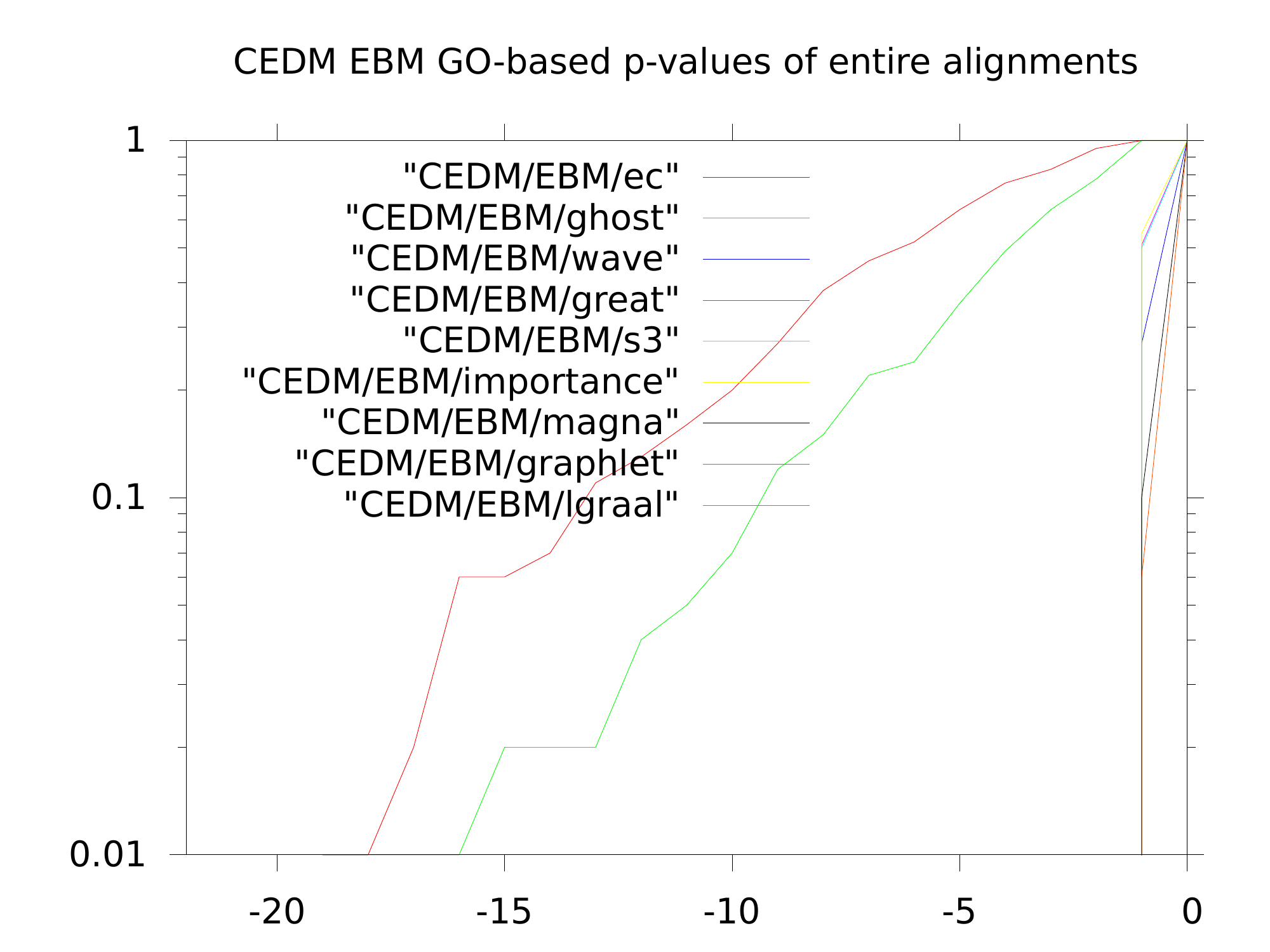}
\includegraphics[width=0.24 \textwidth]{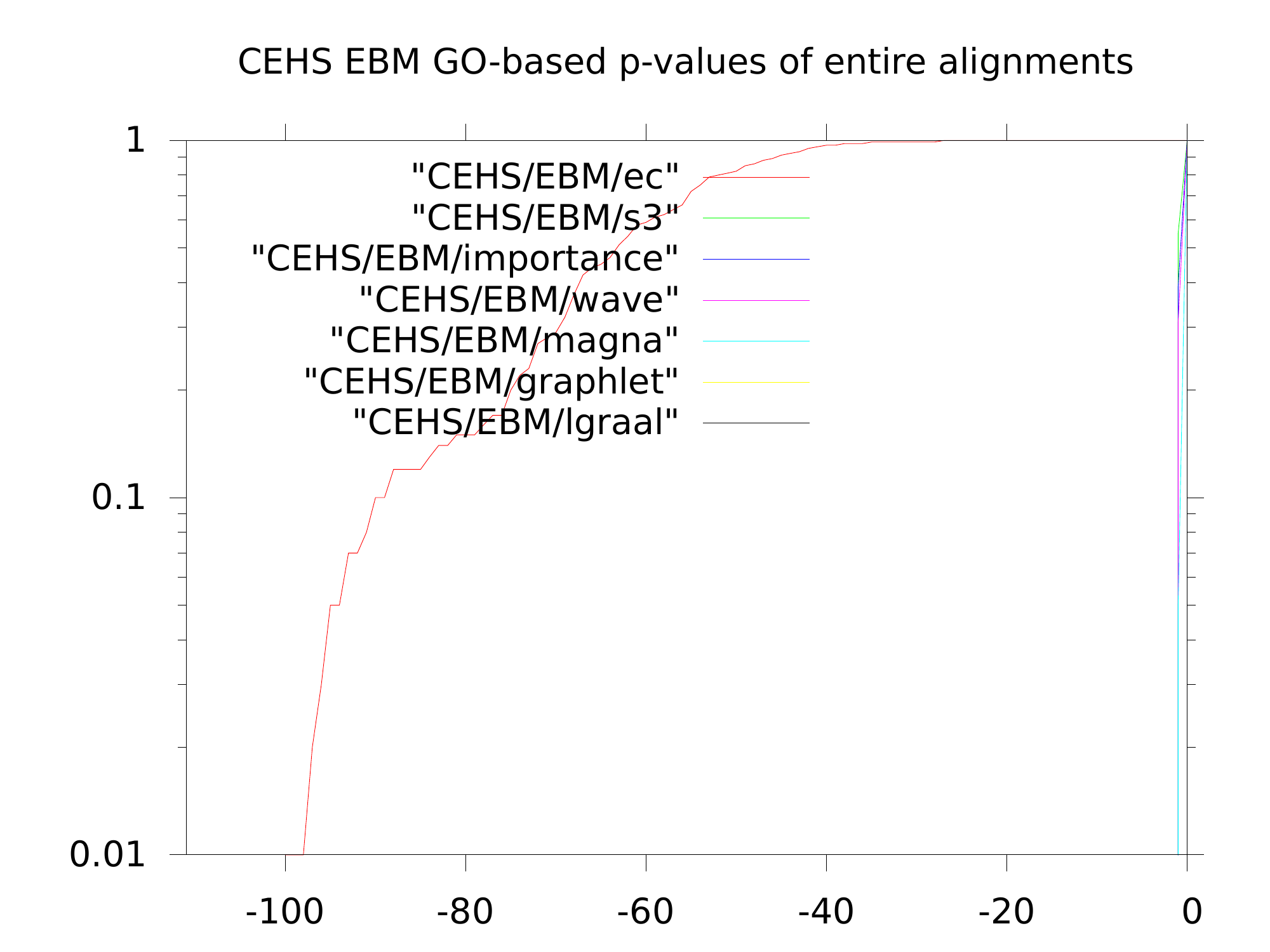}
\includegraphics[width=0.24 \textwidth]{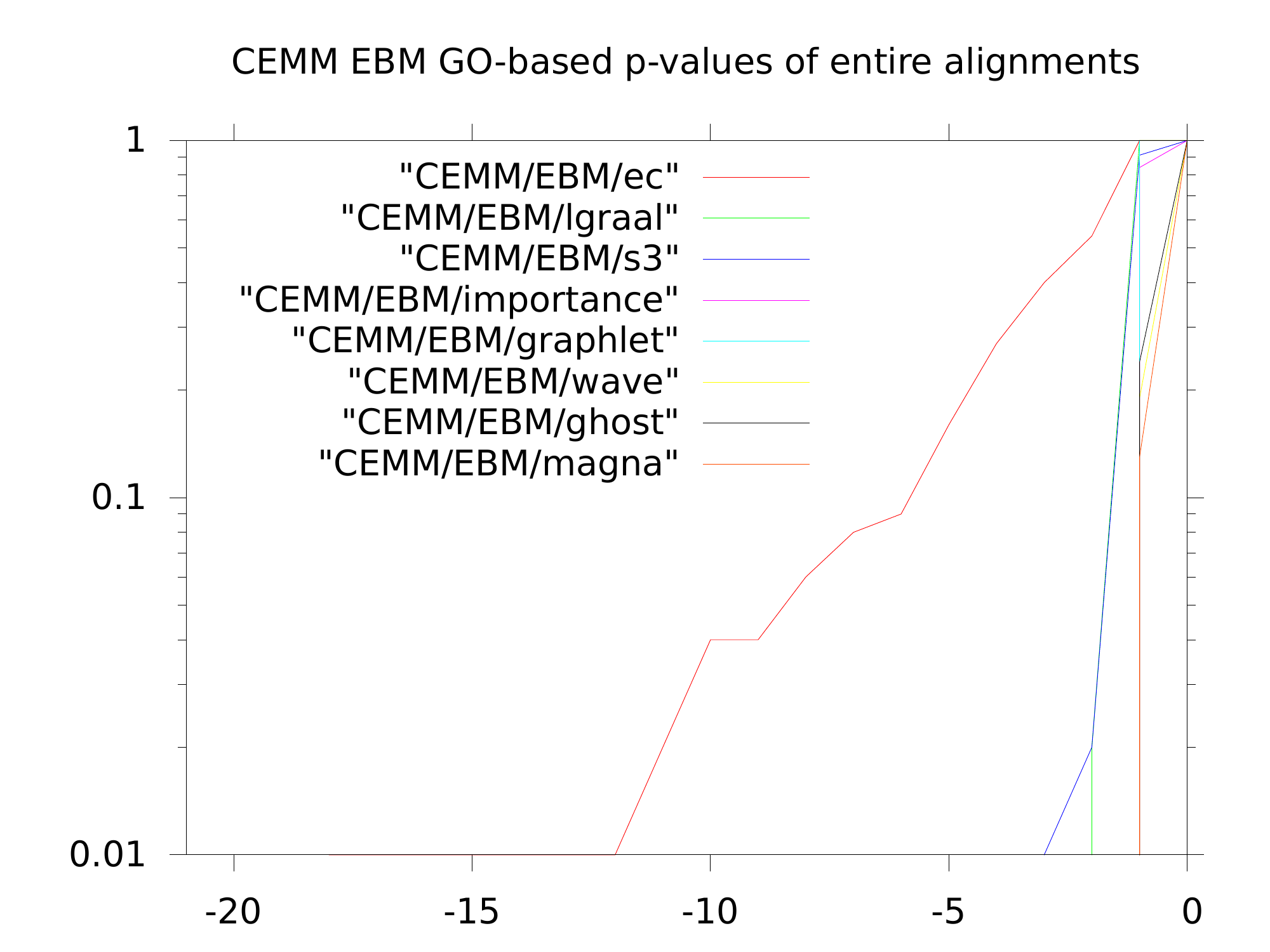}
\includegraphics[width=0.24 \textwidth]{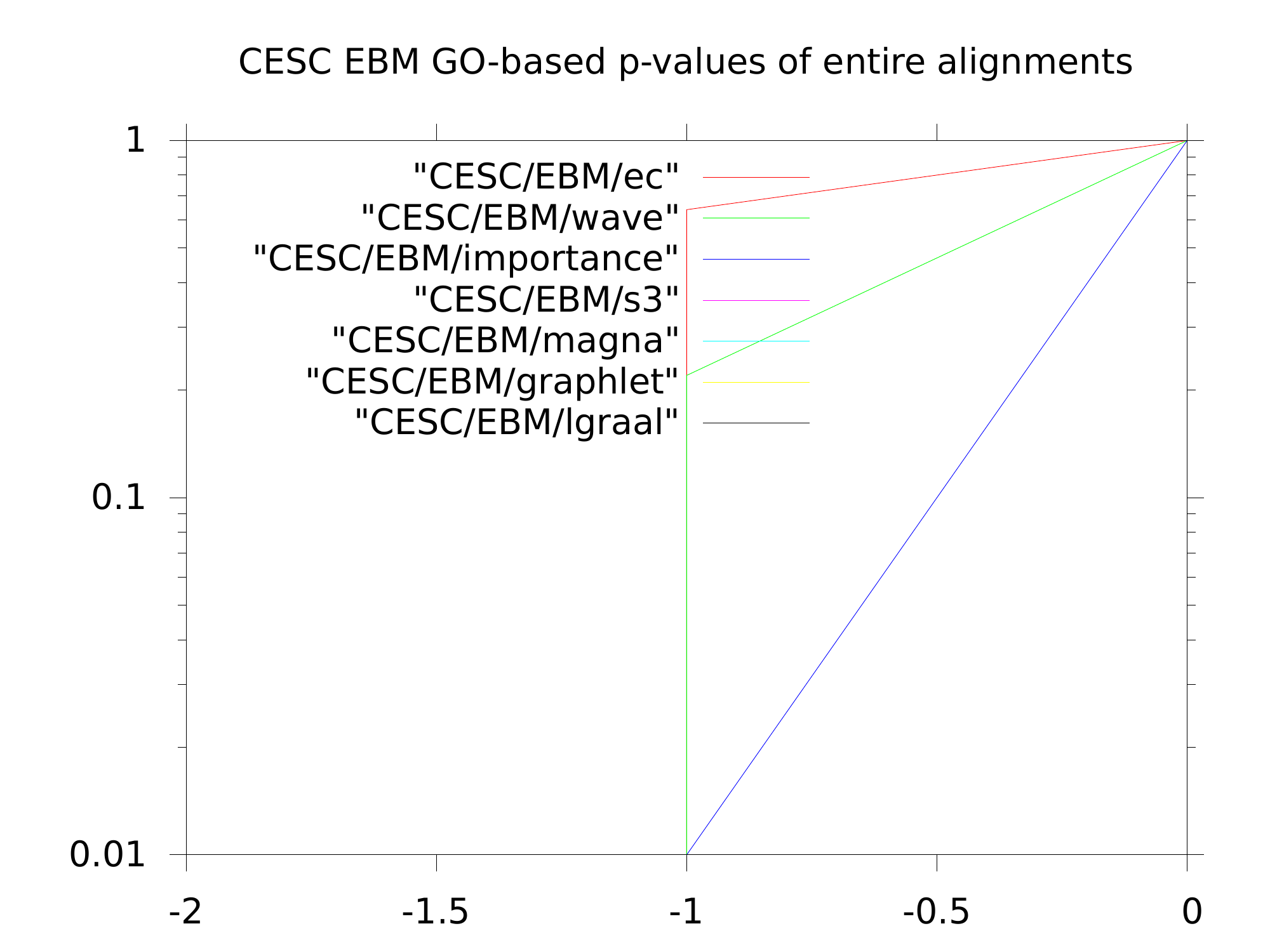}
\includegraphics[width=0.24 \textwidth]{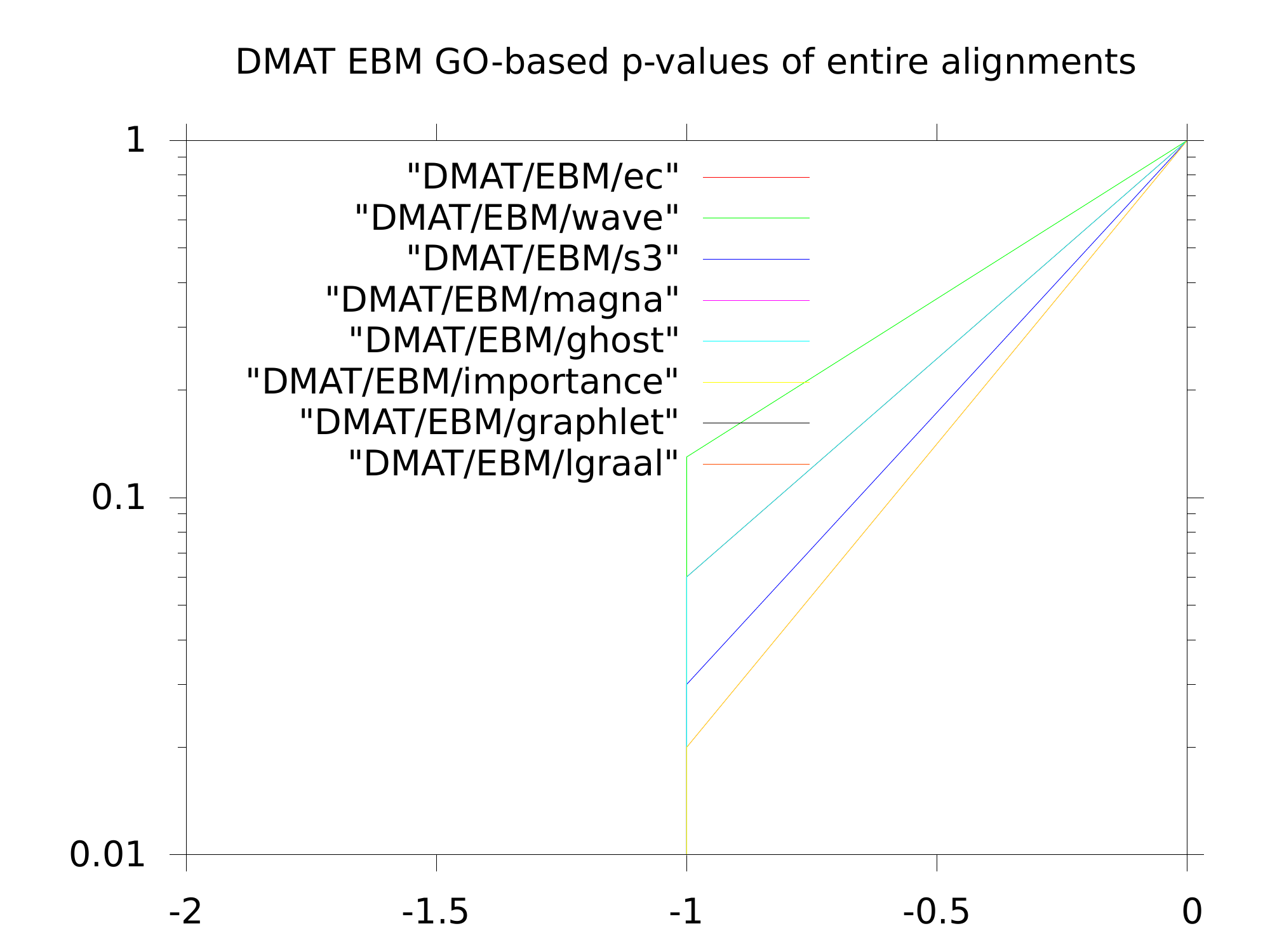}
\includegraphics[width=0.24 \textwidth]{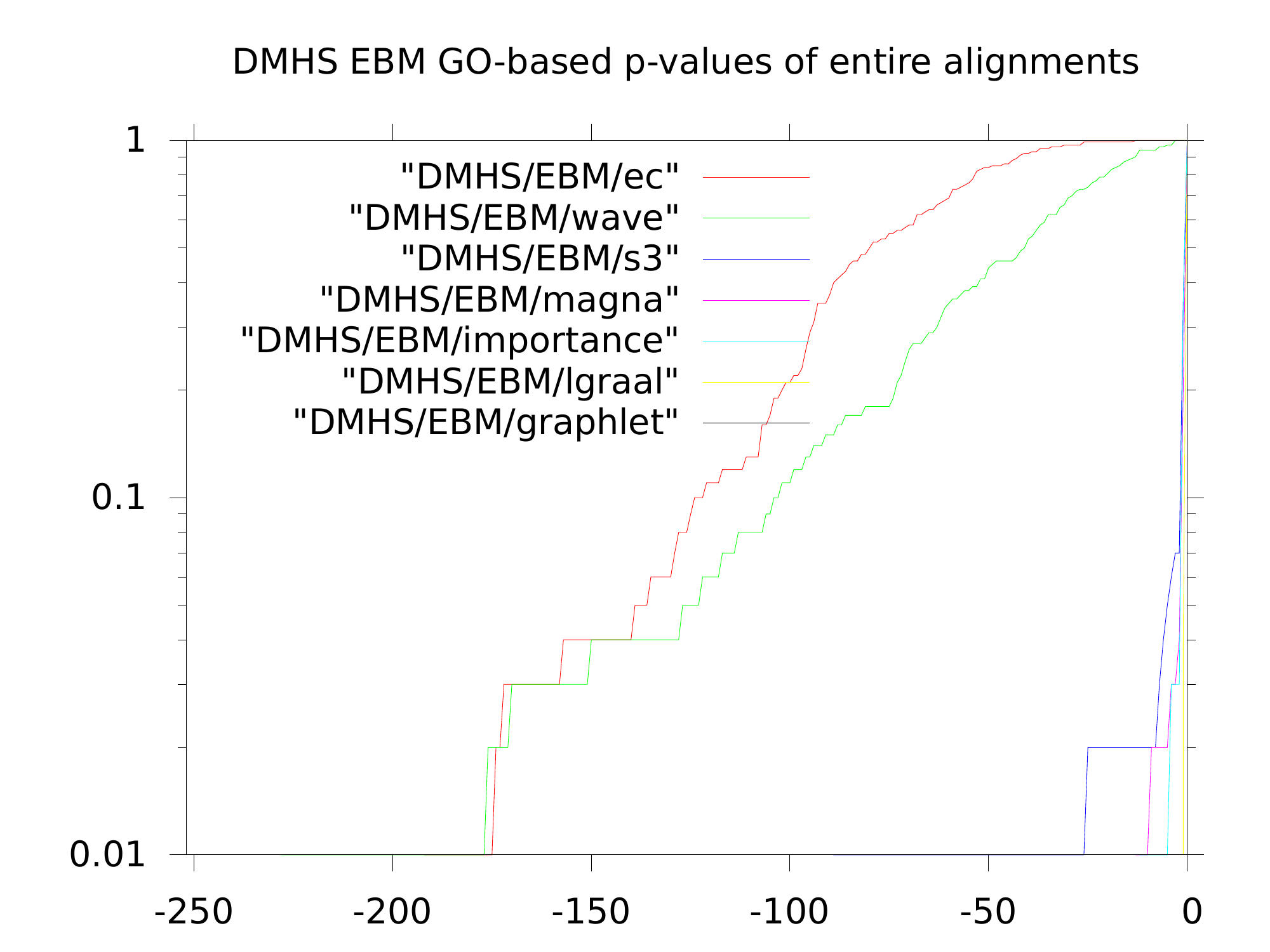}
\includegraphics[width=0.24 \textwidth]{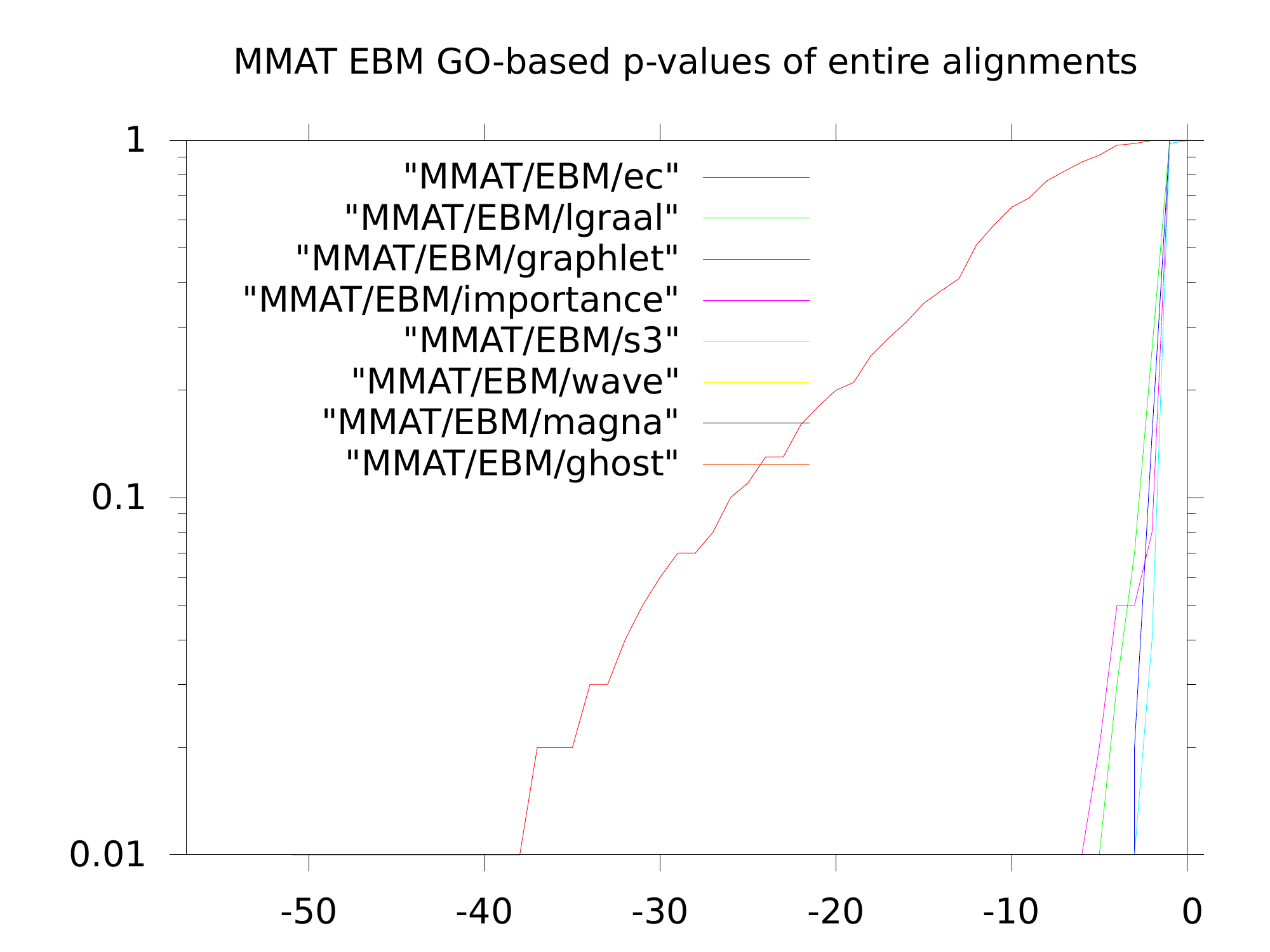}
\includegraphics[width=0.24 \textwidth]{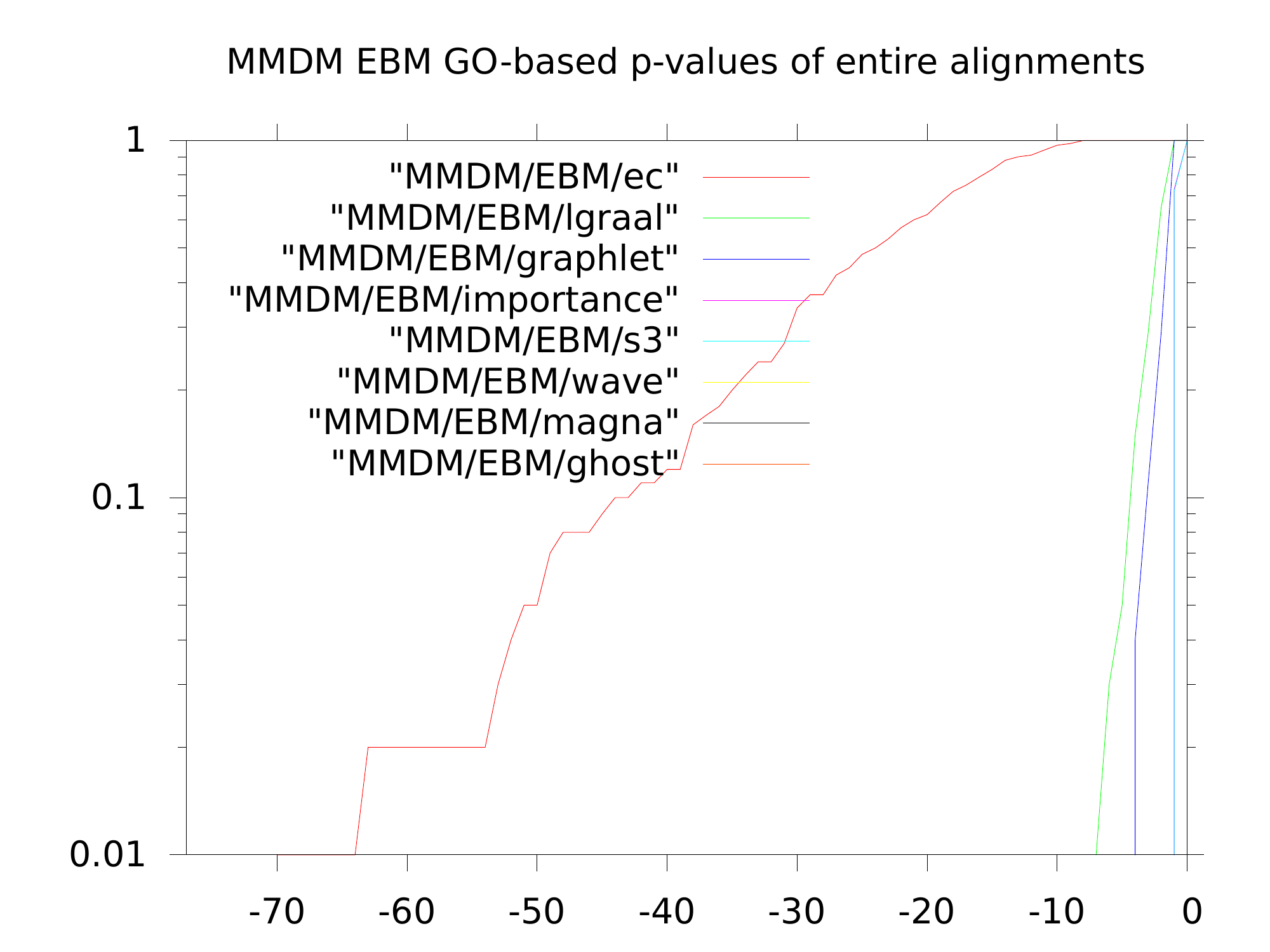}
\includegraphics[width=0.24 \textwidth]{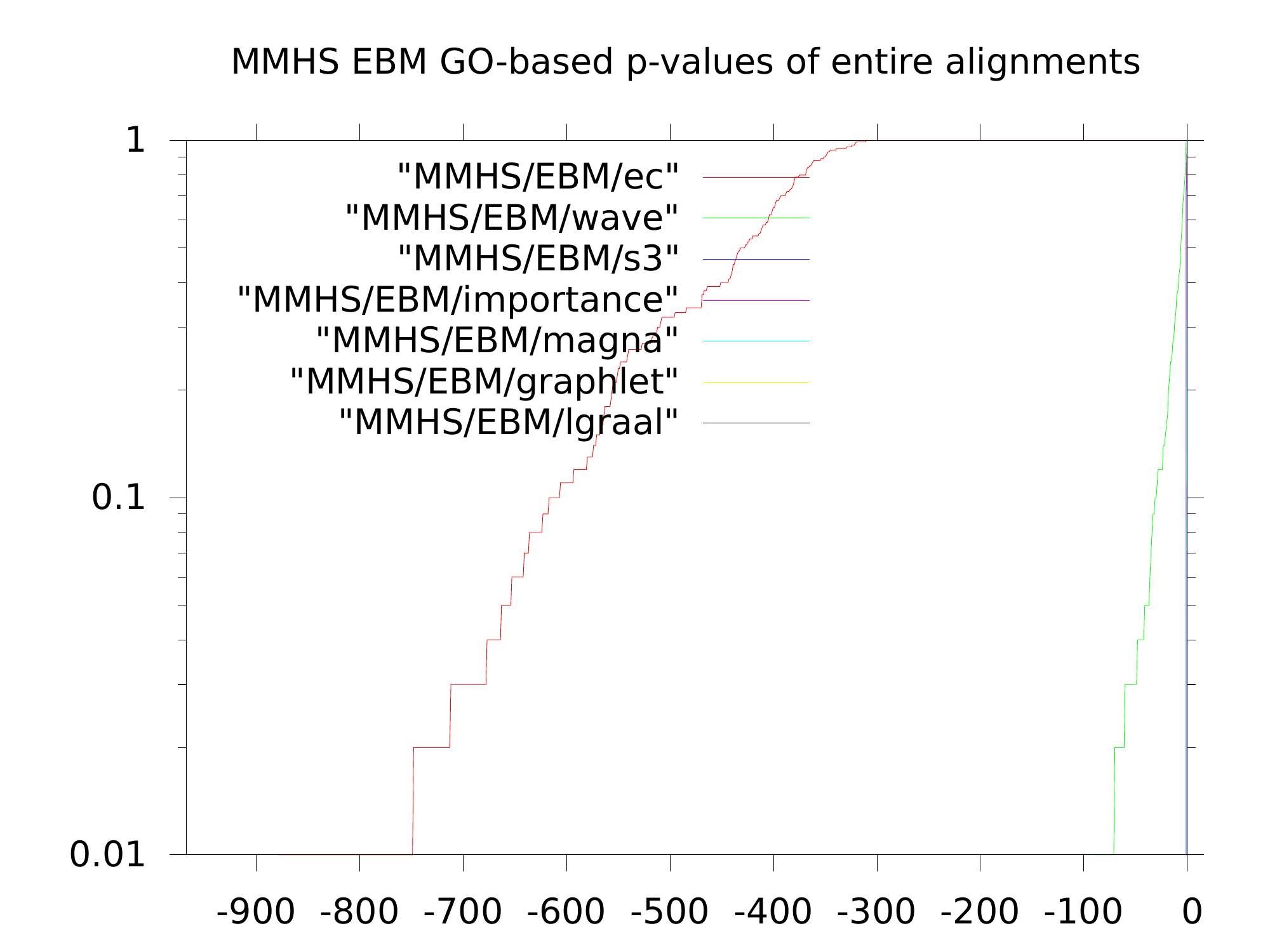}
\includegraphics[width=0.24 \textwidth]{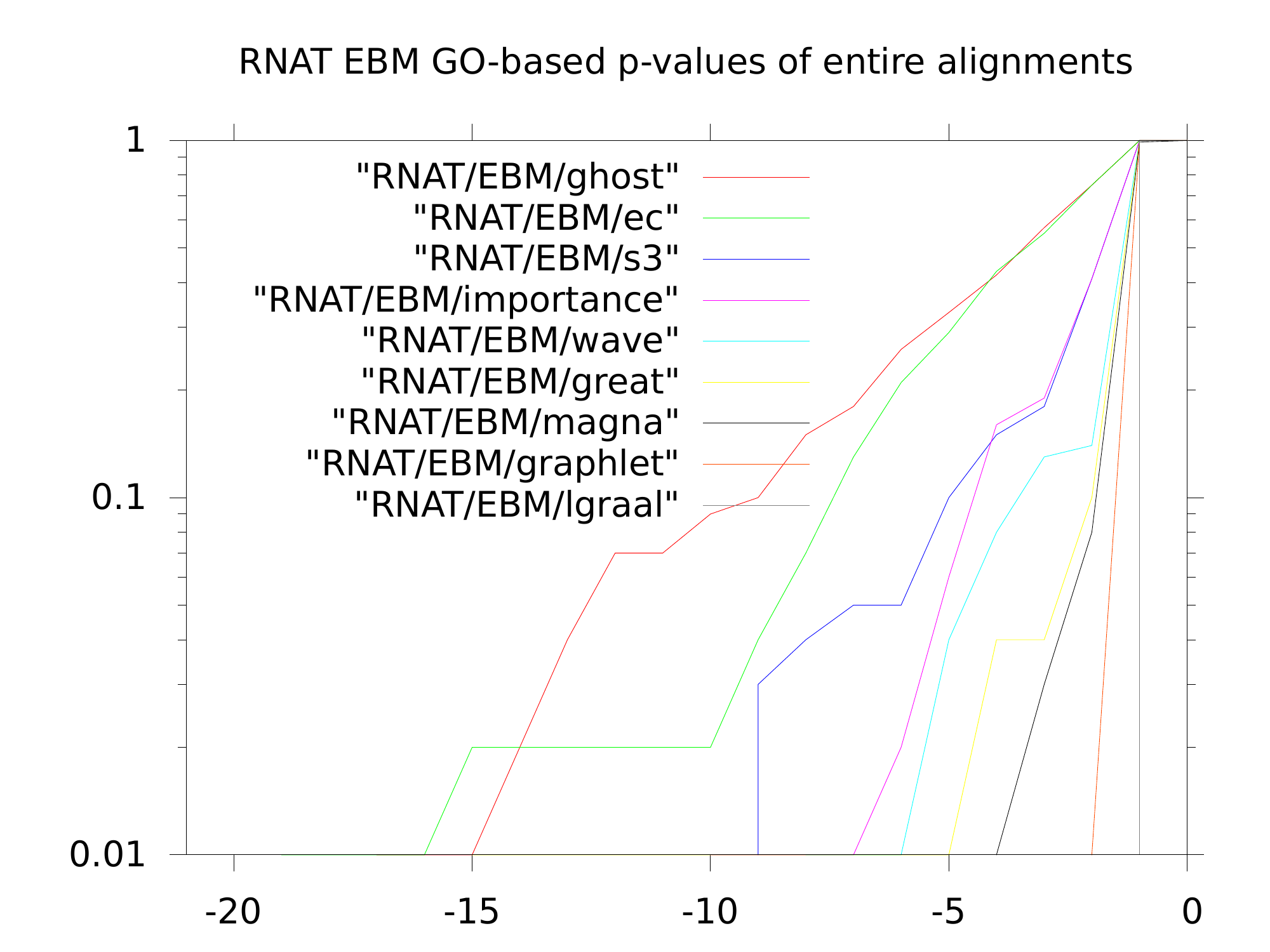}
\includegraphics[width=0.24 \textwidth]{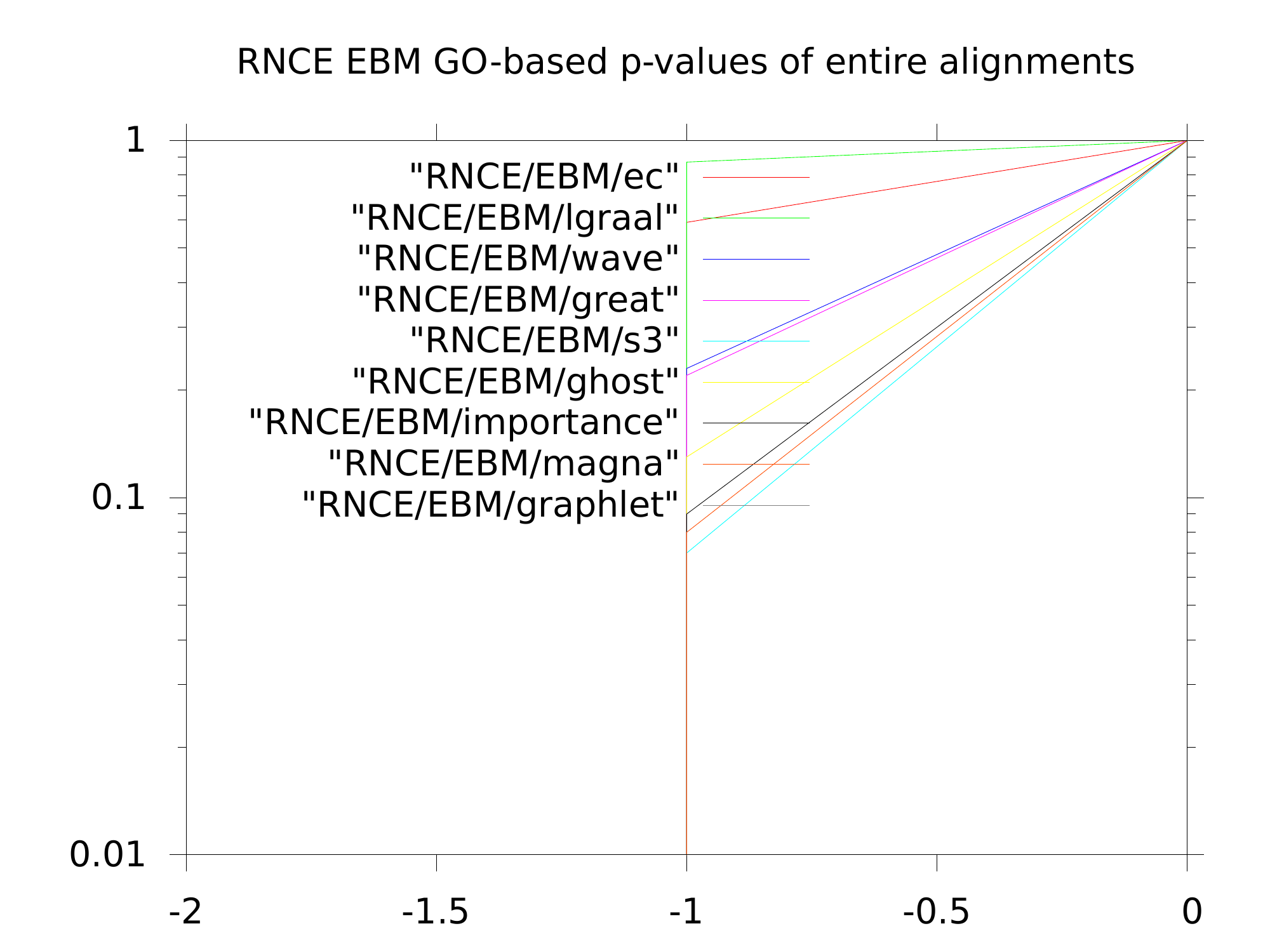}
\includegraphics[width=0.24 \textwidth]{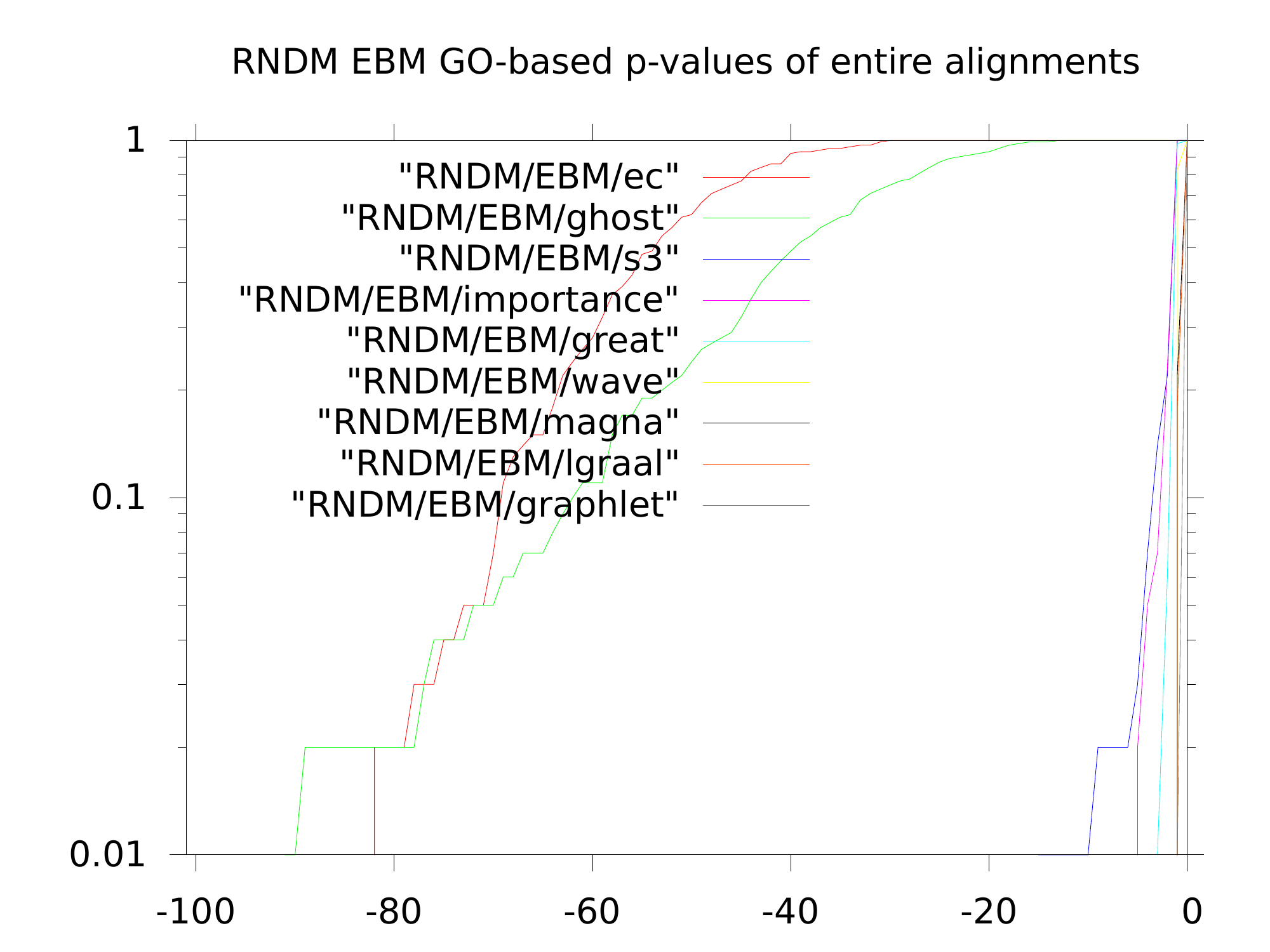}
\includegraphics[width=0.24 \textwidth]{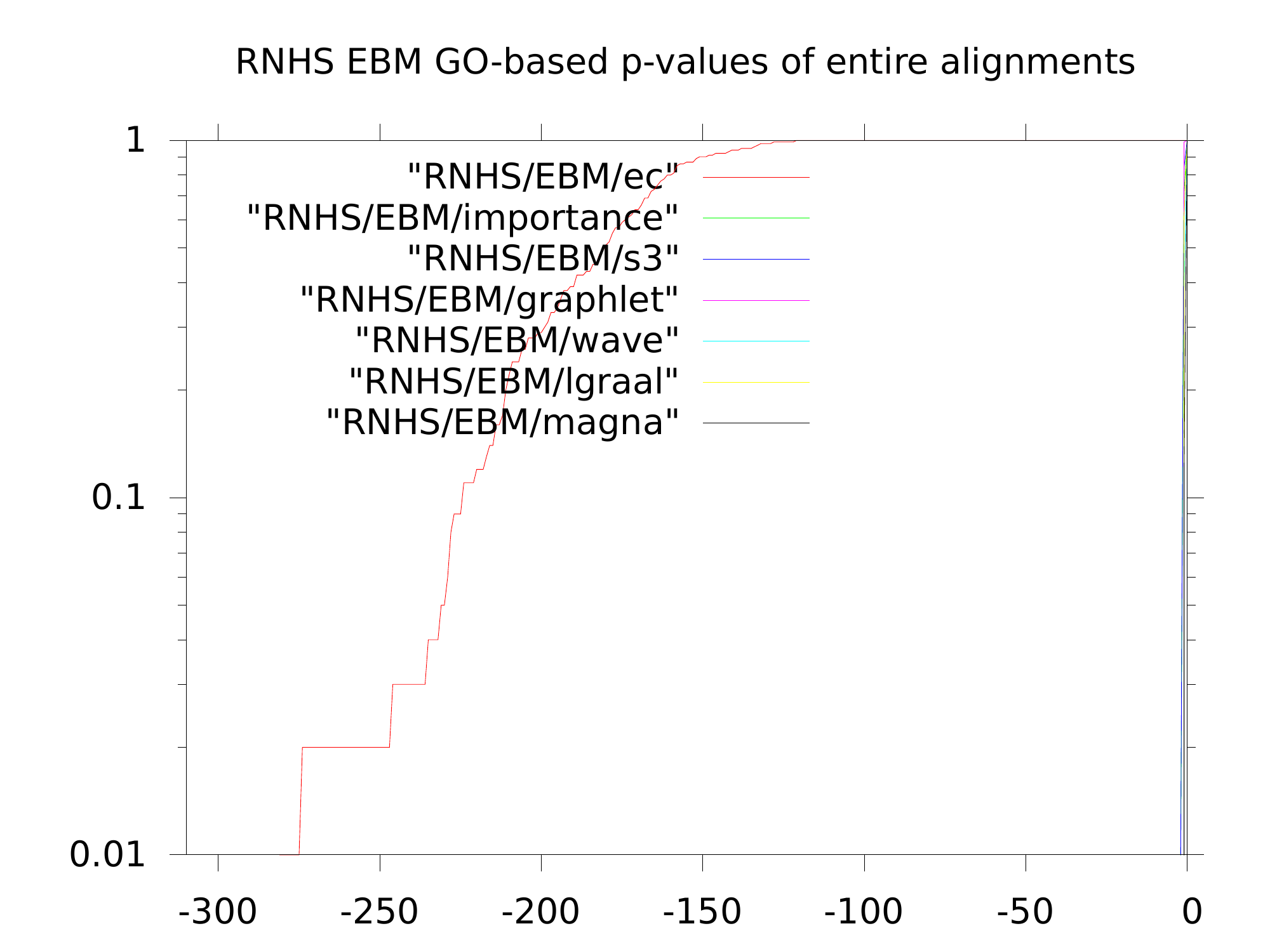}
\includegraphics[width=0.24 \textwidth]{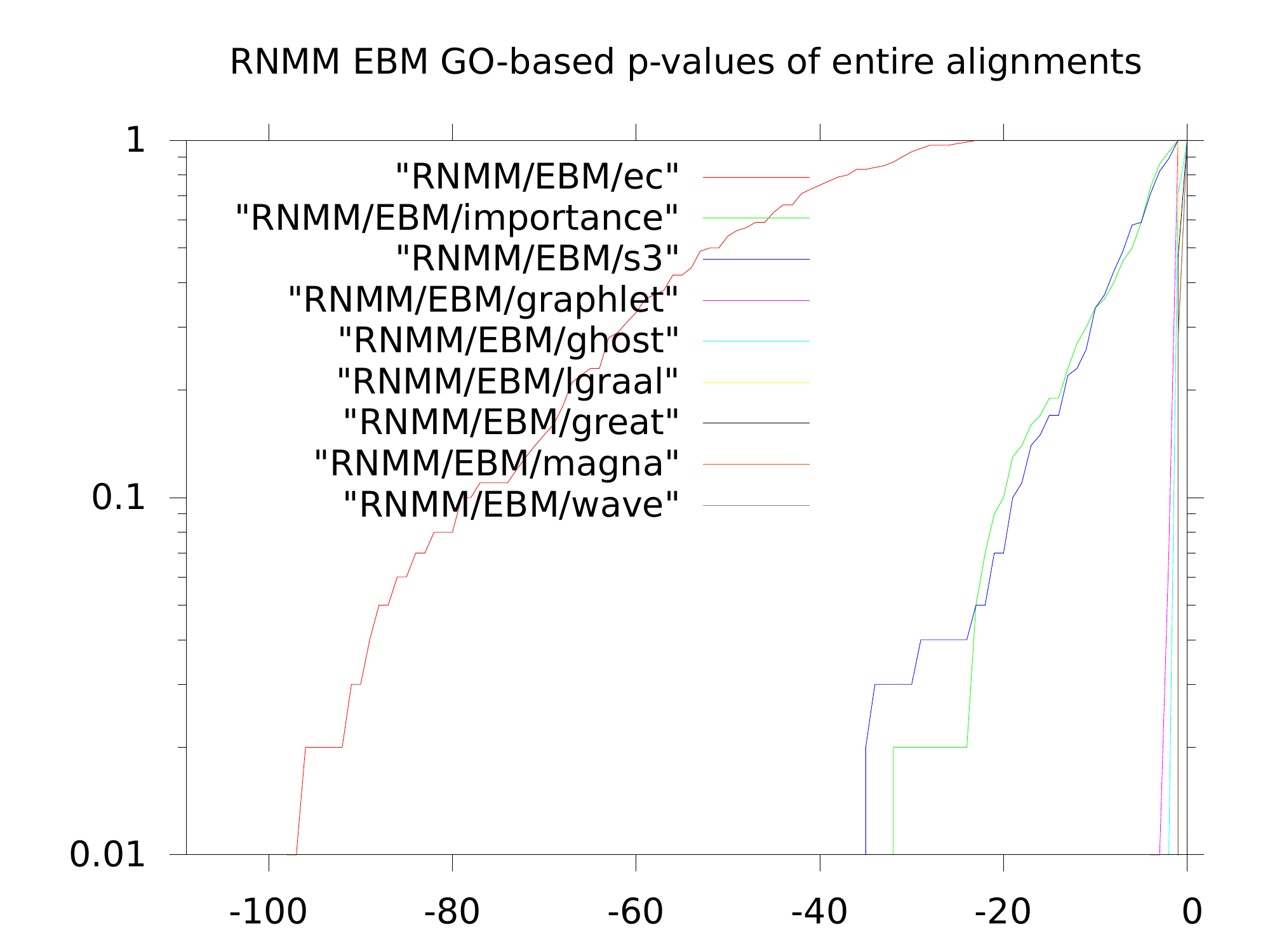}
\includegraphics[width=0.24 \textwidth]{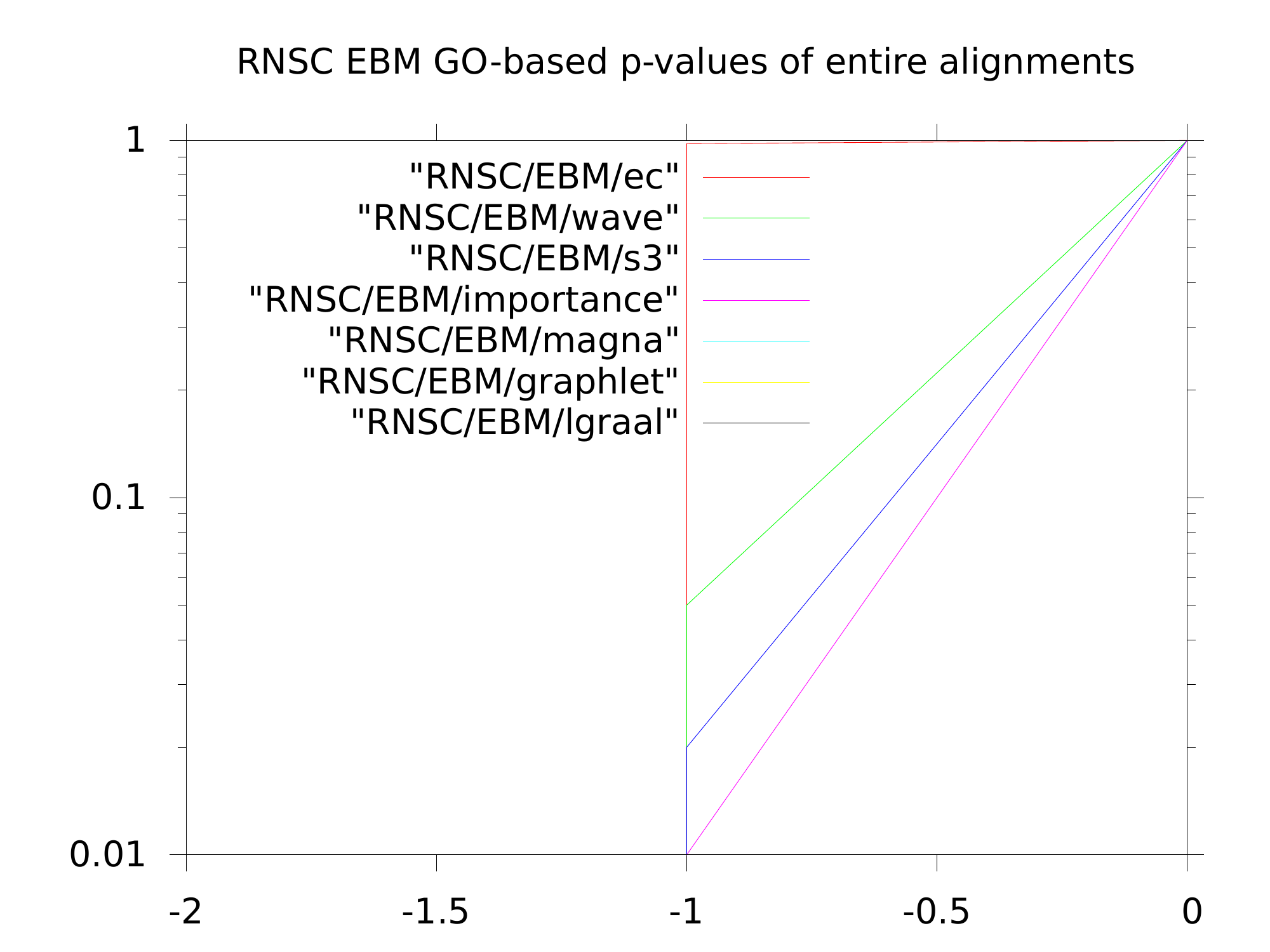}
\includegraphics[width=0.24 \textwidth]{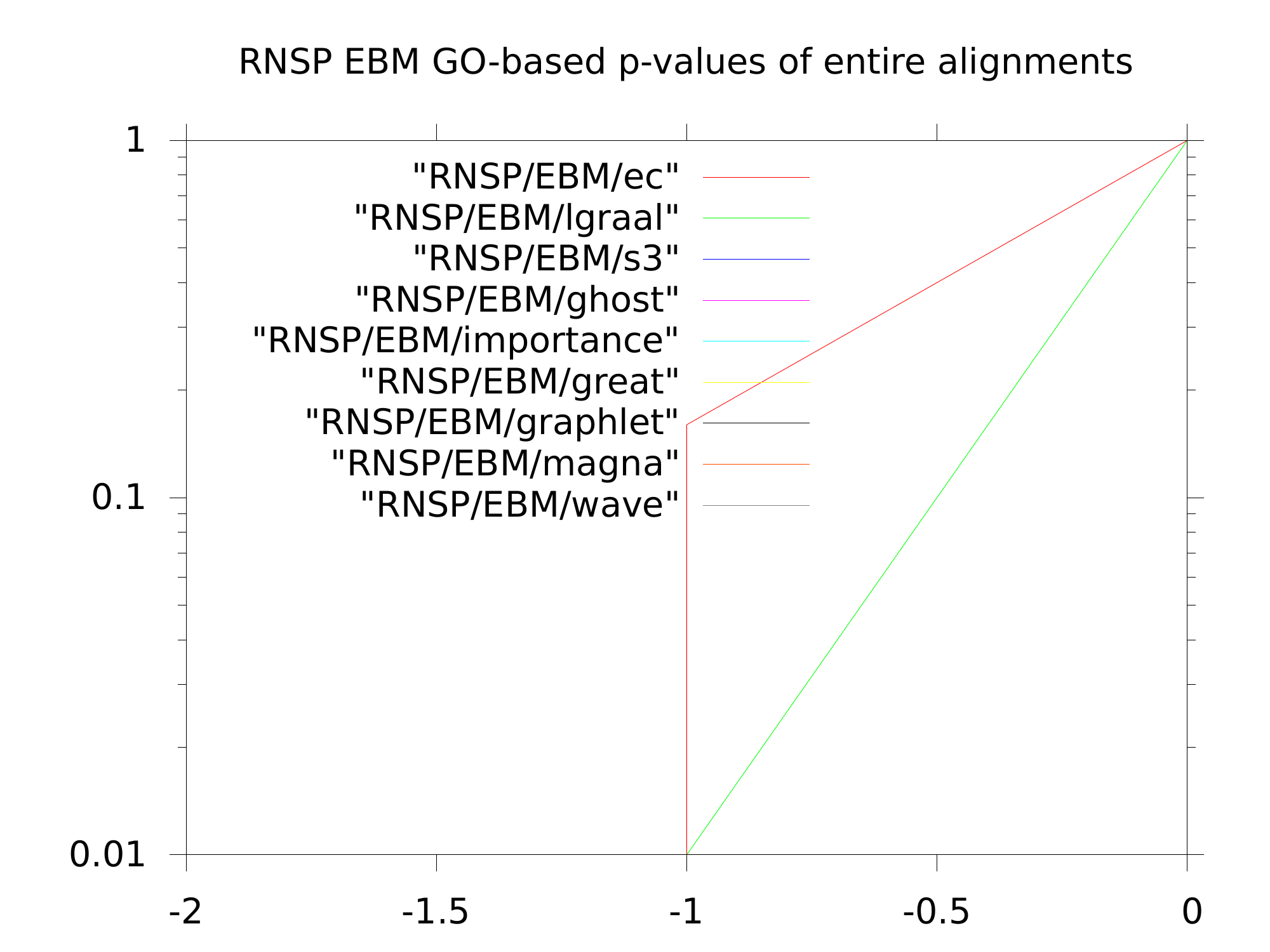}
\includegraphics[width=0.24 \textwidth]{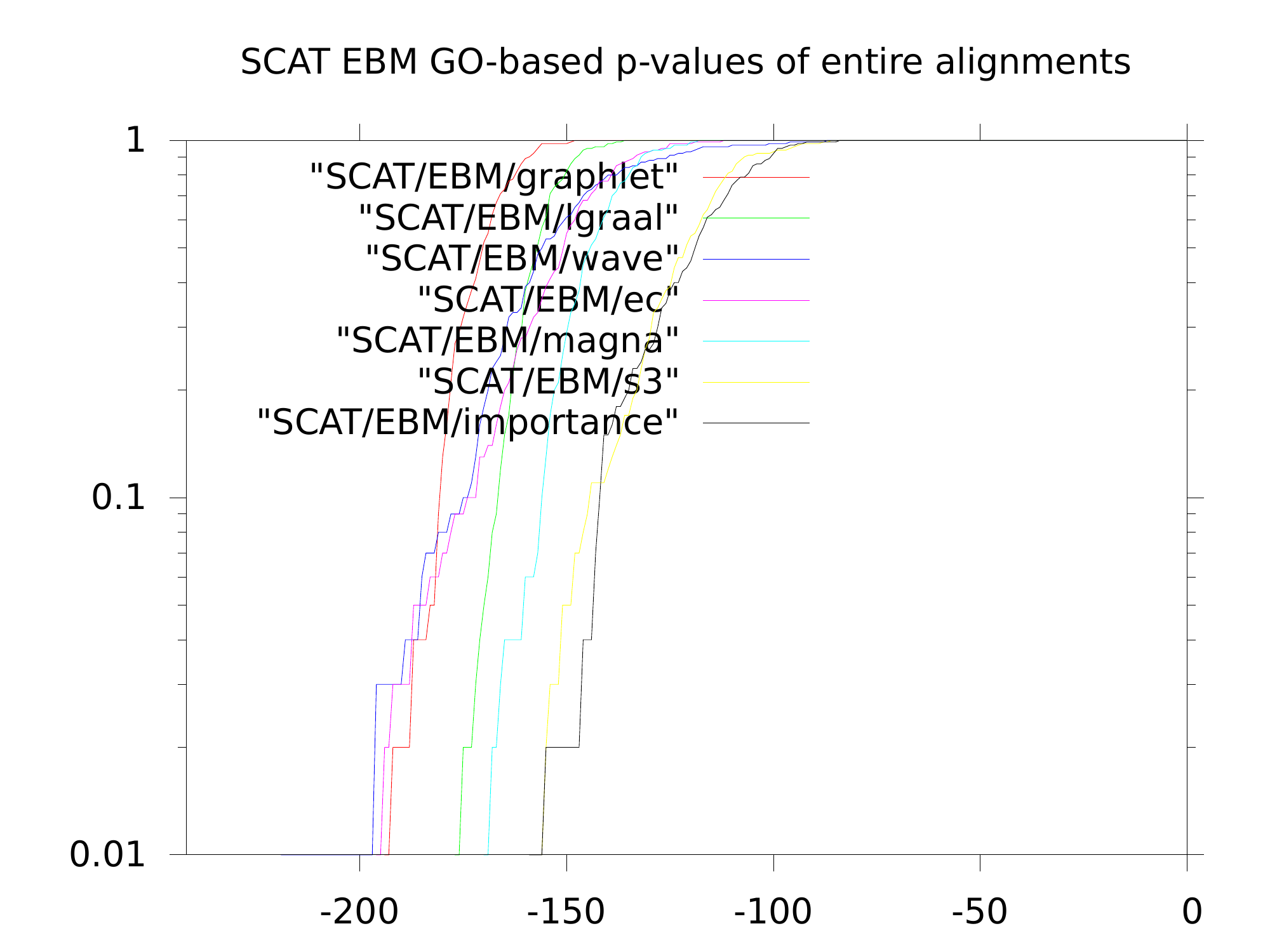}
\includegraphics[width=0.24 \textwidth]{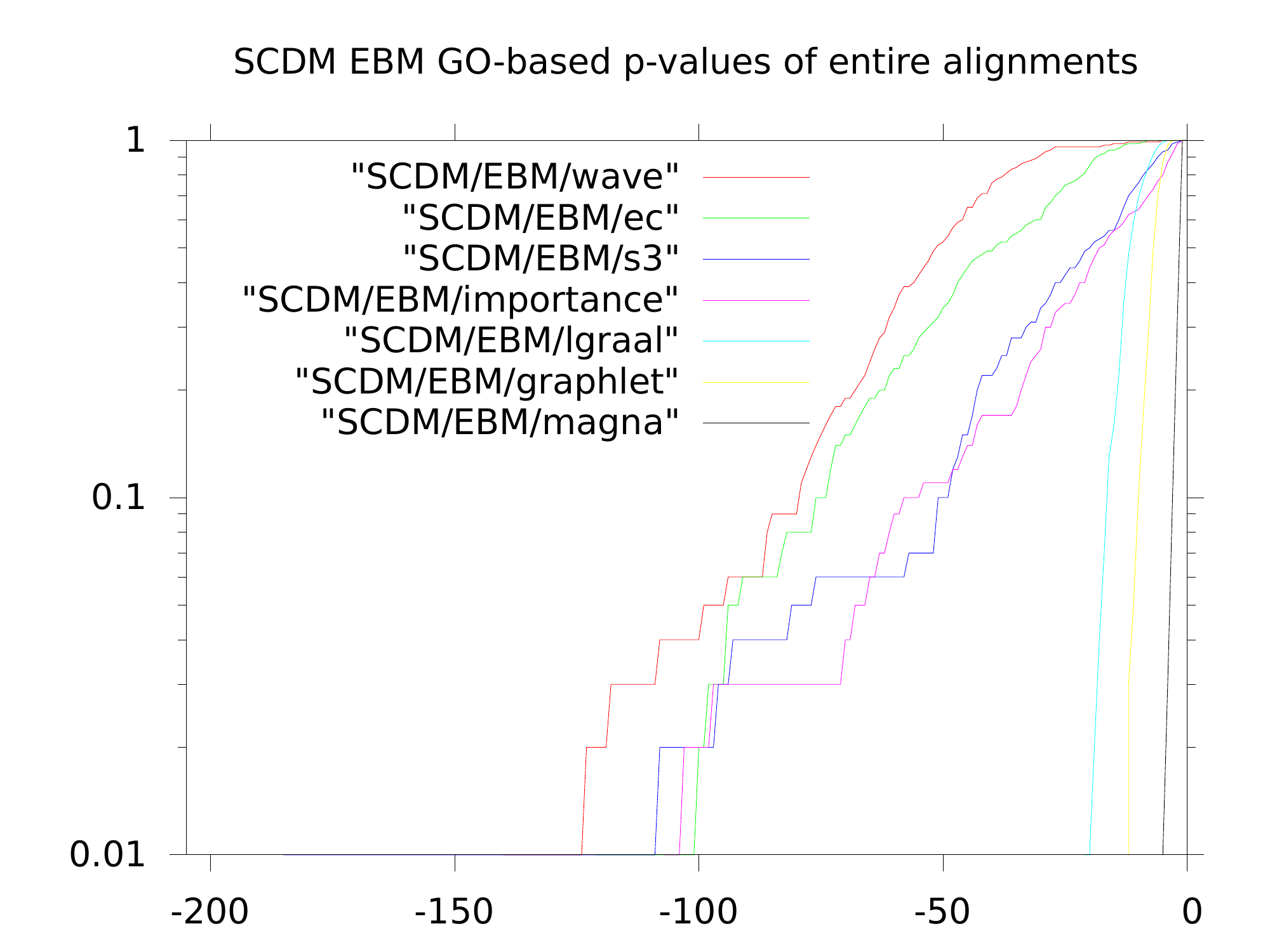}
\includegraphics[width=0.24 \textwidth]{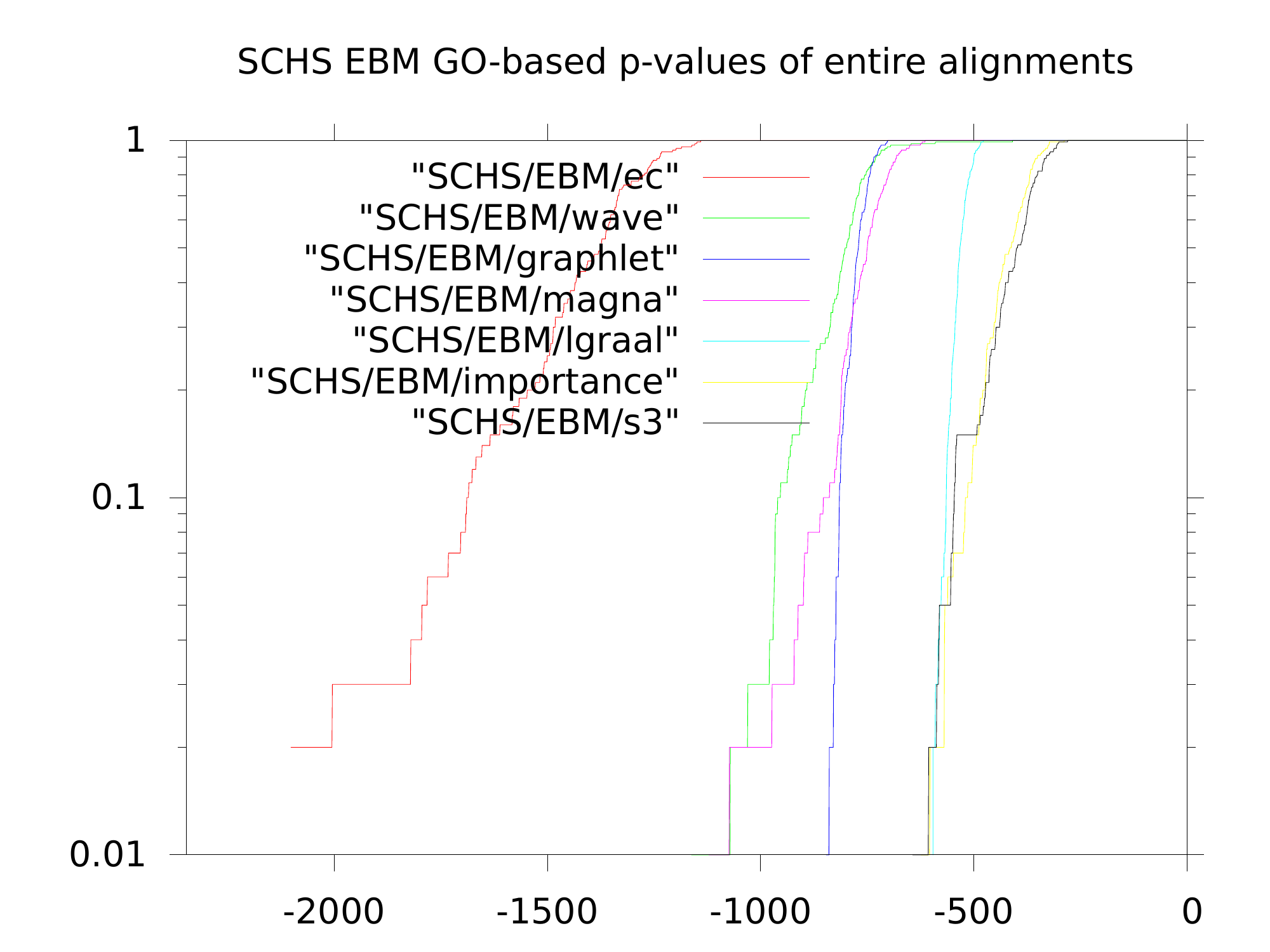}
\includegraphics[width=0.24 \textwidth]{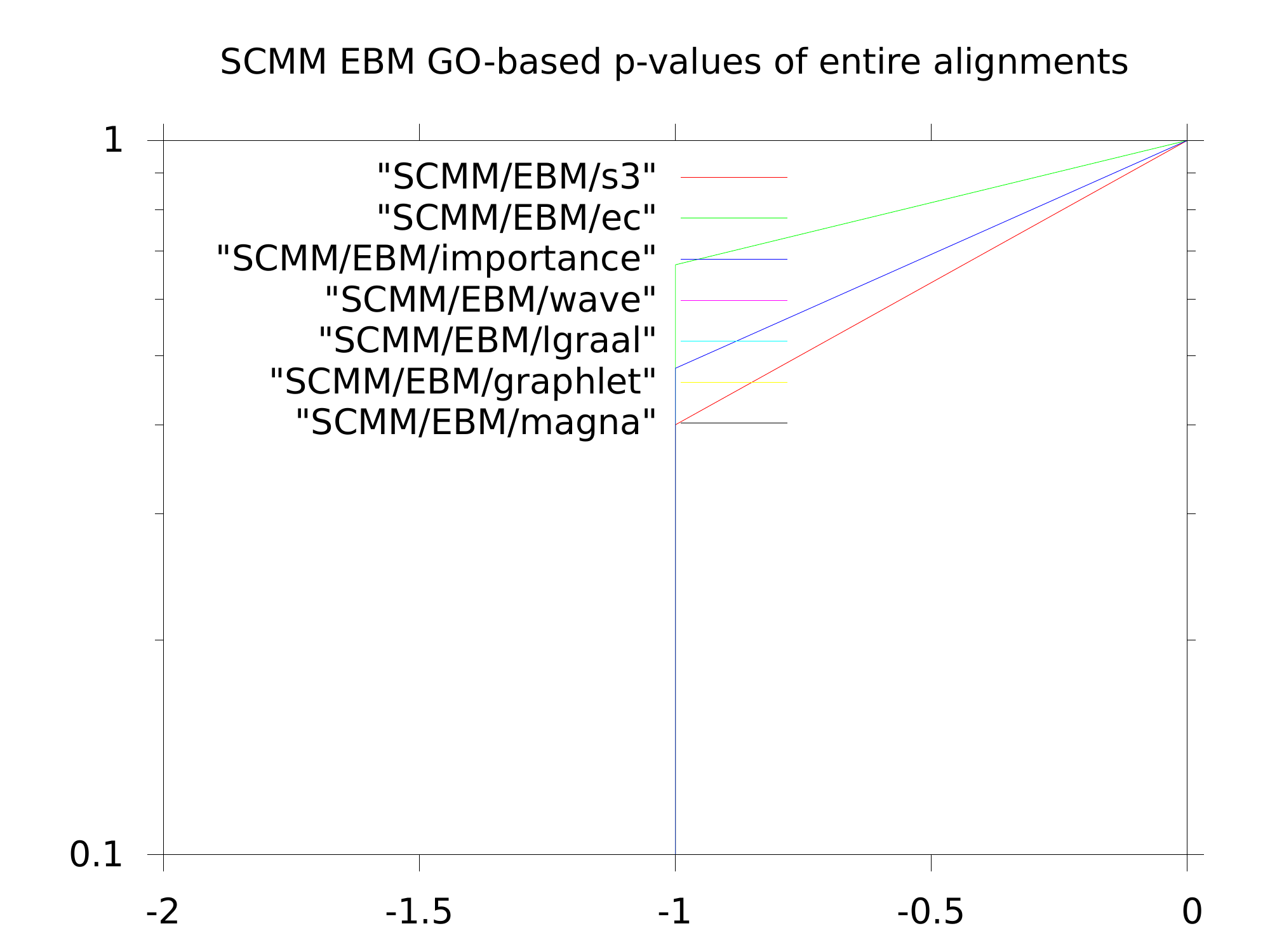}
\includegraphics[width=0.24 \textwidth]{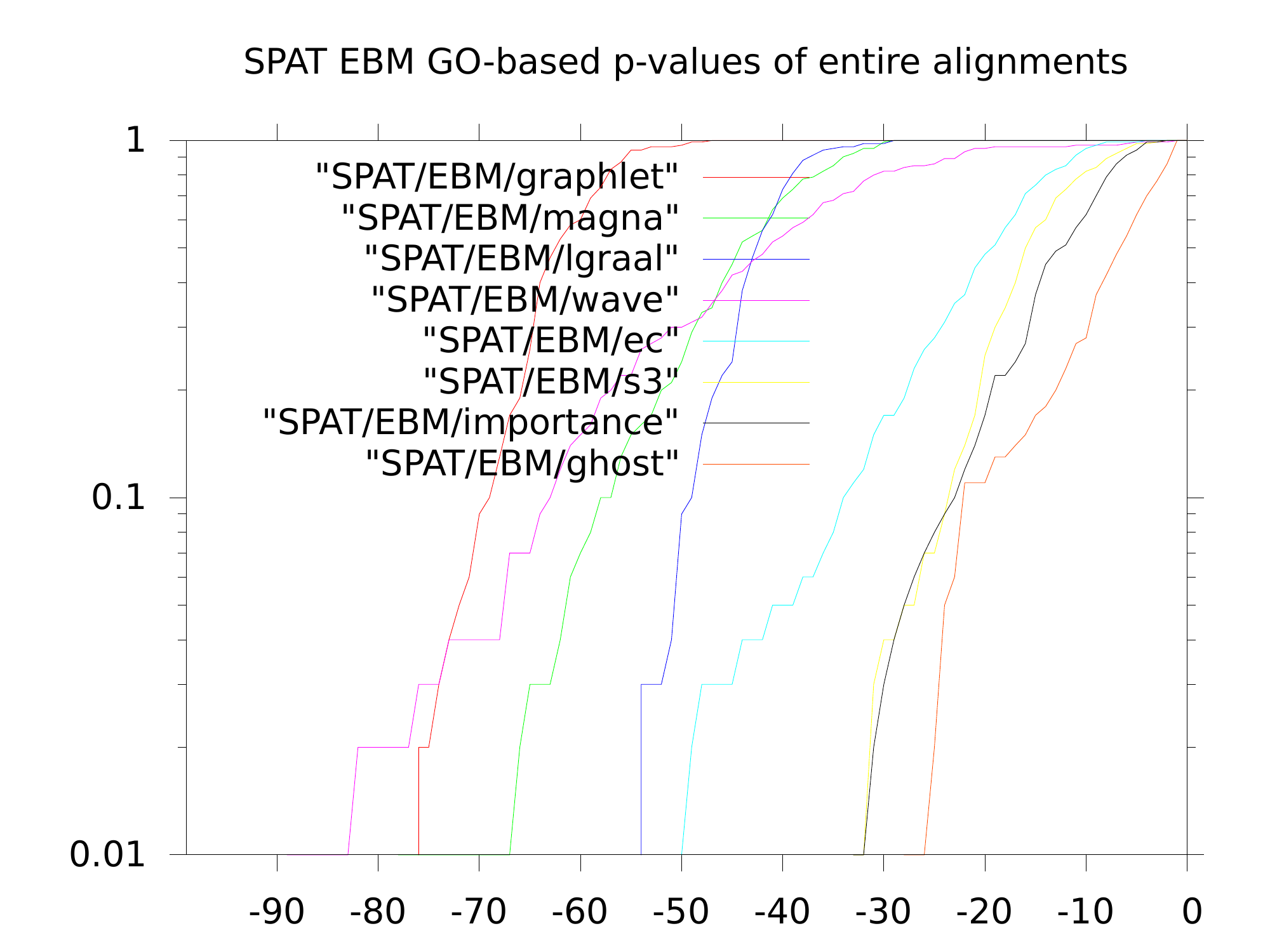}
\includegraphics[width=0.24 \textwidth]{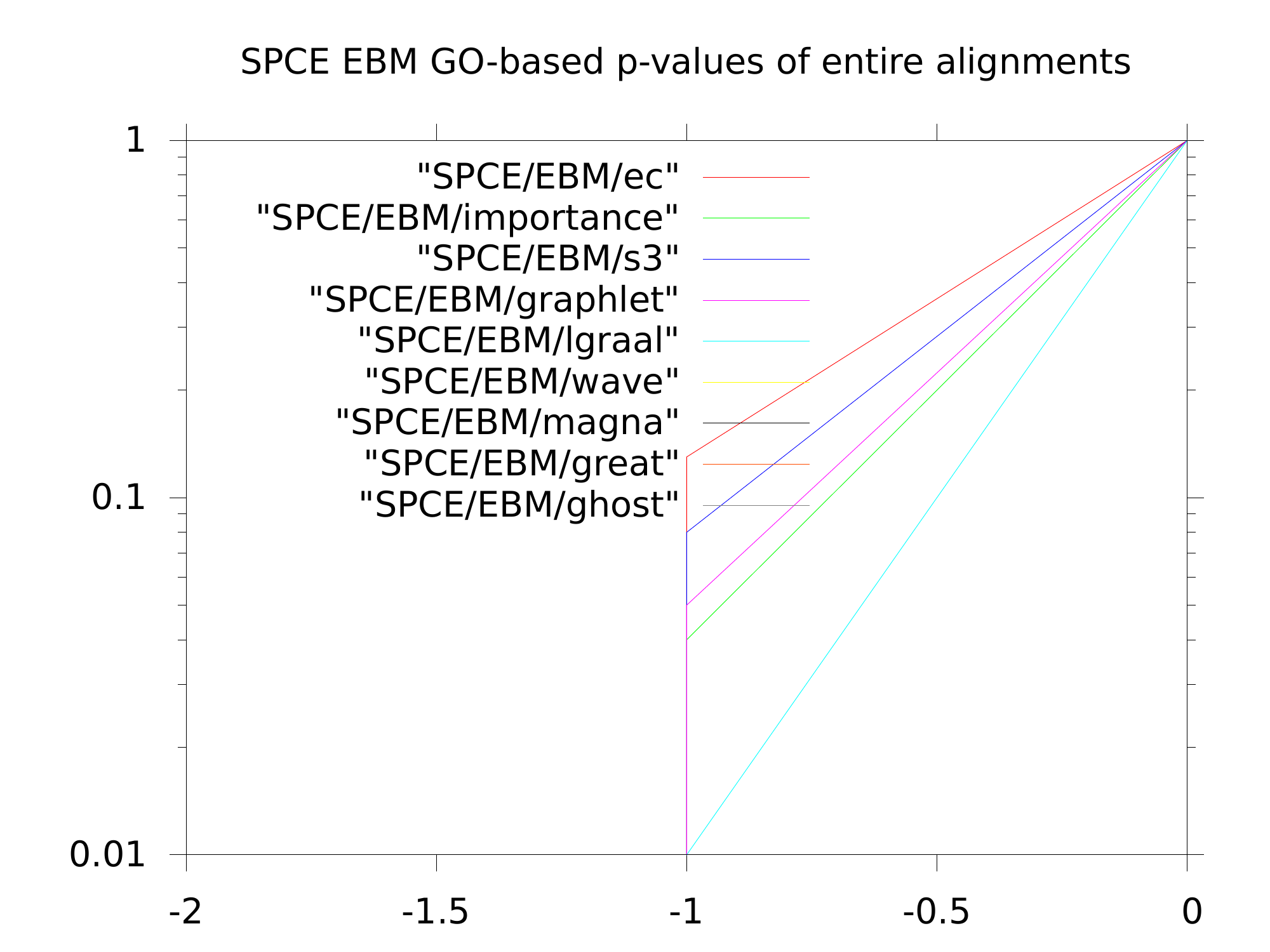}
\includegraphics[width=0.24 \textwidth]{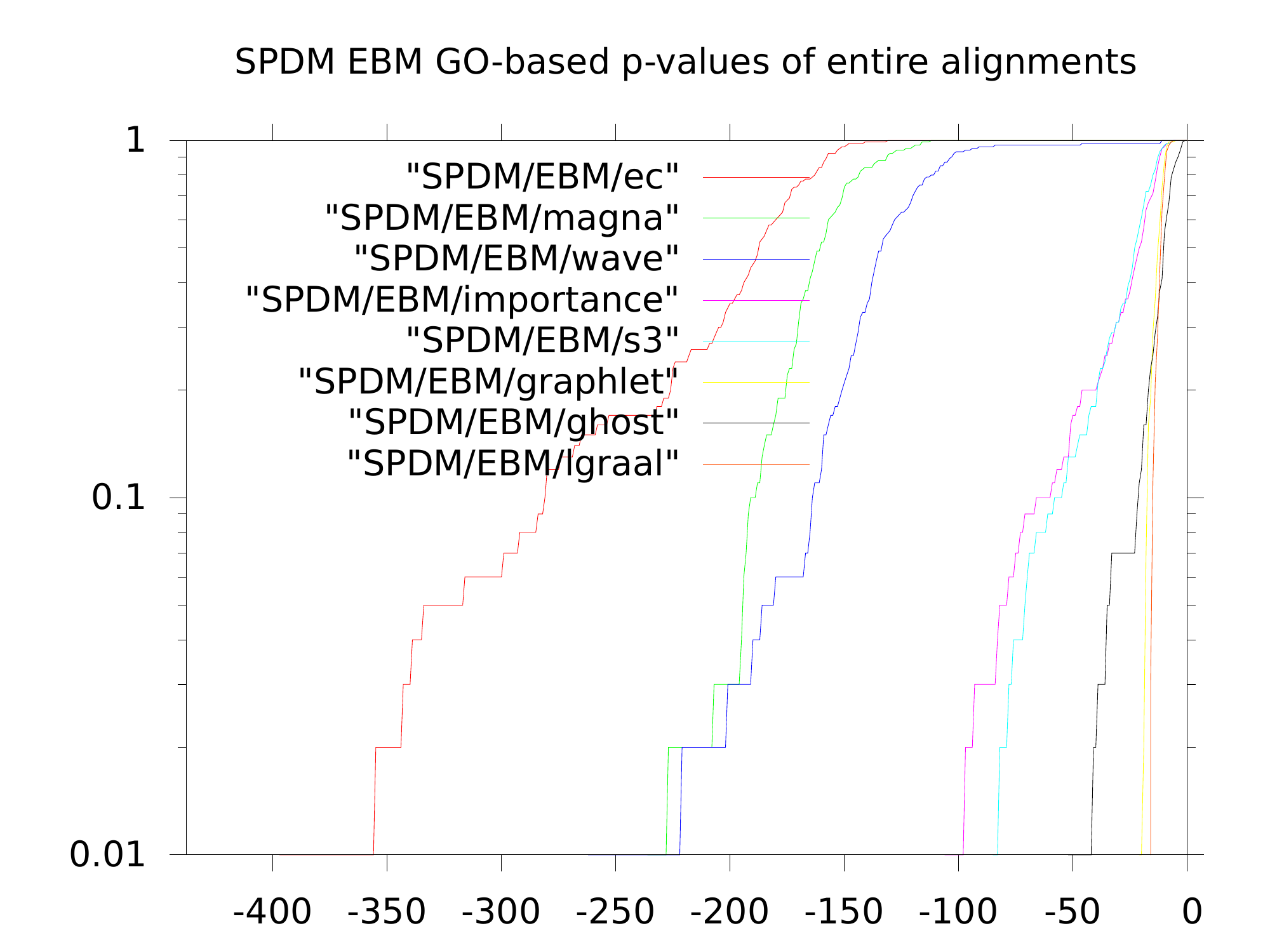}
\includegraphics[width=0.24 \textwidth]{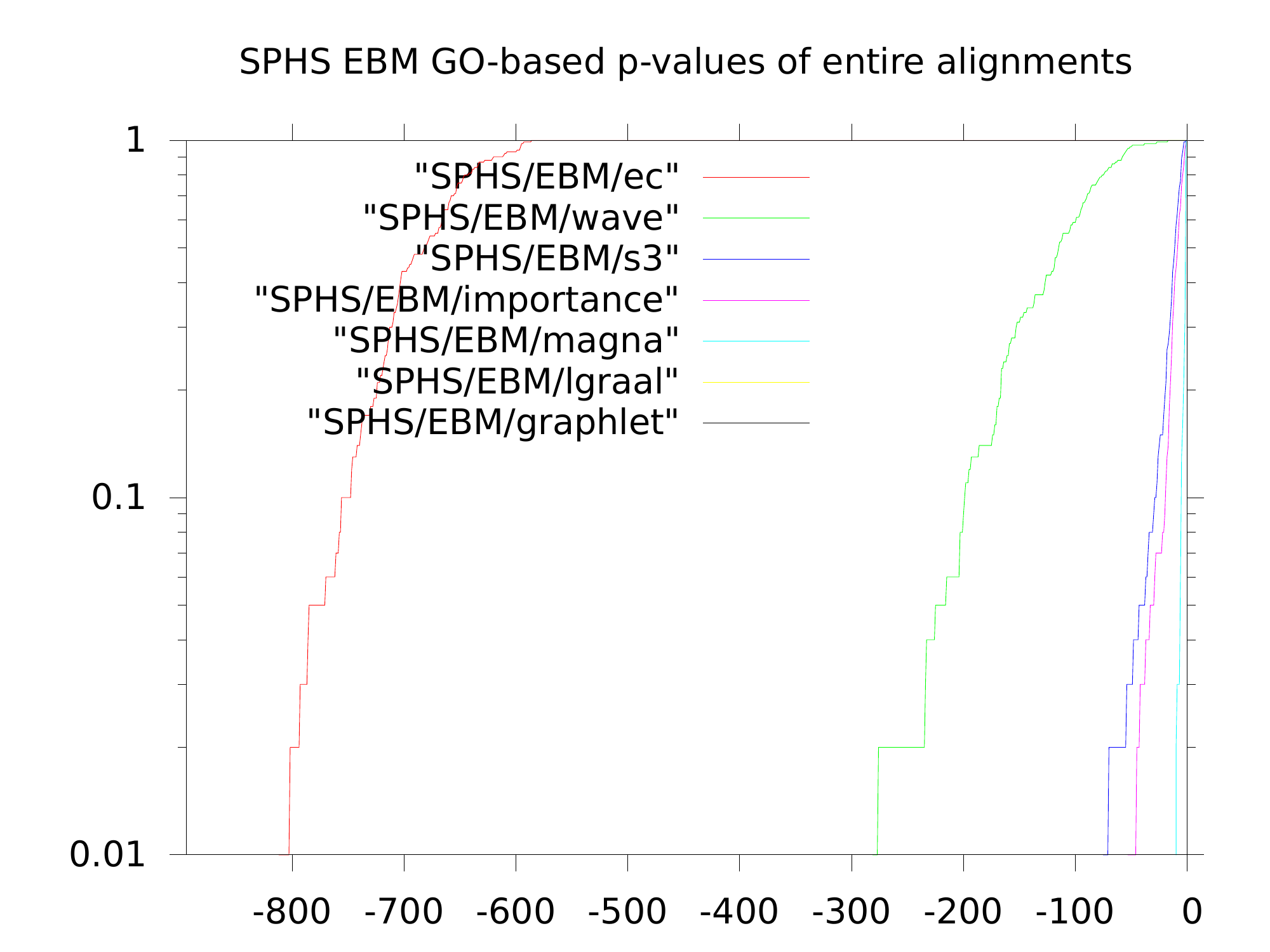}
\includegraphics[width=0.24 \textwidth]{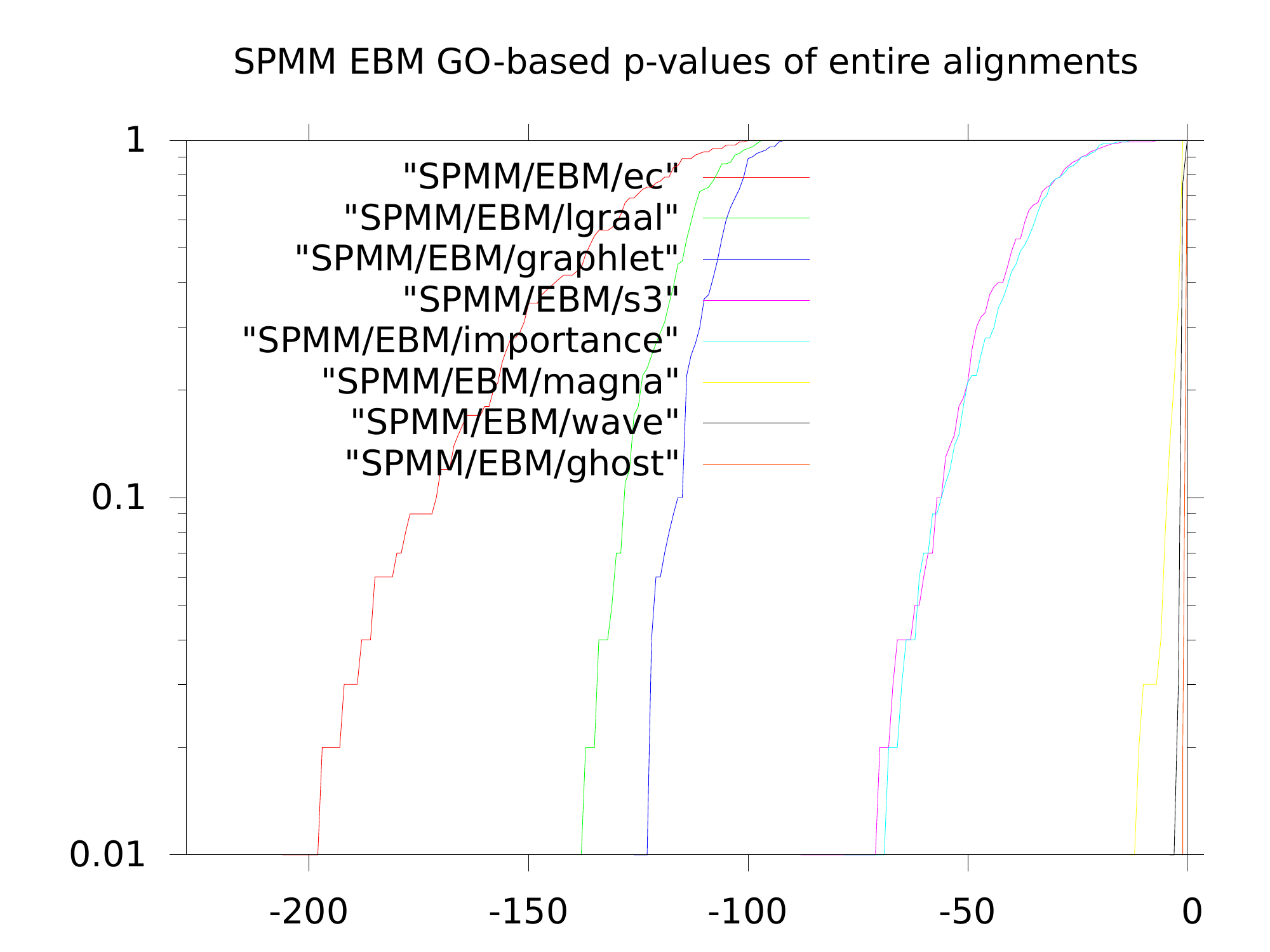}
\includegraphics[width=0.24 \textwidth]{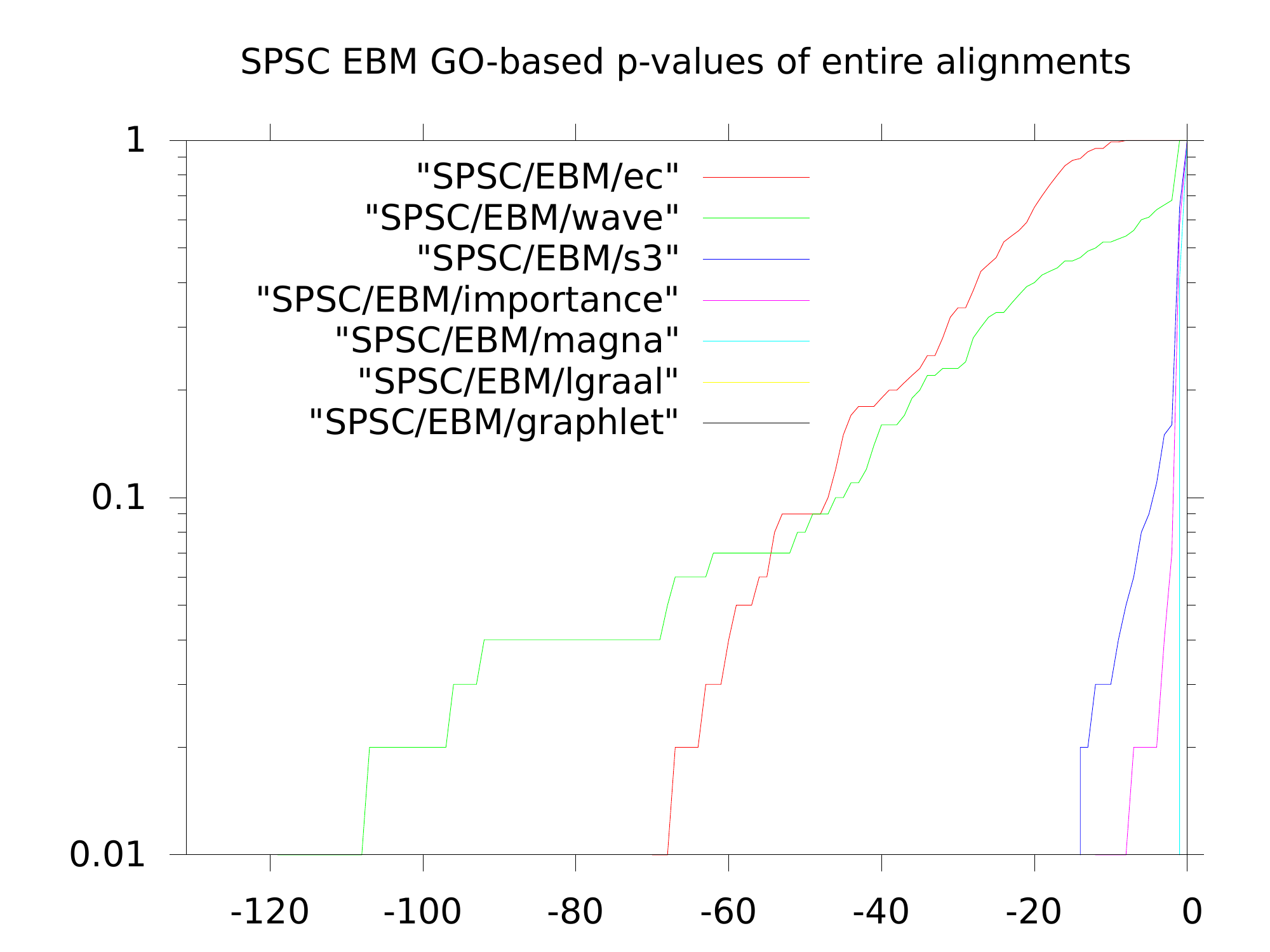}
\caption{Similar to Figure \ref{fig:EBMhist}, but for all network pairs (alphabetical order).
Note the horizontal axis is different for each pair.}
    \label{fig:EBMhistALL}
\end{figure}

\subsubsection{Degree distribution of recovered vs. non-recovered orthologs}\label{sec:orthoDegree}
\begin{figure}[tb]
    \centering
    \includegraphics[width=0.4 \textwidth]{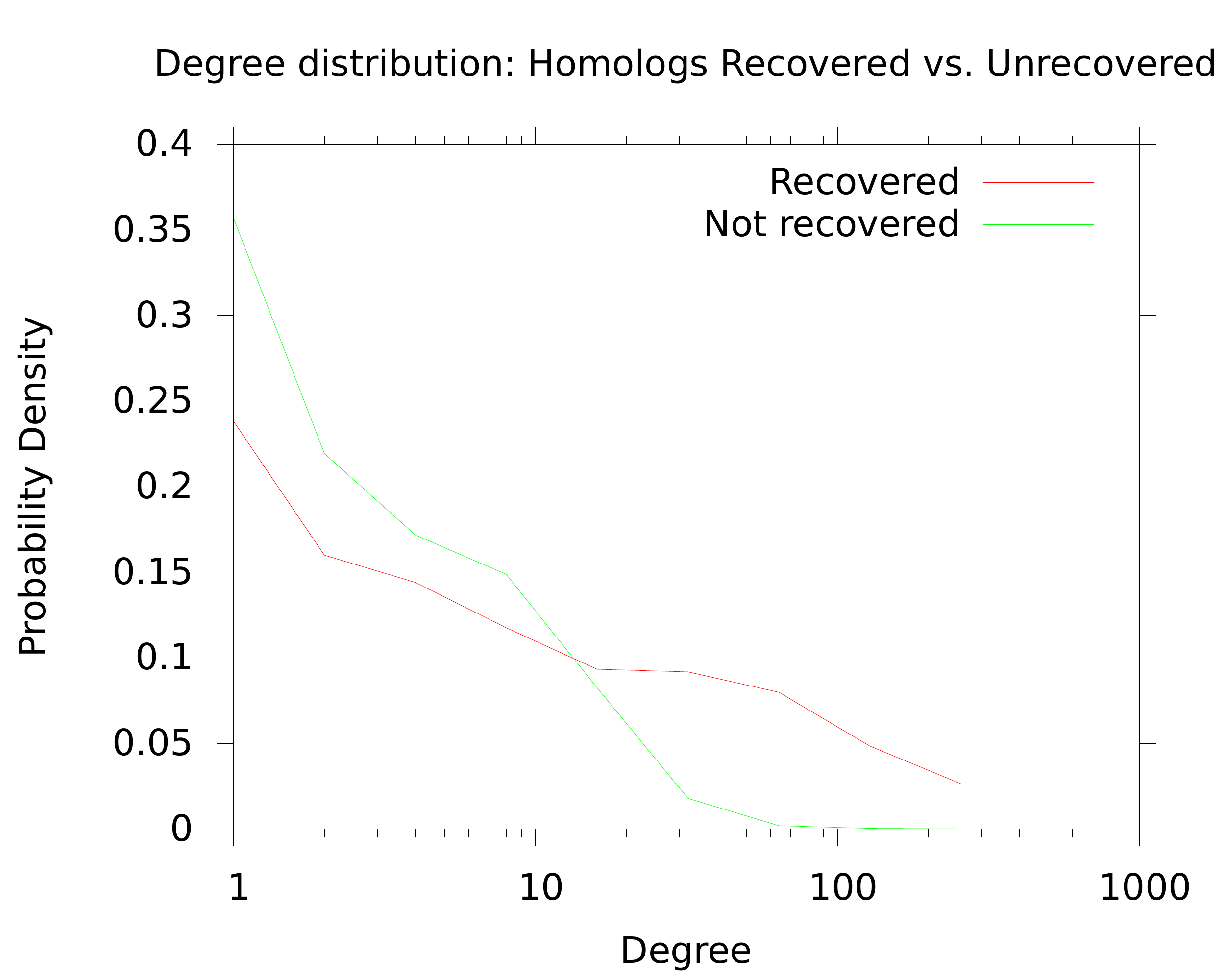}
    \caption{{\bf Degree distribution of recovered vs. non-recovered orthologs} We see that orthologs that were recovered have a degree distribution skewed towards the right (ie., higher degree on average) compared to those that were not recovered, which reinforces the conclusion that higher density regions are easier to align ``correctly''.
    }
    \label{fig:degree-orthologs-recovered}
\end{figure}

Figure \ref{fig:degree-orthologs-recovered} plots the degree distribution of orthologs that were recovered, vs. those that were not. We see that correctly aligned orthologs tend to have significantly higher degree than those that were not aligned: almost a third of non-recovered orthologs have degree 1, whereas less than a quarter of recovered ones have degree 1; the curves cross at about degree 15. The total density above degree 40 is more than an order of magnitude higher for recovered orthologs than non-recovered ones, again demonstrating that network alignments improve with increasing edge density.

\end{document}